\documentclass[11pt]{article}

\usepackage{mathtools}
\usepackage{amsmath}
\RequirePackage{amssymb}
\RequirePackage{amsthm}
\usepackage{amsmath,scalefnt}
\usepackage{footnote}
\usepackage{fmtcount}
\usepackage{threeparttable}
\usepackage{caption}
\usepackage{subcaption}
\usepackage{color}
\usepackage{tikz}
\usetikzlibrary{positioning,chains,fit,shapes,calc}
\usetikzlibrary{arrows}
\tikzstyle{vertex}=[circle, draw, inner sep=0pt, minimum size=6pt]
 \usepackage{enumitem}
 \usepackage{graphics,graphicx}

\newcommand{\arrowchem}[1]{\xrightarrow{#1}}
\newcommand{\arrowschem}[2]{\xrightleftharpoons[#1]{#2}}

\newcommand{\abs}[1]{\left\vert #1 \right\vert}
\newcommand{\norm}[1]{\left\Vert #1 \right\Vert}
\newcommand{\normpQ}[1]{\norm{#1}_{p,Q}}
\newcommand{\MpQ}{M_{p,Q}}

\newcommand{\notforthesis}[1]{}

	\newtheorem{definition}{Definition}
	\newtheorem{claim}{Claim}
	\newtheorem{lemma}{Lemma}

	 \newtheorem{proof of proposition}{Proof of Proposition}
	\newtheorem{proof of claim}{Proof of Claim}
	\newtheorem{proposition}{Proposition} 
	\newtheorem{theorem}{Theorem}
	\newtheorem{notation}{Notation}
        \newtheorem{corollary}{Corollary}
        \newtheorem{remark}{Remark}

\newcommand{\diag}{\mbox{diag}\,}

\renewcommand{\r}{{\mathbb R}}
\renewcommand{\a}{{\alpha}}
\renewcommand{\l}{{\lambda}}
\renewcommand{\b}{{\beta}}
\renewcommand{\d}{{\delta}}
\renewcommand{\o}{{\omega}}
\newcommand{\e}{{\epsilon}}
\newcommand{\ee}{\end{equation}}
\newcommand{\bal}{\begin{aligned}}
\newcommand{\eal}{\end{aligned}}
\newcommand{\bi}{\begin{itemize}}
\newcommand{\ei}{\end{itemize}}
\newcommand{\ben}{\begin{enumerate}}
\newcommand{\een}{\end{enumerate}}
\newcommand{\beqn}{\begin{eqnarray*}}
\newcommand{\eeqn}{\end{eqnarray*}}

\newcommand{\be}[1]{\begin{equation}\label{#1}}
\newcommand{\bp}{\begin{proof}}
\newcommand{\ep}{\end{proof}}
\newcommand{\bremark}{\begin{remark}\rm } 
\newcommand{\eremark}{\end{remark}}
\newcommand{\blem}{\begin{lemma}}
\newcommand{\elem}{\end{lemma}}
\newcommand{\bclaim}{\begin{claim}}
\newcommand{\eclaim}{\end{claim}}
\newcommand{\bnote}{\begin{notation}}
\newcommand{\enote}{\end{notation}}

\newcommand{\bthm}{\begin{theorem}}
\newcommand{\ethm}{\end{theorem}}
\newcommand{\bprop}{\begin{proposition}}
\newcommand{\eprop}{\end{proposition}}
\newcommand{\bcor}{\begin{corollary}}
\newcommand{\ecor}{\end{corollary}}
\newcommand{\dis}{\displaystyle}
\newcommand{\lt}{\left}
\newcommand{\rt}{\right}

\newcommand{\fract}[2]{#1/#2} 

\title{Some remarks on spatial uniformity of solutions of
    reaction-diffusion PDE's and a related synchronization problem for ODE's}

\author{{Zahra Aminzare and Eduardo D. Sontag}\\ \\
{\small Department of Mathematics, Rutgers University,}\\
{\small Piscataway, NJ 08854-8019 USA}}

\begin{document} 
\maketitle

%------------------------------------------------------------------------------------------------
\begin{abstract}
%------------------------------------------------------------------------------------------------

In this note, we present a condition which guarantees spatial uniformity for
the asymptotic behavior of the solutions of a reaction-diffusion PDE with
Neumann boundary conditions in one dimension, using the Jacobian matrix of the
reaction term and the first Dirichlet eigenvalue of the Laplacian operator on
the given spatial domain. We also derive an analog of this PDE result for the
synchronization of a network of identical ODE models coupled by diffusion
terms. 
\end{abstract}

%------------------------------------------------------------------------------------------------
\section{Introduction} \label{introduction}
%------------------------------------------------------------------------------------------------

Global stability is a central research topic in dynamical systems theory. 
Stability properties are typically defined in terms of attraction to an
invariant set, for example to an equilibrium or a periodic orbit, often coupled
with a Lyapunov stability requirement that trajectories that start near the
attractor must stay close to the attractor for all future times.

A far stronger requirement than attraction to a pre-specified target set is
to ask that any two trajectories should (exponentially, and with no
overshoot) converge to each other, or, in more abstract mathematical terms,
that the flow be a contraction in the state space. 
While this requirement will be less likely to be satisfied for a given system,
it is sometimes comparatively easier to check.
Indeed, checking stability properties often involves constructing an
appropriate Lyapunov function, which, in turn, requires a priori knowledge of
the attractor location.
In contrast, contraction-based methods, discussed here, do
not require the prior knowledge of attractors.  Instead, one checks an
infinitesimal property, that is to say, a property of the vector field
defining the system, which guarantees exponential contractivity of the
induced flow.

It is useful to first discuss the relatively trivial case of linear
time-invariant systems of differential equations $\dot x = Ax$, 
with Euclidean norm.
 Since differences of solutions are also solutions,
contractivity amounts simply to the requirement that there exist a positive
number $c$ such that, for all solutions,
$\abs{x(t)}\leq e^{-ct}\abs{x(0)}$, 
where $|\cdot|$ refers to the Euclidean norm.
This is clearly equivalent to the requirement that $A+A^T$ be a
negative definite matrix.
In Lyapunov-function terms, $x^TPx$ is a Lyapunov function for the system,
when $P=I$. 

This property is of course stronger than merely asymptotic stability of
the zero equilibrium of $\dot x = Ax$, that is, that $A$ be a Hurwitz matrix
(all eigenvalues with negative real part).
Of course, asymptotic stability is equivalent to the existence of some positive
definite matrix $P$ (but not necessarily the identity) so that $x^TPx$
is a Lyapunov function, and this can be interpreted, as remarked later, as a
contractivity property with respect to a weighted Euclidean norm associated to
$P$. This simple example with linear systems already illustrates why
an appropriate choice of norms when defining ``contractivity'' is critical; even
for linear systems, contractivity is not a topological, but is instead a metric
property: it depends on the norm being used, in close analogy to the choice of
an appropriate Lyapunov function.

The proper tool for characterizing contractivity for nonlinear systems
is provided by the matrix measures, also called logarithmic norms
(see e.g.~\cite{michelbook,desoer-vidyasagar-siam}), of the Jacobian of the
vector field, evaluated at all possible states.
This idea is a classical one, and can be traced back at least to work of
D.C.~Lewis in the 1940s, see \cite{Lewis1949,Hartman1961}.
Dahlquist's 1958 thesis under H\"ormander
(see~\cite{dahlquist} for a journal paper) used matrix
measures to show contractivity of differential equations, and more generally
of differential inequalities, the latter applied to the analysis of
convergence of numerical schemes for solving differential equations.
Several authors have independently rediscovered the basic ideas.
For example, in the 1960s,
Demidovi{\v{c}}~\cite{Demidovich1961,Demidovich1967} established basic
convergence results with respect to Euclidean norms, as did
Yoshizawa~\cite{Yoshizawa1966,Yoshizawa1975}. 
In control theory, the field attracted much attention after the work
of Lohmiller and Slotine~\cite{Loh_Slo_98}, and especially a string of 
follow-up papers by Slotine and collaborators, see for example
\cite{Loh_Slo_00,slotine_wang,slotine_stochastic,russo_slotine}.
These papers showed the power of contraction techniques for the study of not
merely stability, but also observer problems, nonlinear regulation, and
synchronization and consensus problems in complex networks.
(See also the work by Nijmejer and coworkers~\cite{pavlov_book}.) 
We refer the reader especially to the careful
historical analysis given in \cite{Jouffroy}.
Other very useful historical references are~\cite{Pav_Pog_Wou_Nij} and the
survey~\cite{Soderlind}.

In this paper, we establish new results for synchronization of diffusively interconnected and identical components, described by nonlinear differential equations. Specifically, we consider interconnected systems 
$\dot{x_i}= f(x_i,t) + \sum_{j\in N(i)} D(x_j-x_i)$,
where the $i$th subsystem (or ``agent'') has state $x_i(t)$, 
and show that the difference between any two states goes to zero exponentially, 
i.e., $\forall i,j$, $(x_i-x_j)(t)\to0$ as $t\to\infty$.
An interconnection graph provides the adjacency structure, and the indices in
$N(i)$ represent the  ``neighbors'' of the $i$th subsystem.
The matrix $D$ is a diagonal matrix of diffusion strengths.
The analysis of synchrony in networks of identical components is a
long-standing problem in different fields of science and engineering as well
as in mathematics.  In biology, the synchronization phenomenon is exhibited at
the physiological level, for example in neuronal interactions, in the generation
of circadian rhythms, or in the emergence of organized bursting in pancreatic
beta-cells,
\cite{Gray,Vries_Sherman_Zhu, Czeisler_etal, Sherman_Rinzel_Keizer, Chow_Kopell, Lewis_Rinzel}.
It is also exhibited at the population level, for example
in the simultaneous flashing of fireflies,
\cite{Smith_synch, Strogatz_Stewart}.
In engineering, one finds applications of synchronization ideas in areas as
varied as robotics or autonomous vehicles,
\cite{Nijmeijer_Rodriguez, Pettersen_Gravdahl_Nijmeijer}.
Synchronization results based on contraction-based techniques, most by using measures derived from $L^2$ or
weighted $L^2$ norms, have been developed, see for example
\cite{Loh_Slo_98, Arcak, slotine, Russo_Bernardo, slotine_wang}.
For non $L^2$ norms, current results are partial, applying only to certain
types of graphs.  Finding general statements and proofs is still an open
problem.

The convergence to uniform solutions in reaction-diffusion partial
differential equations
$
\fract{\partial u}{\partial t} = {F(u, t)}+D\Delta u
$
where $u=u(\o, t)$, is a 
formal analogue
 of the synchronization of ODE
systems.  In the analogy, we think of $u(\omega,\cdot)$ as representing an individual system
or agent (the index ``$i$'' in the synchronization problem)
whose state is described at time $t$ by $u=u(\o, t)$.
(So $u=u(\o, t)$ plays the role of $x_i(t)$.  We use ``$u$'' to denote the
state, instead of $x$, so as to be consistent with standard PDE notations.)
Questions of convergence to uniform solutions in reaction-diffusion PDE's are
also a classical topic of research.
The ``symmetry breaking'' phenomenon of diffusion-induced, or Turing,
instability refers to the case where a dynamic equilibrium $\bar u$ of the
non-diffusing ODE system $\fract{d u}{d t}=F(u, t)$ is stable, but, at least for some diagonal
positive matrices $D$, the corresponding uniform state $u(\omega)=\bar u$ is
unstable for the PDE system $\fract{\partial u}{\partial t}=F(u, t)+D\Delta u$.
This phenomenon has been studied at least since Turing's seminal work on
pattern formation in morphogenesis~\cite{Turing}, where he argued that
chemicals might react and diffuse so as result in heterogeneous spatial
patterns.
Subsequent work by Gierer and Meinhardt~\cite{GiererMeinhardt1972,Gierer1981}
produced a molecularly plausible minimal model, using two substances that
combine local autocatalysis and long-ranging inhibition.
Since that early work, a variety of processes in physics, chemistry, biology,
and many other areas have been studied from the point of view of diffusive
instabilities, and the mathematics of the process has been extensively studied
\cite{
Othmer1969,
Segel1972,
Cross1978,
Conway1978,
Othmer1980,
murray2002,
Borckmans1995,
Castets1990,
keshet,
maini2000}.
Most past work has focused on local stability analysis, through the analysis
of the instability of nonuniform spatial modes of the linearized PDE.
Nonlinear, global, results are usually proved under strong constraints on
diffusion constants as they compare to the growth of the reaction part.
Contraction techniques add a useful set of tools to that analysis.
As with synchronization, for non-Euclidean norms we only provide results
  in special cases, the general problem being open.

After presenting some mathematical tools in Section
\ref{preliminaries}, we will 
revisit, in the current context,
the biochemical example described in
\cite{ddv07,Russo} and the Goodwin example studied in \cite{Ingalls, Arcak}
 in Section \ref{Motivation}. Next, in Section \ref{Synchronization in PDEs}, we will state and prove the main result of this work: we present a condition which guarantees spatial uniformity for the asymptotic behavior of the solutions of a reaction-diffusion PDE with Neumann boundary conditions in one dimension. We also present some conditions which guarantees contractivity of the solutions of a reaction-diffusion PDE with Dirichlet boundary conditions.
We may think of convergence to spatially uniform solutions as a sort of
  ``synchronization'' of independent ``agents'', one at each spatial location,
  and each evolving according to a dynamics specified by an ODE.
In that interpretation, our work is related to a large literature on
synchronization of discrete groups of agents connected by diffusion, whose
interconnections are specified by an undirected graph.  In that spirit,
in Section \ref{Synchronization in a system of ODEs} we derive an
  analog of our PDE result to the synchronization of a network of identical
  ODE models coupled by diffusion terms through different types of graphs, 
including line, complete, and star graphs, and
Cartesian products of such graphs.

%------------------------------------------------------------------------------------------------
\smallskip
\section{Preliminaries}\label{preliminaries}
%------------------------------------------------------------------------------------------------

We now define and state elementary properties of
logarithmic Lipschitz constants.
(For applications to ODE's, we will always take $X=Y$ in the definitions to
follow.)

\begin{definition}\cite{Soderlind}
 Let $(X,\|\cdot\|_X)$ be a normed space and $f\colon Y\to X$ be a function, where $Y\subseteq X$. The least upper bound (lub) Lipschitz constant of $f$ induced by the norm $\|\cdot\|_X$, on $Y$, is defined by
\[
L_{Y,X}[f]=\displaystyle\sup_{u\neq v\in Y}\frac{\|f(u)-f(v)\|_X}{\|u-v\|_X}.
\]
Note that $L_{Y,X}[f]<\infty$ if and only if $f$ is Lipschitz on $Y$.
\end{definition}

\begin{definition}\label{def:M}
\cite{Soderlind}
Let $(X,\|\cdot\|_X)$ be a normed space and $f\colon Y \to X$ be a Lipschitz function. The least upper bound (lub) logarithmic Lipschitz constant of $f$ induced by the norm $\|\cdot\|_X$, on $Y\subseteq X$, is defined by
 \[M_{Y,X}[f]=\displaystyle\lim_{h\to0+}\frac{1}{h}\left(L_{Y,X}[I+hf]-1\right),\]
 or equivalently, it is equal to
\begin{equation}\label{defM}\nonumber
\displaystyle\lim_{h\to0^+}\sup_{u\neq v\in Y}\frac{1}{h}\left(\frac{\|u-v+h(f(u)-f(v))\|_X}{\|u-v\|_X}-1\right).
\end{equation}
 If $X=Y$, we write $M_X$ instead of $M_{X, X}$.
\end{definition}

\bnote Under the conditions of Definition \ref{def:M}, let $M^{\pm}_{Y,X}$ denote 
\begin{equation}\label{defM+}\nonumber
\displaystyle\sup_{u\neq v\in Y}\lim_{h\to0^{\pm}}\frac{1}{h}\left(\frac{\|u-v+h(f(u)-f(v))\|_X}{\|u-v\|_X}-1\right).
\end{equation}
If $X=Y$, we write $M^{\pm}_X$ instead of $M^{\pm}_{X, X}$.
\enote

\bremark\cite{Deimling, Soderlind}
Another way to define $M^{\pm}$ is by the concept of semi inner product which is in fact the generalization of inner product to non Hilbert spaces. 
Let $(X,\|\cdot\|_X)$ be a normed space. For $x_1, x_2\in X$, the right and left semi inner products are defined by
\begin{equation}\label{semi-inner}\nonumber
(x_1, x_2)_{\pm}=\displaystyle\|x_1\|_X\lim_{h\to 0^{\pm}}\frac{1}{h}\left(\|x_1+hx_2\|_X-\|x_1\|_X\right).
\end{equation}
In particular, when $\|\cdot\|_X$ is induced by a true inner product $(\cdot, \cdot)$, (for example when $X$ is a Hilbert space), 
then $(\cdot, \cdot)_-=(\cdot, \cdot)_+=(\cdot, \cdot)$.

Using this definition, 
\[M_{Y, X}^{\pm}[f]=\displaystyle\sup_{u\neq v\in Y}\frac{(u-v,f(u)-f(v))_{\pm}}{\|u-v\|_X^2}.\]
\eremark

The following elementary properties of semi inner products are consequences of
the properties of norms. 
See \cite{Deimling, Soderlind} for a proof.

\begin{proposition}\label{properties_semi_inner}
For $x, y, z\in X$ and $\alpha\geq 0$, 
\begin{enumerate}
\item $(x, -y)_{\pm}=-(x, y)_{\mp}$;
\item $(x, \alpha y)_{\pm}=\alpha(x, y)_{\pm}$;
\item $(x, y)_-+(x, z)_{\pm}\leq(x, y+z)_{\pm}\leq (x,y)_++(x,z)_{\pm}$.
\end{enumerate}
\end{proposition}

\bremark\label{M+<M}
For any operator $f\colon Y\subset X\to X$:
\[
M^-_{Y, X}[f]\leq M^+_{Y, X}[f]\leq M_{Y, X}[f].
\]
However, $M^-[f] = M^+[f] = M[f]$ if the norm is induced by an inner product.

For linear $f$, one has the reverse
of the second
 inequality as well, so
$M^+_{Y, X}[f]= M_{Y, X}[f]$.
See \cite{aminzare-sontag} for a detailed proof.
When identifying a linear operator $f:\r^n\rightarrow\r^n$ with its matrix
  representation $A$ with respect to the canonical basis, we write
``$\mu(A)$'' instead of $M^+_{X}[f]$,
and call $M$ or $\mu$ a ``matrix measure''.
\end{remark}

\bremark
For a linear operator $f$, $M$ and $M^+$ can be written as follows:
\begin{equation}\label{defM-linear}
M_{Y, X}[f]=\displaystyle\lim_{h\to0^+}\sup_{u\neq 0\in Y}\frac{1}{h}\left(\frac{\|u+hf(u)\|_X}{\|u\|_X}-1\right)
\end{equation}
and 
\begin{equation}\label{defM+-linear}
M^+_{Y, X}[f]=\displaystyle\sup_{u\neq 0\in Y}\lim_{h\to0^+}\frac{1}{h}\left(\frac{\|u+hf(u)\|_X}{\|u\|_X}-1\right).
\end{equation}
\eremark

\begin{notation}
In this work, for $(X, \|\cdot\|_X)=(\r^n, \|\cdot\|_{p})$, where
 $\|\cdot\|_p$ is the $L^p$ norm on $\r^n$, 
for some $1\leq p \leq \infty$, we sometimes use the notation ``$M_{p}$'' instead of
$M_X$ for the (lub) logarithmic Lipschitz constant,
and by ``$M_{p,Q}$'' we denote
the (lub) logarithmic Lipschitz constant
induced by the weighted $L^p$ norm, $\|u\|_{p,Q}:= \|Qu\|_p$ on $\r^n$,
where $Q$ is a fixed nonsingular matrix.
 Note that $M_{p,Q}[A] = M_p[QAQ^{-1}]$. 
 \end{notation}
 
\bremark
 In Table \ref{tab-mu}, the algebraic expression of the least upper bound logarithmic Lipschitz constant induced by the $L^p$ norm for $p=1,2,$ and $\infty$ are shown for matrices. For proofs, see for instance \cite{Desoer}.
\eremark

\newcommand{\spec}{\mbox{spec}}
\begin{table}[htdp]
%{\scriptsize{
\caption{\scshape Standard matrix measures for a real $n\times n$ matrix, $A=[a_{ij}]$.}
\begin{center}
\begin{tabular}{|c|c|}
\hline
vector norm, $\|\cdot\|$ & induced matrix measure, $M[A]$\\
\hline
$\|x\|_1=\displaystyle\sum_{i=1}^{n}\abs{x_i}$ & $M_1[A]=\displaystyle\max_{j}\left(a_{jj}+\displaystyle\sum_{i\neq j}\abs{a_{ij}}\right)$\\
\hline
$\|x\|_2=\left(\displaystyle\sum_{i=1}^{n}\abs{x_i}^2\right)^{\frac{1}{2}}$ & $M_2[A]=\displaystyle\max_{\lambda\in\spec{\frac{1}{2}(A+A^T)}}
\lambda$\\
\hline
$\|x\|_{\infty}=\displaystyle\max_{1\leq i\leq n} \abs{x_i}$ & $M_{\infty}[A]=\displaystyle\max_{i}\left(a_{ii}+\displaystyle\sum_{i\neq j}\abs{a_{ij}}\right)$\\
\hline
\end{tabular}
\end{center}
\label{tab-mu}
%}}
\end{table}%

The following subadditivity property is key to diffusive interconnection
analysis.

\begin{proposition}\cite{Soderlind}\label{subadd}
 Let $(X,\|\cdot\|_X)$ be a normed space. For any $f$, $g\colon Y\to X$ and any $Y\subseteq X$:
\begin{enumerate}
\item $M^+_{Y, X}[f+g]\leq M^+_{Y, X}[f]+M^+_{Y, X}[g]$;
\item $M^+_{Y, X}[\alpha f]=\alpha M_{Y, X}[f]$ for $\alpha\geq0$.
\end{enumerate}
\end{proposition}

The (lub) logarithmic Lipschitz constant makes sense even if $f$ is not
differentiable.  However, the constant can be tightly estimated, for
differentiable mappings on convex subsets of finite-dimensional spaces, by
means of Jacobians.

\begin{lemma} \cite{Soderlind2}\label{key3}
For any given norm on $X=\r^n$, let $M$ be the (lub) logarithmic Lipschitz
constant induced by this norm. Let $Y$ be a connected subset of $X=\r^n$. Then
for any (globally) Lipschitz and continuously differentiable function 
$f\colon Y\to \r^n$, 
\[
\displaystyle\sup_{x\in Y}M_{X}[J_f(x)]\leq M_{Y,X}[f] \,
\]
Moreover, if $Y$ is convex, then
\[
\displaystyle\sup_{x\in Y}M_{X}[J_f(x)]= M_{Y,X}[f]\,.
\]
Note that for any $x\in Y$, $J_f(x)\colon X\to X$. Therefore, we use $M_X$ instead of $M_{X,X}$, as we said in Definition \ref{def:M}.
\end{lemma}

We also recall a notion of generalized derivative, that can be used
  when taking derivatives of norms (which are not differentiable).

\begin{definition}
The upper left and right Dini derivatives for any continuous function, $\Psi\colon[0, \infty)\to\r$, are defined by
\[
\begin{array}{rcl}
\left(D^{\pm}\Psi\right)(t)=\displaystyle\limsup_{h\to 0^{\pm}}\frac{1}{h}\left(\Psi(t+h)-\Psi(t)\right).
 \end{array}
\]
Note that $D^+\Psi$ and/or $D^-\Psi$ might be infinite.
\end{definition}

The following Lemma from \cite{Deimling}, indicates the relation between the Dini derivative and the semi inner product. 
\begin{lemma} \label{M+Dini-linear}
For any bounded linear operator $A\colon X\to X$, and any solution $u\colon [0,T)\to X$ of $\displaystyle\frac{du}{dt}=Au$,
 \begin{equation*}
 D^{+}\|u(t)\|_X=\displaystyle \frac{(u(t), Au(t))_{+}}{\|u(t)\|_X^2}\|u(t)\|_X\leq M_X[A]\|u(t)\|_X,
   \end{equation*}
  for all $t\in [0, T)$.
\end{lemma}

{In this note, we will use the following general result,
which estimates
rates of contraction (or expansion) among any two functions, even
functions that are not solutions of the same system of ODEs (see comment on
observers to follow):
}

\blem\label{Dini_M}
Let $\left(X,\|\cdot\|_X\right)$ be a normed space and $G\colon Y\times [0, \infty) \to X$ be a
 $C^1$ function, where $Y\subseteq X$. Suppose $u, v\colon [0,
    \infty)\to Y$ satisfy 
\[
(\dot{u}-\dot{v})(t)\;=\;G_t(u(t))-G_t(v(t)), 
\]
where $\dot{u}=\frac{du(t)}{dt}$ and $G_t(u)=G(u, t)$. Let 
\[
c := \displaystyle\sup_{\tau\in [0, \infty)}M_{Y, X}\;[{G_{\tau}}].
\]
Then for all $t\in [0, \infty)$, 
 \begin{equation}\label{Dini-u-b}
  \|u(t)-v(t)\|_X\;\leq\; e^{ct}\|u(0)-v(0)\|_X.
  \ee
\elem

\bp
Using the definition of Dini derivative, we have
(dropping the argument $t$ for simplicity):
\beqn
&&\!\!\!\!\!\!\!\!\!\!\!\!\!\!\!\!
D^{+}\|(u-v)(t)\|\\
&=&\limsup_{h\to 0^+}\frac{1}{h}\;\left(\|(u-v)(t+h)\|_X-\|(u-v)(t)\|_X\right)\\
&=&\limsup_{h\to 0^+}\frac{1}{h}\left(\|u-v+h(\dot{u}-\dot{v})\|_X-\|u-v\|_X\right)\\
&=&\lim_{h\to0^+}\frac{1}{h}\left(\|u-v+h(G_t(u)-G_t(v))\|_X-\|u-v\|_X\right)\\
&\leq& M_{Y, X}^+[G_t]\;\|(u-v)(t)\|_X \quad (\mbox{by definition of $M^+$})\\
&\leq& M_{Y, X}[G_t]\;\|(u-v)(t)\|_X \quad (\mbox{by Remark \ref{M+<M}})\\
&\leq& \dis\sup_{\tau} M_{Y, X}[{G_{\tau}}]\;\|(u-v)(t)\|_X.
\eeqn
The third equality holds because since every norm possesses right (and also left) G\^ateaux-differentials, the limit exists. 
Using Gronwall's Lemma for Dini derivatives (see e.g.~\cite{Lorenz}, Appendix
A), we obtain~\eqref{Dini-u-b}, 
where $c := \displaystyle\sup_{t\in [0, \infty)}M_{Y, X}\;[G_t]$.
\ep

\bremark\label{M[f]:M[J]}
In the finite-dimensional case, Lemma \ref{Dini_M} can be verified in terms of
Jacobians.
Indeed, suppose that $X=\r^n$, and that $Y$ is a 
convex subset
of $\r^n$.  Then, by Lemma \ref{key3}, 
\[
c \;=\; \tilde{c} :=\displaystyle\sup_{(t, w)\in [0, \infty)\times Y}M_{X}\;[J_{G_t}(w)] \,.
\]
Therefore,
\be{}\nonumber
  \|u(t)-v(t)\|_X\;\leq\; e^{\tilde{c}t}\|u(0)-v(0)\|_X.
  \ee

In fact, in the finite-dimensional case, a more direct proof of
  Lemma~\ref{Dini_M} can instead be given.  We sketch it next.
Let $z(t) = u(t) - v(t)$. We have that
  \[
   \dot{z}(t) = A(t) z(t),
   \]
where $A(t) = \dis\int_0^1 \dis\frac{\partial f}{\partial x} \lt(s u(t) + (1-s) v(t) \rt) \; ds.$
Now, by subadditivity of matrix measures, which, by continuity, extends
to integrals, we have:
\[
M[A(t)]  \leq \dis \sup_{w \in V} M\lt[\dis\frac{\partial f}{\partial x}(w)\rt]
\,.
\]
Applying Coppel's inequality, (see e. g. \cite{Vidyasagar_book}), gives the
result.  
\eremark

%------------------------------------------------------------------------------------------------
\smallskip
\section{Motivation} \label{Motivation}
%------------------------------------------------------------------------------------------------

%------------------------------------------------------------------------------------------------
\smallskip
\subsection*{Biochemical model} \label{Biochemical model}
%------------------------------------------------------------------------------------------------
{As a motivation
we will 
revisit, in the current context,
the biochemical example described in
\cite{ddv07,Russo} and \cite{aminzare-sontag}.
}
 A typical biochemical reaction is one in which an enzyme $X$ (whose concentration is quantified by the non-zero variable $x=x(\omega,t), \omega\in[0,1]$) binds to a substrate $S$ (whose concentration is quantified by $s=s(\omega,t)\geq 0$), to produce a complex $Y$ (whose concentration is quantified by $y=y(\omega,t)\geq0$), and the enzyme is subject to degradation and dilution (at rate $\delta x$, where $\delta>0$) and production according to an external signal $z=z(\omega,t)\geq 0$. An entirely analogous system can be used to model a transcription factor binding to a promoter, as well as many other biological process of interest. The complete system of chemical reactions is given by the following diagram:
\[
0 \arrowchem{z} X \arrowchem{\delta} 0\,,\quad
X+S \arrowschem{k_1}{k_2} Y.
\]
We let the domain $\Omega$ represent the part of the cytoplasm where these chemicals are free to diffuse. Taking equal diffusion constants for $S$ and $Y$ (which is reasonable since typically $S$ and $Y$ have approximately the same size), a natural model is given by a reaction diffusion system 
\beqn\label{eq1:example}
\dis\frac{\partial x}{\partial t}&=&z(t)-\delta x+k_1y-k_2 s x+d_1\Delta x\\
\dis\frac{\partial y}{\partial t}&=&-k_1y+k_2sx+d_2\Delta y\\
\dis\frac{\partial s}{\partial t}&=&k_1y-k_2sx+d_2\Delta s,
\eeqn
subject to the Neumann boundary condition, $\frac{\partial x}{\partial \o} (0,t)= \frac{\partial x}{\partial \o} (1,t) =0$, etc.
{ If we assume that initially $S$ and $Y$ are uniformly distributed, i.e. $(y+s)(\omega, 0)=S_Y$, it follows that 
$\frac{\partial}{\partial t}(y+s)(\omega, t)=\frac{\partial^2}{\partial \omega^2}(y+s)(\omega, t)$, so $(y+s)(\omega, t)=(y+s)(\omega, 0)=S_Y$ is a constant.
} Thus, we can study the following reduced system: 
\be{example_reduced}
\bal
\dis\frac{\partial x}{\partial t}&=z(t)-\delta x+k_1y-k_2(S_Y-y)x+d_1\Delta x\\
\dis\frac{\partial y}{\partial t}&=-k_1y+k_2(S_Y-y)x+d_2\Delta y.
\eal
\ee
Note that $(x(\o,t), y(\o,t))\in V=[0, \infty)\times[0, S_Y]$ for all $\o\in(0,1)$, and $t\geq0$ ($V$ is convex and forward-invariant), and $S_Y$, $k_1$, $k_2$, $\delta$, $d_1$, and $d_2$ are arbitrary positive constants.

Let $J_{F_t}$ be the Jacobian of $F_t(x,y):=(z(t)-\delta x+k_1y-k_2(S_Y-y)x, -k_1y+k_2(S_Y-y)x)^T$:
\[
J_{F_t}(x,y)\;=\;\left(\begin{array}{cc}-\delta-k_2(S_Y-y) & k_1+k_2x \\k_2(S_Y-y) & -(k_1+k_2x)\end{array}\right).
\]
In \cite{Russo}, it has been shown that $\sup_{t}\sup_{(x,y)\in V}M_{1,Q} [J_{F_t}(x,y)]<0$, for some non-identity, positive diagonal matrix $Q$. 
In \cite{aminzare-sontag}, it has been shown that for any $p>1$, and any positive diagonal $Q$, $\sup_{t}\sup_{(x,y)\in V}M_{p,Q} [J_{F_t}(x,y)]\geq0$. 
Here, we will show that not only $\sup_{t}\sup_{(x,y)\in V}M_{2,Q} [J_{F_t}(x,y)]\geq0$, but 
 \be{M2>0}
 \dis\sup_{t}\sup_{(x,y)\in V}M_{2,Q} [J_{F_t}(x,y)-\l D]\geq0,
 \ee
  for any positive diagonal matrix $Q$ and any $\l>0$:
 
 Without loss of generality we assume $Q=\diag(1,q)$. Then 
 \[QJ_{F_t}(x,y)Q^{-1}\;=\;\displaystyle\left(\begin{array}{cc}-\delta-a & \displaystyle\frac{b}{q} \\aq & -b\end{array}\right),\]
 where $a=k_2(S_Y-y)\in[0,k_2S_Y]$ and $b=k_1+k_2x\in[k_1,\infty)$. 
 By definition of $M_{2,Q}$, we know that, $M_{2, Q}\lt[J_{F_t}(x,y)-\l D\rt]\;=\;\l_{\max}\{R\}$,
where $\l_{\max}\{R\}$ denotes the largest eigenvalue of 
\[R:=\dis\frac{1}{2}\lt(Q(J_{F_t}(x,y)-\l D)Q^{-1}+\lt(Q(J_{F_t}(x,y)-\l D)Q^{-1}\rt)^T\rt).\]

 A simple calculation shows that the eigenvalues of $R$ are as follows:
 \[\l_{\pm}\;=\;-(\d+a+b+(d_1+d_2)\l)\pm\sqrt{\lt((d+a+d_1\l)-(b+d_2\l)\rt)^2+\lt(aq+\dis\frac{b}{q}\rt)^2}.\]
We can pick $x=x^*$ large enough (i.e. $b$ large enough) and $y=y^*=S_Y$ (i.e. $a=0$), such that $\l_+>0$ and hence by Table \ref{tab-mu}, $M_{2,Q}\lt[J_{F_t}(x^*,y^*)-\l D\rt]>0$.

We will get back to this example later (Remark \ref{synch_biochemical_example_better_rate} below) and study the behavior of its solutions.

%------------------------------------------------------------------------------------------------
\smallskip
\subsection*{Goodwin Oscillator}
%------------------------------------------------------------------------------------------------

In 1965, Brian Goodwin proposed a differential equation model, that describes the generic model of an oscillating autoregulatory gene, and studied its oscillatory behavior \cite{Goodwin1965}.
 The following systems of ODEs is a variant of Goodwin's model \cite{Thron}:
 \be{}
\bal
\dot{x}&=\dis\frac{a}{k+z(t)}-bx\\
\dot{y}&=\a x-\beta y\\
\dot{z}&=\gamma y-\dis\frac{\delta z}{k_M+z}.
\eal
\ee
The model, sketched in Fig.~\ref{fig:Goodwin}, shows a single gene with mRNA, $X$, which is translated into an enzyme $Y$, which in turn, catalyses production of a metabolite, $Z$. But the metabolite inhibits the expression of the gene. 

%fig
\begin{figure}[htdp]
\begin{center}
\caption{Goodwin Oscillator: a single gene}
\includegraphics[scale=.34]{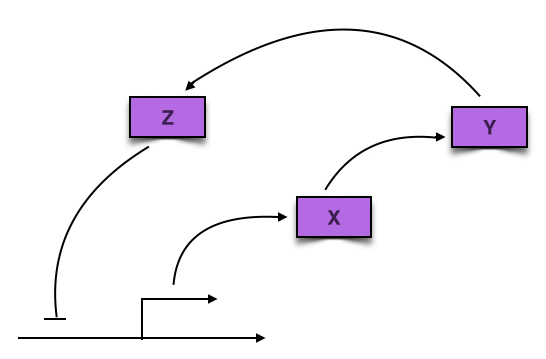}
\label{fig:Goodwin}
\end{center}
\end{figure}

We now assume a continuous model where species diffuse in space. This example has been studied in \cite{Arcak}. The following system of PDEs, subject to Neumann boundary conditions, describe the evolution of $X$, $Y$, and $Z$ on $(0,1)\times [0,\infty)$: 
\be{Goodwin_PDE}
\bal
\dis\frac{\partial x}{\partial t}&\;=\;\dis\frac{a}{k+z}-b\;x+d_1\Delta x\\
\dis\frac{\partial y}{\partial t}&\;=\;\a\; x-\beta \;y+d_2\Delta y\\
\dis\frac{\partial z}{\partial t}&\;=\;\gamma\; y- \dis\frac{\delta z}{k_M+z} +d_3\Delta z
\eal
\ee

Figure \ref{Goodwin_no_diffusion} provides plots of solutions $x$, $y$, and
$z$ of (\ref{Goodwin_PDE}), using the following
parameter values from the textbook \cite{Ingalls}: 
 \be{parameter}
 a=150,\; k=1,\; b=\a=\beta=\gamma=0.2,\; \delta=15,\; K_M=1.
 \ee
which oscillate when there is no diffusion
($d_1=d_2=d_3=0$).  
\begin{figure}[htdp]
\begin{center}
\caption{Goodwin Oscillator, no diffusion {(parameters as in Eq. (\ref{parameter}))}}
\begin{subfigure}[b]{0.3\textwidth}
\includegraphics[scale=.21]{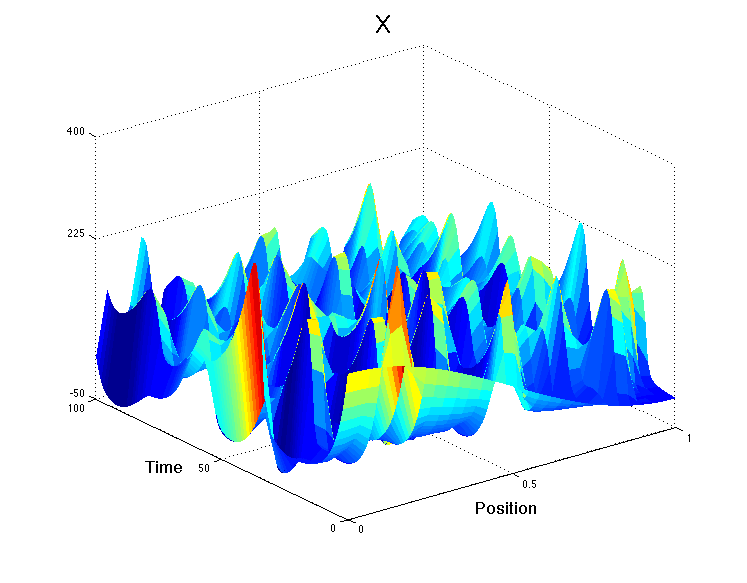}
\end{subfigure}
\quad
\begin{subfigure}[b]{0.3\textwidth}
\includegraphics[scale=.21]{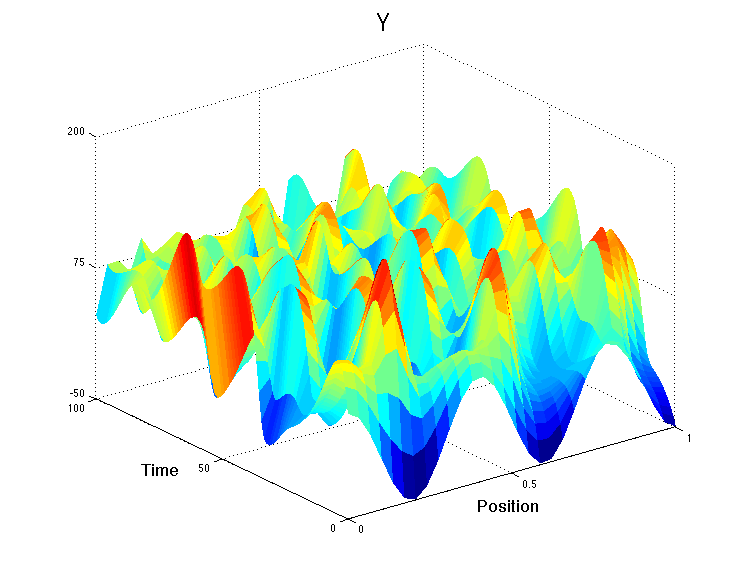}
\end{subfigure}
\quad
\begin{subfigure}[b]{0.3\textwidth}
\includegraphics[scale=.21]{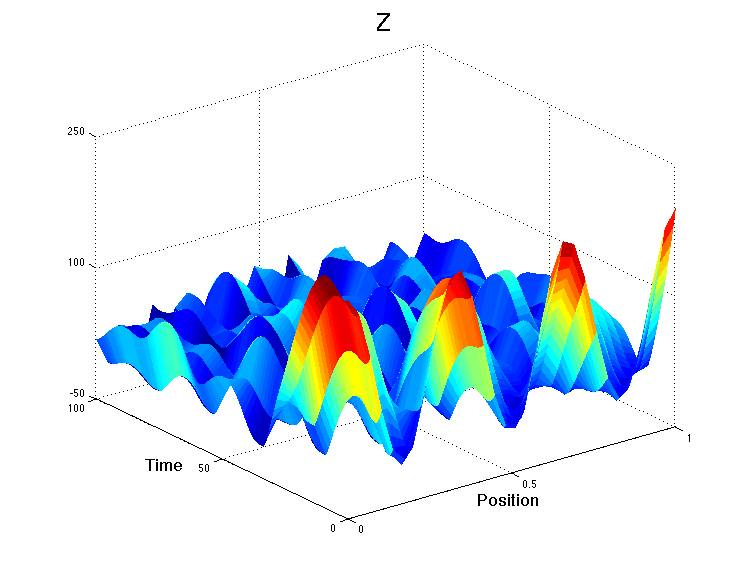}
\end{subfigure}
\label{Goodwin_no_diffusion}
\end{center}
\end{figure}

Figure~\ref{Goodwin_X_diffuces} shows the spatially uniformity of the
solutions of (\ref{Goodwin_PDE}), for the same parameter values and initial
conditions as in Figure \ref{Goodwin_no_diffusion}, when ${2.2}/{\pi^2}<d_1$, 
{$d_1=0.3$,}
and $d_2=d_3=0$.

In this work we provide a condition for synchronization when only $X$ diffuses, i.e. the diffusion matrix is $D=\diag(d_1, 0, 0)$.
{We will get back to this example later in Section \ref{Synchronization in PDEs}, Remark \ref{goodwin_compare_with_arcak_othmer}.}

\begin{figure}[htdp]
\begin{center}
\caption{Goodwin Oscillator, X diffuses {(parameters as in Eq. (\ref{parameter}))}}
\begin{subfigure}[b]{0.3\textwidth}
\includegraphics[scale=.21]{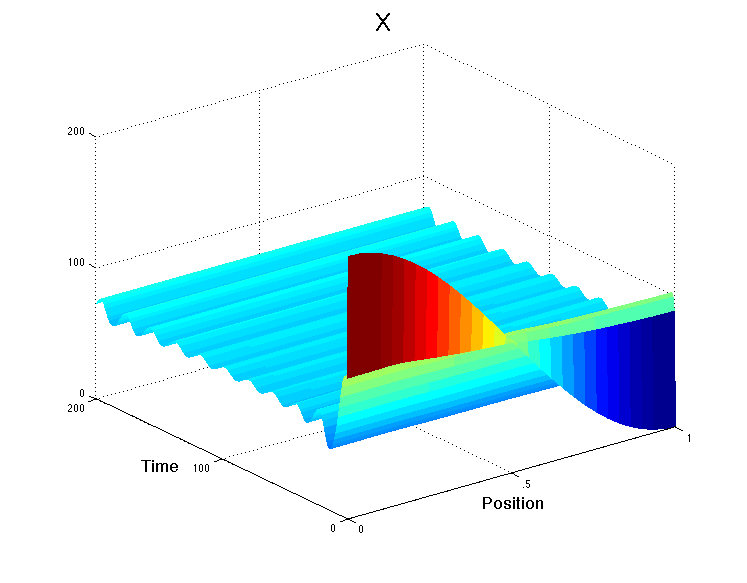}
\end{subfigure}
\quad
\begin{subfigure}[b]{0.3\textwidth}
\includegraphics[scale=.21]{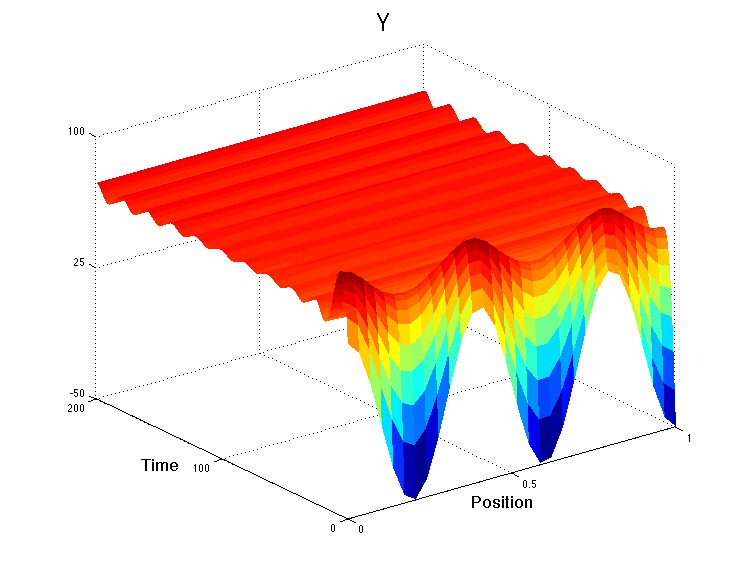}
\end{subfigure}
\quad
\begin{subfigure}[b]{0.3\textwidth}
\includegraphics[scale=.21]{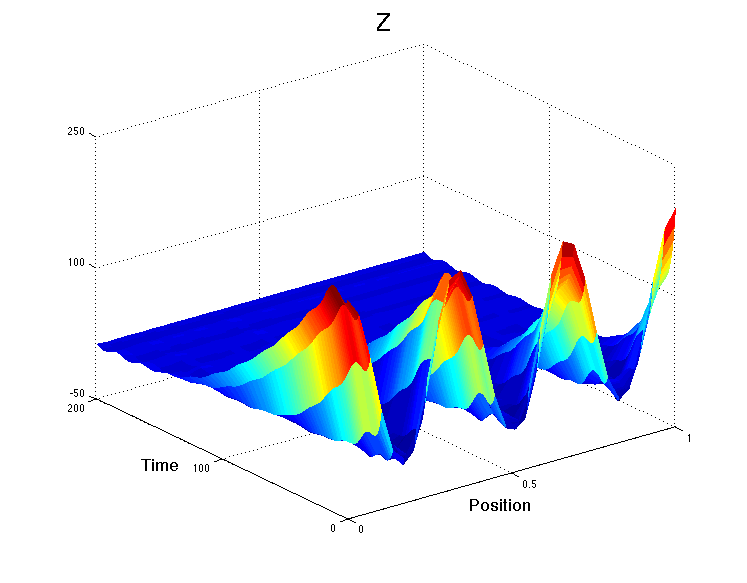}
\end{subfigure}
\label{Goodwin_X_diffuces}
\end{center}
\end{figure}

%------------------------------------------------------------------------------------------------
\section{Convergence to uniform solutions in PDEs}\label{Synchronization in PDEs}
%------------------------------------------------------------------------------------------------

In this section we review the existing results for synchronization and state and prove the new result. 

We study reaction-diffusion PDE systems of the general form:
\be{}\nonumber
\bal
\displaystyle\frac{\partial u_1}{\partial t}(\omega, t)&=F_1(u(\omega, t), t)+d_1\Delta u_1(\omega, t)\\
&\vdots\\
\displaystyle\frac{\partial u_n}{\partial t}(\omega, t)&=F_n(u(\omega, t), t)+d_n\Delta u_n(\omega, t)\\
\eal
\ee
which can be written as the following closed form:
\begin{equation} \label{re-di}
\bal
&\displaystyle\frac{\partial u}{\partial t}(\omega, t)=F_t(u(\omega, t))+D\Delta u(\omega, t)\\
\eal
\end{equation}

subject to the Neumann boundary condition:
\begin{equation} \label{NBC}
\frac{\partial u_i}{\partial\mathbf{n}}(\xi,t)=0\quad \forall\xi\in\partial\Omega,\;\;\forall t\in[0,\infty), \;\;\forall i=1,\ldots, n,
\end{equation} 

{or subject to the Dirichlet boundary condition:}
\begin{equation} \label{DBC}
u_i(\xi,t)=0\quad \forall\xi\in\partial\Omega,\;\;\forall t\in[0,\infty), \;\;\forall i=1,\ldots, n,
\end{equation} 
 where 
 \begin{itemize}
\item $F_t(x)=F(x,t)$ and $F\colon V\times [0, \infty)\to\r^n$ is a (globally) Lipschitz vector field with components $F_i$:
\[F(x, t)\;=\;(F_1(x, t), \cdots, F_n(x, t))^T,\] for some functions $F_i\colon V\times [0, \infty)\to\r$, where $V$ is a convex subset of $\r^n$.
\item $D=\diag(d_1,\ldots, d_n)$ with $d_i\geq0$, and $d_j>0$ for some $j$, which we call the diffusion matrix.
\item $\Omega$ is a bounded domain in $\r^m$ with smooth boundary $\partial\Omega$ and outward normal $\mathbf{n}$. 

\item  $\dis\frac{\partial u}{\partial\mathbf{n}}= \lt(\dis\frac{\partial u_1}{\partial\mathbf{n}}, \ldots, \dis\frac{\partial u_n}{\partial\mathbf{n}}\rt)^T$.
\end{itemize}

In biology, a PDE system of this form describes individuals (particles,
chemical species, etc.) of $n$ different types, with respective abundances
$u_i(\omega,t )$ at time $t$ and location $\omega \in \Omega $, that can react instantaneously,
guided by the interaction rules encoded into the vector field $F$, and can
diffuse due to random motion.

\begin{definition}
By a solution of the PDE
\begin{equation*}
\displaystyle\frac{\partial u}{\partial t}(\omega, t)=F_t(u(\omega, t))+D\Delta u(\omega, t)
\end{equation*}
\begin{equation*} 
\frac{\partial u}{\partial\mathbf{n}}(\xi,t)=0\quad \forall\xi\in\partial\Omega,\;\;\forall t\in[0,\infty),
\end{equation*}
 on an interval $[0, T)$, where $0<T\leq\infty$, we mean a function $u=(u_1, \cdots, u_n)^T$, with $u\colon \displaystyle\bar\Omega\times [0,T)\to V$, such that:
\begin{enumerate}
 \item for each $\omega\in\bar\Omega$, $u(\omega, \cdot)$ is continuously differentiable;
 \item for each $t\in[0,T)$, $u(\cdot, t)$ is in $\mathbf{Y}^{(n)}_V$, 
 where 
 \[
 {\mathbf{Y}^{(n)}_V}\;=\;\left\{\begin{array}{c}\displaystyle v\colon\bar\Omega\to V,\quad v=(v_1,\cdots, v_n), \\v_i\in C^2_{\r}\left(\bar\Omega\right),\quad\dis\frac{\partial v_i}{\partial\mathbf{n}}(\xi)=0,\; \forall\xi\in\partial\Omega,\;\;\forall i\end{array}\right\}
 \]
and $C^2_{\r}\left(\bar\Omega\right)$ is the set of twice continuously differentiable functions $\bar\Omega\to \r$; and
 \item for each $\omega\in\bar\Omega$, and each $t\in [0,T)$, $u$ satisfies the above PDE.
\end{enumerate}
\end{definition}

\begin{definition}
By a solution of the PDE
\begin{equation*}
\displaystyle\frac{\partial u}{\partial t}(\omega, t)=F_t(u(\omega, t))+D\Delta u(\omega, t)
\end{equation*}
\begin{equation*} 
u(\xi,t)=0\quad \forall\xi\in\partial\Omega,\;\;\forall t\in[0,\infty),
\end{equation*}
 on an interval $[0, T)$, where $0<T\leq\infty$, we mean a function $u=(u_1, \cdots, u_n)^T$, with $u\colon \displaystyle\bar\Omega\times [0,T)\to V$, such that:
\begin{enumerate}
 \item for each $\omega\in\bar\Omega$, $u(\omega, \cdot)$ is continuously differentiable;
 \item for each $t\in[0,T)$, $u(\cdot, t)$ is in $\mathbf{Y}^{(d)}_V$, 
 where 
 \[\mathbf{Y}^{(d)}_V\;=\;\left\{\begin{array}{c}\displaystyle v\colon\bar\Omega\to V,\quad v=(v_1,\cdots, v_n), \\v_i\in C^2_{\r}\left(\bar\Omega\right),\quad\dis v_i(\xi)=0,\; \forall\xi\in\partial\Omega,\;\;\forall i\end{array}\right\}\]
and $C^2_{\r}\left(\bar\Omega\right)$ is the set of twice continuously differentiable functions $\bar\Omega\to \r$; and
 \item for each $\omega\in\bar\Omega$, and each $t\in [0,T)$, $u$ satisfies the above PDE.
\end{enumerate}
\end{definition}

Under the additional assumptions that $F(x,t)$ is twice continuously differentiable with respect to $x$ and continuous with respect to $t$, theorems on existence and uniqueness for PDEs such as (\ref{re-di}) can be found in standard references, e.g. \cite{Smith, Cantrell}. One must impose appropriate conditions on the vector field, on the boundary of $V$, to insure invariance of $V$. Convexity of $V$ insures that the Laplacian also preserves $V$. Since we are interested here in estimates relating pairs of solutions, we will not deal with existence and well-posedness. Our results will refer to solutions already assumed to exist.

Let $Q=\diag(q_1,\ldots, q_n)$ be a positive diagonal matrix and $1\leq p\leq \infty$. Let 
\be{def:V}
V_p\;:=\;(V, \|\cdot\|_{p,Q})
\ee
 be a normed space, where for any $x=(x_1,\cdots, x_n)^T\in \r^n$, $\|x\|_{p,Q}$ is defined as follows: 
\be{}
\bal
&\|x\|_{p,Q}\;=\; \lt(\dis\sum_{i=1}^nq_i^p\abs{x_i}^p\rt)^{\frac{1}{p}} &  \quad& 1\leq p<\infty \\
&\|x\|_{\infty,Q}\;=\; \dis\max_{1\leq i\leq n} q_i\abs{x_i}& \quad & p=\infty.
\eal
\ee

In this section, by $M_{V_p}[\cdot]$ or $M_{p,Q}[\cdot]$, we mean (lub) logarithmic Lipschitz constant induced by $\|\cdot\|_{p,Q}$ on $V$. 

\begin{definition}
We say that the reaction diffusion PDE (\ref{re-di}) is contractive, if for any two solutions $u, v$ of (\ref{re-di}), subject to the Neumann or Dirichlet boundary condition, $\|(u-v)(\cdot, t)\|\to 0$ as $t\to\infty$.
\end{definition}

\begin{definition}
We say that the reaction diffusion PDE (\ref{re-di}) synchronizes, if for any solution $u$ of (\ref{re-di}), subject to the Neumann or Dirichlet boundary condition, there exists $\bar{u}(t)$ such that $\|u(\cdot, t) - \bar{u}(t)\|\to 0$ as $t\to\infty$, or equivalently $\|\nabla u(\cdot, t)\|\to 0$ as $t\to\infty$.
\end{definition}

The following theorem, from \cite{aminzare-sontag}, provides a sufficient condition for contractivity of the reaction diffusion PDE (\ref{re-di}) subject to the Neumann boundary condition (\ref{NBC}). Then, in Remark \ref{contraction->synchronization} below, we show that how contractivity of the reaction diffusion PDE implies synchronization.

\begin{theorem}\label{contraction}
Consider the reaction diffusion PDE (\ref{re-di}) subject to the Neumann boundary condition (\ref{NBC}). Let $c=\displaystyle\sup_{t\in[0, \infty)}\MpQ[F_t]$ for some $1\leq p\leq\infty$, and some positive diagonal matrix $Q$. Then for every two solutions $u, v$ of the PDE (\ref{re-di}) subject to the Neumann boundary condition (\ref{NBC}) and all $t\in[0,T)$:
\[\normpQ{u(\cdot, t)-v( \cdot, t)}\leq e^{ct}\normpQ{u(\cdot, 0)-v(\cdot, 0)}.\]
\end{theorem}

\bremark\label{contraction->synchronization}
Under the conditions of Theorem~\ref{contraction}, if $c<0$, any solution $u$ of the PDE (\ref{re-di}) with $u(\o, 0)= u_0(\o)$ exponentially converges to the spatially uniform solution $\bar{u}(t)$ which is itself the solution of the following ODE system:
\be{ubar}
\bal
\dot{x}&= F(x, t), \\
x(0) &= \dis\frac{1}{\abs{\Omega}} \dis\int_{\Omega} u_0(\o) \; d\o. 
\eal
\ee
{But, note that the condition $c<0$
rules out any interesting non-equilibrium behavior. 
For instance in Goodwin's oscillatory system, $c<0$ kills out the oscillation.
So we look for a weaker condition than $c<0$, that guarantees spatial uniform
convergence result (which is a weaker property than contraction) while keeps interesting non-equilibrium behavior, like oscillatory in Goodwin example.}
\end{remark}

Recall \cite{Henrot} that for any bounded, open subset  $\Omega\subset\mathbb{R}^m$, there
exists a sequence of positive eigenvalues 
$0\leq \lambda_1^{(n)} \leq\lambda_2^{(n)}\leq\ldots$ (going to $\infty$, superscript $(n)$ for Neumann) and a sequence
of corresponding orthonormal eigenfunctions: 
$\phi_1^{(n)} , \phi_2^{(n)} , \ldots$ (defining a Hilbert basis of $L^2(\Omega)$)
satisfying the following Neumann eigenvalue problem:
\be{Neumann_eigenvalue_problem}
\bal
-\Delta \phi_i^{(n)} =\lambda_i^{(n)} \phi_i^{(n)} &\quad  \mbox{in $\Omega$} \\
\nabla\phi_i^{(n)}\cdot\mathbf{n}=0  &\quad  \mbox{on $\partial \Omega$}
\eal
\ee
Note that the first eigenvalue is always zero, $\l_1=0$, and the corresponding
eigenfunction is a nonzero constant 
 ($\phi(\o)= 1/\sqrt{\abs{\Omega}}$). 

The following re-phasing of a theorem from \cite{Arcak}, 
 provides a sufficient condition
on $F$ and $D$ using the Jacobian matrix of the reaction term and the second
Neumann eigenvalue of the Laplacian operator on the given spatial domain to insure the convergence of trajectories, in this case to their
space averages in weighted $L^2$ norms. The proof in \cite{Arcak} is based on the use of a quadratic
Lyapunov function, which is appropriate for Hilbert spaces. 
We have translated the result to the language of contractions.
(Actually, the result in \cite{Arcak} is stronger, in that it allows for
non-diagonal diffusion and also non-diagonal weighting matrices $Q$, by
substituting these assumptions by a commutativity type of condition.)

\bthm\label{Arcak}
Consider the reaction-diffusion system (\ref{re-di}). Let 
\[c:=\displaystyle\sup_{(x, t)\in V\times [0, \infty)}M_{2,Q}\lt[J_F(x, t)-\lambda_2^{(n)} D\rt],\] 
where $Q$ is a positive diagonal matrix. 
Then
 \begin{equation}\label{eq2:cont-weighted-L2}
\|u(\cdot, t)-\tilde{u}(t)\|_{2,Q}\;\leq\;e^{c t}\|u(\cdot, 0)-\tilde{u}(0)\|_{2,Q}.
\end{equation}
where $\tilde{u}(t)= \dis\frac{1}{\abs{\Omega}} \dis\int_{\Omega} u(\o,t) \; d\o $. 
\ethm

Note that when $c<0$, the reaction-diffusion system (\ref{re-di}) synchronize. As we discussed in the  biochemical example earlier, 
\[\displaystyle\sup_{(x, t)\in V\times [0, \infty)}M_{2,Q}\lt[J_F(x, t)-\lambda_2^{(n)} D\rt]\geq0,\]
therefore, conditions given in \cite{Arcak} do not hold
for the biochemical example. 

A generalization of Theorem~\ref{Arcak} to spatially-varying diffusion is given in \cite{aminzare_shafi_arcak_sontag_bookchapter2013}. 

We next prove an analogous result to Theorem \ref{Arcak} for any norm but restricted to the linear operators $F$, $F(u, t)= A(t) u$, where for any $t$, $A(t)\in\r^{n\times n}$.

\bthm\label{linear_F_PDE}
For a given norm $\|\cdot\|$ in $\r^n$, consider the reaction-diffusion system (\ref{re-di}), for a linear operator $F$. Let 
\[c:=\displaystyle\sup_{(x, t)\in V\times [0, \infty)}M\lt[J_F(x, t)-\lambda_2^{(n)} D\rt],\] 
where $M$ is the logarithmic norm induced by $\|\cdot\|.$ Then for any $\o\in\Omega$
and any $t\geq0$, 
\be{}\nonumber
\bal
\|u(\o, t)-\bar{u}(t)\|&\leq\dis\sum_{i\geq2} \lt\|\a_i(t)  \phi_i(\o)\rt\|\\
& \leq e^{ct}  \dis\sum_{i\geq2} \lt\|\a_i(0)  \phi_i(\o)\rt\|.
\eal
\ee
where $\bar{u}(t)$ is the solution of the system (\ref{ubar}) with $u_0(\o)=u(\o,0)$,
 and $\a_i(t)=\int_{\Omega} u(\o,t)\phi_i^{(n)}(\o)\;d\o$.
In particular, when $c<0$, 
 \[
 \|u(\o, t)-\bar{u}(t)\|\to 0\quad\mbox{exponentially, as $t\to\infty$}.
\]
\ethm

\bp
We first show that the solution of Equation (\ref{ubar}), $\bar u$, is equal to $\tilde{u}(t)=\frac{1}{\abs{\Omega}} \int_{\Omega} u(\o,t) \; d\o$. Note that both $\bar u$ and $\tilde {u}$ satisfy $\dot x=A(t)x$.
In addition, by the definition, $\bar u(0) = \tilde{u}(0)= \frac{1}{\abs{\Omega}} \int_{\Omega} u(\o,0) \; d\o$. Therefore, by uniqueness of the solutions of ODEs, $\bar u(t) = \tilde {u} (t)$.  
The solution $u(\o, t)$ can be written as follows:
\be{u:expansion}
u(\o, t) \;=\; \dis\sum_{i\geq1}  \phi_i (\o)\a_i(t)
\ee
where for any $t$, $\a_i(t)=\int_{\Omega} u(\o,t)\phi_i(\o)\;d\o\in\r^n$ and $\phi_i$'s are the eigenfunctions of (\ref{Neumann_eigenvalue_problem}). 

\textbf{Claim $1$.} 
\be{u-ubar:thm:additive_F_PDE}
u(\o, t)-\bar u(t) \;=\; \dis\sum_{i\geq2} \a_i(t) \phi_i^{(n)} (\o).
\ee
Using the expansion of $u$ as in (\ref{u:expansion}), we have
\beqn
u(\o, t)-\bar{u} (t) &=& \a_1(t)\phi_1^{(n)}(\o)-\bar{u}(t) + \dis\sum_{i\geq2}  \phi_i ^{(n)}(\o)\a_i(t).
\eeqn
Multiplying both sides of the above equality by the constant eigenfunction $\phi_1$ and taking integral over $\Omega$, by orthonormality of $\phi_i$'s, we get:
\[
\dis\int_{\Omega} \lt(u(\o, t)-\bar{u} (t)\rt)\; d\o = \a_1(t).
\]
We showed that $\bar{u}(t)=\frac{1}{\abs{\Omega}} \int_{\Omega} u(\o,t) \; d\o$, hence $\a_1(t)=0$.
This proves Claim $1$. 

\textbf{Claim $2$.}
Fix $\o\in\Omega$. Then for any $i\geq1$, 
\beqn
\dot{\a}_i(t) &=& (A(t)-\l_i^{(n)}D) \a_i(t).
\eeqn
Using the expansion of $u$ as in (\ref{u:expansion}) and after omitting the arguments $\o, t$ for simplicity, we have:
\beqn
\dis\sum_{i\geq1} \dot{\a_i} \phi_i^{(n)} &=& \dot{u}
= A(t) u +D\Delta u \\
&= &A(t)\lt(\dis\sum_{i\geq1} \a_i \phi_i^{(n)}\rt) + D\Delta \lt(\dis\sum_{i\geq1} \a_i \phi_i^{(n)} \rt)\\
&=&\dis\sum_{i\geq1} \lt(A(t)-\l_i^{(n)}D\rt) \a_i \phi_i^{(n)}.
\eeqn
Multiplying both sides of the above equality by $\phi_i$ and taking integral over $\Omega$, by orthonormality of $\phi_i$'s
we get:
\[\dot{\a}_i(t) \;=\; (A(t)-\l_i^{(n)}D) \a_i(t).
\]
This proves Claim $2$. 

If for any $t$, $M\lt[ A(t)-\lambda_2^{(n)} D\rt]\leq c$, then for any $t$ and any $i>2$, $M\lt[A(t)-\lambda_i^{(n)} D\rt]\leq c$ too. Then by Claim $2$ and Lemma \ref{Dini_M}:
\be{}
\|\a_i(t)\| \leq e^{ct}\|\a_i(0)\|.
\ee
Using the above inequality and triangle inequality in Equation (\ref{u-ubar:thm:additive_F_PDE}), for any $\o\in\Omega$ and any $t$, we get the following inequality:
\beqn
\|u(\o, t)-\bar{u}(t)\|&\leq& \dis\sum_{i\geq2} \lt\|\a_i(t)  \phi_i^{(n)}(\o)\rt\|\\
& \leq& e^{ct}  \dis\sum_{i\geq2} \lt\|\a_i(0)  \phi_i^{(n)}(\o)\rt\|.
\eeqn
Specifically, when $c<0$, $\|u(\o, t)-\bar{u}(t)\| \to0$, exponentially as $t\to\infty$. 
\ep

We next prove an analogues result to Theorem \ref{contraction} (restricted to $p=1$), for reaction diffusion PDE (\ref{re-di}) subject to the Dirichlet boundary condition (\ref{DBC}).

Recall \cite{Henrot} that for any bounded, open subset  $\Omega\subset\mathbb{R}^m$, there
exists a sequence of positive eigenvalues 
$0< \lambda^{(d)}_1 \leq\lambda^{(d)}_2\leq\ldots$ (going to $\infty$, superscript $(d)$ for Dirichlet) and a sequence
of corresponding orthonormal eigenfunctions: 
$\phi^{(d)}_1 , \phi^{(d)}_2 , \ldots$ (defining a Hilbert basis of $L^2(\Omega)$)
satisfying the following Dirichlet eigenvalue problem:
\be{Dirichlet_eigenvalue_problem}
\bal
-\Delta \phi^{(d)}_i &=\lambda^{(d)}_i \phi^{(d)}_i \quad  \mbox{in $\Omega$} \\
\phi^{(d)}_i&=0  \quad  \mbox{on $\partial \Omega$}.
\eal
\ee
Let us assume that
$\Omega$ is a connected open set. Then the first eigenvalue $\l^{(d)}_1$ is simple
and the first eigenfunction $\phi^{(d)}_1$ has a constant sign on $\Omega$. Without loss of generality, $\phi^{(d)}_1$ can be assumed to be everywhere
positive on $\Omega$.

Let 
\[
{Y^{(d)}}\;:=\;\lt\{v\colon\bar\Omega\to\r^n,\; v\in\lt(C_{\r}^2\lt(\bar\Omega\rt)\rt)^n,\;v(\xi)=0,\;{\xi\in\partial\Omega}\rt\}.
\]

Then, for any $v=\lt(v_1,\cdots,v_n\rt)^T\in Y^{(d)}$, $v_i$ can be written as follows:
\be{expansion_vi}
v_i\;=\; \dis\sum_{k=1}^{\infty} \langle v_i,\phi^{(d)}_k \rangle \;\phi^{(d)}_k,
\ee
where $\langle v_i,\phi^{(d)}_k \rangle = \dis\int_{\Omega}v_i\;\phi^{(d)}_k.$

Now consider the following weighted norm on $Y^{(d)}$:

For $Q=\diag(q_1,\ldots, q_n)$, a positive diagonal matrix, and $\phi=\phi^{(d)}_1\geq0$, we define
\be{weighted_norm}
\|v\|_{1,\phi,Q} \;:=\; \dis\sum_{i=1}^n q_i\dis\int_{\Omega}\phi(\o)\abs{v_i(\o)}\;d\o, \quad \mbox{$v=(v_1,\ldots, v_n)^T$},
\ee
and let
\be{mathcal_V1}
{Y^{(d)}_{1,\phi}}:= \lt(Y^{(d)}, \|\cdot\|_{1,\phi,Q}\rt).
\ee

 In this section, by $M_{Y^{(d)}_{1,\phi}} [\cdot]$, we mean the (lub) logarithmic Lipschitz constant induced by $\|\cdot\|_{1,\phi,Q}$ on $Y^{(d)}$. 

Now we state and prove the following theorem which provides sufficient condition for contractivity of the reaction diffusion PDE (\ref{re-di}) subject to the Dirichlet boundary condition (\ref{DBC}):

\bthm\label{thm-dpde-contraction}
Consider the reaction diffusion PDE (\ref{re-di}) subject to the Dirichlet boundary condition (\ref{DBC}). Let 
\[c_1=\dis\sup_{(x,t)} M_{V_1}\lt[J_{F_t}(x)-\l^{(d)}_1 D\rt],\]
 where $M_{V_1}$ is the logarithmic norm induced by $Q$-weighted $L^1$ norm for a positive diagonal matrix $Q$ on $V$, and $\l_1^{(d)}$ is the first Dirichlet eigenvalue of $-\Delta$ on $\Omega$. Let $u(\omega,t)$ and $v(\omega,t)$ be two solutions of 
(\ref{re-di}) and (\ref{DBC}). Then 
\be{thm}
\|(u-v)(\cdot, t)\|_{1,\phi, Q}\leq e^{c_1t}\|(u-v)(\cdot, 0)\|_{1,\phi, Q}
\ee
where $\phi=\phi^{(d)}_1\geq0$ is the eigenfunction corresponding to $\l^{(d)}_1$.
\ethm 

To prove Theorem \ref{thm-dpde-contraction}, we need the following lemmas:
  
 \blem\label{claim2:thm:npde:1d}
Let $\Omega$ be an open {subset of $\r^m$}. 
Let $\mathcal{A}$ denote an $n \times n$ diagonal matrix of operators on $Y^{(d)}$ with the operators $d_i\Delta$ on the diagonal. Let $\Lambda^{(d)}$ denote an $n\times n$ diagonal matrix of operators on $Y^{(d)}$ with operators 
\[\Lambda_i^{(d)}(\psi)(\o):= \l_1^{(d)} d_i \psi_i(\o)\]
 on the diagonal.
  Then, for $p=1$,
\be{ineq:claim2:thm:dpde}
M^+_{Y^{(d)}_{1,\phi}}\lt[\mathcal{A}+\Lambda^{(d)}\rt]= 0,
\ee
where $M^+_{Y^{(d)}_{1,\phi}}$ is induced by $\|\cdot\|_{1, \phi, Q}$.
\elem

See Appendix for the proof. 

 \blem\label{M+:reaction}
 Let $G\colon\r^n\to\r^n$ be a (globally) Lipschitz function and define $\hat{G}\colon Y^{(d)}\to\r^n$ as follows:
 \[\hat{G}(u)(\o) := G(u(\o)).\]
 Then, 
 \be{eq:M+:reaction}
 M^+_{Y^{(d)}_{1,\phi}} [\hat{G}]\leq M^+_{V_1} [G] = \dis\sup_x M^+_{V_1} [J_G(x)]. 
 \ee 
 \elem
 
In \cite[Lemma 7]{aminzare-sontag}, we proved a simpler version of this lemma, namely, we showed that Equation (\ref{eq:M+:reaction}) holds when $\phi=1$, i.e. there is no weighted norm in the space. One can modify the proof of \cite[Lemma 7]{aminzare-sontag} to get Equation (\ref{eq:M+:reaction}). See Appendix for more details.

%proof of theorem
\textbf{Proof of Theorem \ref{thm-dpde-contraction}.} 
Suppose that $u$ is a solution of Equation (\ref{re-di}) defined on $\Omega\times [0, T)$. Note that for any $t$, $u(\cdot, t)\in Y^{(d)}$.

Define $\hat{u}\colon [0, T)\to Y^{(d)}$ by
\[\hat{u}(t)(\o) \;:=\; u(\o, t).\]
Also define $\mathcal{H}_t\colon Y^{(d)}\to\r^n$ as follows: for any $\psi\in Y^{(d)}$ and any $\o\in\Omega$:
\[\mathcal{H}_t(\psi)(\o) \;:=\; F_t(\psi(\o)).\]
Let $\mathcal{A}$ denote an $n \times n$ diagonal matrix of operators on $Y^{(d)}$ with the operators $d_i\Delta$ on the diagonal.
Then 
\be{PDE-ODE-Dirichlet}
\dis\frac{\partial\hat{u}}{\partial t}(t)\;=\; \lt(\mathcal{H}_t+\mathcal{A}\rt)\lt(\hat{u}(t)\rt).
\ee

Suppose $u$ and $v$ are two solutions of Equation (\ref{re-di}). By Lemma \ref{Dini_M}, we have:
\be{ineq:Dini:Dirichlet}
\bal
D^+\|(\hat{u}-\hat{v})(t)\|_{1,\phi,Q}
&\leq M^+_{Y^{(d)}_{1,\phi}}[\mathcal{H}_t+\mathcal{A}]\|(\hat{u}-\hat{v})(t)\|_{1, \phi,Q}.
\eal
\ee
 Let $\Lambda^{(d)}$ be as in Lemma \ref{claim2:thm:npde:1d}.
By subadditivity of $M^+$, Proposition \ref{subadd}, Lemma \ref{claim2:thm:npde:1d} and Lemma \ref{M+:reaction}, we have:
\be{ineq:subadd:Dirichlet:1}
\bal
M^+_{Y^{(d)}_{1,\phi}}[\mathcal{H}_t+\mathcal{A}]&\leq M^+_{Y^{(d)}_{1,\phi}}[\mathcal{H}_t-\Lambda^{(d)}] + M^+_{Y^{(d)}_{1,\phi}}[\mathcal{A}+\Lambda^{(d)}]\\
&\leq M^+_{Y^{(d)}_{1,\phi}}\lt[\mathcal{H}_t-\Lambda^{(d)}\rt]\\
&\leq \dis\sup_{x\in V}M_{V_1}\lt[J_{F_t}(x)-\l_1^{(d)}D\rt]\\
&\leq  \dis\sup_{t\in[0,T)}\sup_{x\in V}M_{V_1}\lt[J_{F_t}(x)-\l_1^{(d)}D\rt]\\
&= c_1.
\eal
\ee

 By (\ref{ineq:Dini:Dirichlet}), (\ref{ineq:subadd:Dirichlet:1}), and Lemma \ref{Dini_M}, we get:
 \[\|(\hat{u}-\hat{v})(t)\|_{1,\phi,Q}\leq e^{c_1t}\|(\hat{u}-\hat{v})(0)\|_{1,\phi,Q}.\] 
 In terms of the PDE (\ref{re-di}), this last estimate can be equivalently written as:
 \[\|(u-v)(\cdot, t)\|_{1,\phi,Q}\leq e^{c_1t}\|(u-v)(\cdot, 0)\|_{1,\phi,Q}.\]
\qed

{Note that unlike in Neumann boundary problems, one cannot conclude synchronization from contraction in the Dirichlet boundary problems unless for any $t$, $F(0,t)=0$:

\bcor\label{cor-dpde-contraction}
Under the conditions of Theorem \ref{thm-dpde-contraction}, if $F(0,t) = 0$, then $v=0$ is a uniformly spatial solution of Equations (\ref{re-di}) and (\ref{DBC}), and therefore, for any solution $u$ of Equations (\ref{re-di}) and (\ref{DBC}),
 \[\|u(\cdot, t)\|_{1,\phi,Q}\leq e^{c_1t}\|u(\cdot, 0)\|_{1,\phi,Q},\]
Hence, when $c_1<0$, the PDE system synchronizes. 
\ecor 
}

{The following theorem provides a sufficient condition for synchronization of reaction diffusion systems subject to the Neumann boundary condition restricted to $1D$ space and $p=1$. The proof is based on the results of Theorem \ref{thm-dpde-contraction}.}

%thm
\bthm\label{thm-1D-synch}
Let $u(\omega,t)$ be a solution of 
\begin{equation} \label{re-di-1D}
\bal
&\displaystyle\frac{\partial u}{\partial t}(\omega, t)=F(u(\omega, t), t)+D\; \dis\frac{\partial^2u}{\partial\omega^2}(\omega, t)\quad \mbox{on $(0,L)$}\\
&\dis\frac{\partial u}{\partial\omega}(0,t)\;=\;\dis\frac{\partial u}{\partial\omega}(L,t)\;=\;0,
\eal
\end{equation} 
defined for all $t\in[0,T)$ for some $0 < T\leq\infty$. In addition, assume that 
$u(\cdot, t)\in C^3(\Omega)$, for all $t\in[0,T)$. Let 
\[
c\;=\;\dis\sup_{t\in[0,T)}\dis\sup_{x\in V}M_{V_1}\lt[J_{F_t}(x)-\dis\frac{\pi^2}{L^2} D\rt],
\]
where $M_{V_1}$ is the logarithmic norm induced by $\|\cdot\|_{1,Q}$ for a positive diagonal matrix $Q$. 
 Then for all $t\in[0,T)$:
\be{thm-1D}
\lt\|\dis\frac{\partial u}{\partial\omega}(\cdot, t)\rt\|_{1,\phi,Q}\leq e^{ct}\lt\|\dis\frac{\partial u}{\partial\omega}(\cdot, 0)\rt\|_{1, \phi,Q},
\ee
where \[\|\cdot\|_{1,\phi, Q}:=\lt\|\sin(\pi\o/L) (\cdot)\rt\|_{1,Q}.\]
\ethm

The significance of Theorem~\ref{thm-1D-synch} lies in the fact that
$\sin(\pi\o/L)$ is nonzero everywhere in the domain (except at the boundary).
In that sense, we have exponential convergence to uniform solutions in a
weighted $L^1$ norm, the weights being specified in $V$ by the matrix $Q$ and
in space by the function $\sin\lt(\pi\o/L\rt)$.
 
\bp
Suppose that $u$ is a solution of Equation (\ref{re-di-1D})
defined on $[0, L]\times [0, T)$. Let $v=\dis\frac{\partial u}{\partial \o}$,
  then by taking $\frac{\partial}{\partial\o}$ 
in both sides
of Equation (\ref{re-di-1D}), we get the following PDE:
\be{PDE-for-v}
\dis\frac{\partial v}{\partial t}=J_{F_t}(u) v + D\Delta v,
\ee
subject to Dirichlet boundary condition:
$v(0)=v(L)=0.$ 

For $\Omega=(0,L)$, the first Dirichlet eigenvalue is $\pi^2/L^2$ and a corresponding eigenfunction is $\sin(\pi\o/L)$. Therefore, 
by Corollary \ref{cor-dpde-contraction}, 
 \[\|{v}(\cdot, t)\|_{1,\phi,Q}\leq e^{ct}\|{v}(\cdot, 0)\|_{1,\phi,Q},\] 
 where $c\;=\;\dis\sup_{t\in[0,T)}\dis\sup_{x\in V}M_{V_1}\lt[J_{F_t}(x)-\dis\frac{\pi^2}{L^2} D\rt].$
 \ep

\bremark
In the case of $\Omega=(0,L)$, $\l_1^{(d)}= \l_2^{(n)}$. 
\eremark

\bremark\label{synch_biochemical_example_better_rate}
In the Biochemical model, we showed that there exists a positive diagonal matrix $Q$ such that 
\[c:= \displaystyle\sup_{x\in V}M_{V}[J_F(x)] < 0,\]
 with norm $\|\cdot\|_{1,Q}$ on $V$. This condition implies that any solution of (\ref{example_reduced}) converges to a uniform solution with rate $c$ (Remark \ref{contraction->synchronization}). Next, we show that by Theorem \ref{thm-1D-synch}, any solution of (\ref{example_reduced}) 
converges to a uniform solution at a better rate than $c$:

 By subadditivity of $M_V$, we have:
\[\dis\sup_{x\in V}M_V[J_F(x)- \pi^2 D]\leq \dis\sup_{x\in V}M_V[J_F(x)] - \pi^2 d, \quad \mbox{where $d=\min\{d_1, d_2\}$}.\]
Therefore, $c_0:=\dis\sup_{x\in V}M_V[J_F(x)-\pi^2 D]< c<0$. Hence, by Theorem \ref{thm-1D-synch}, for any solution $u$ of (\ref{example_reduced}):
 \[
\lt\|\dis\frac{\partial u}{\partial \o}(\cdot, t)\rt\|_{1,\phi, Q}\;\leq \;e^{c_0t}\;\lt\|\dis\frac{\partial u}{\partial \o}(\cdot, 0)\rt\|_{1,\phi,Q},
\] 
where $\phi(\o)= \sin(\pi\o).$

{Figure \ref{biochemical_example_6figures} indicates two different solutions of the Biochemical model, Equation (\ref{example_reduced}), namely $(x_1,y_1)^T$ and $(x_2,y_2)^T$ on $\Omega=(0,2)$ for $z(t)= 20(1+\sin(10t))$, and for the following set of parameters:
\beqn
\delta=20, \; k_1=0.5, \; k_2=5, \; S_Y=0.1, \; d_1=0.001, \; d_2=0.1.
\eeqn

Also, in Figure \ref{biochemical_example_6figures}, the difference between two solutions has been shown that goes to zero as expected. 
}
\begin{figure}[htdp]
\begin{center}
\caption{Two different solutions of (\ref{example_reduced}) and their difference.}
\begin{subfigure}[b]{0.42\textwidth}
\includegraphics[scale=.27]{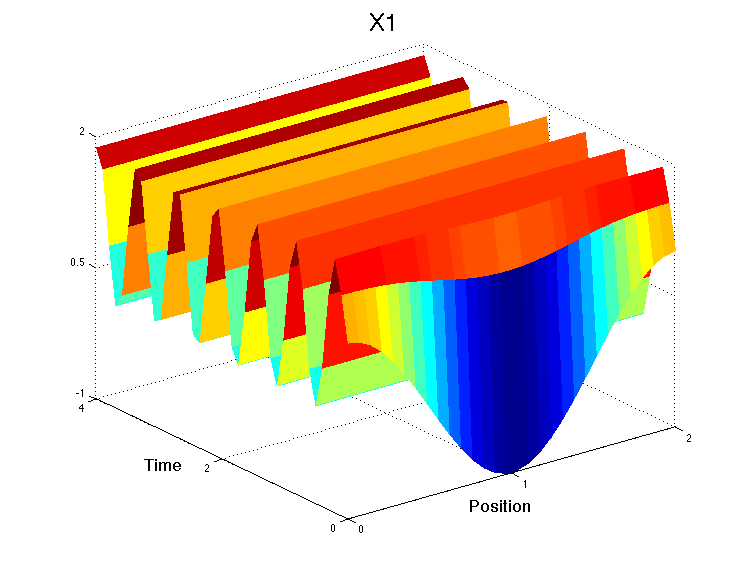}
\end{subfigure}
\qquad
\begin{subfigure}[b]{0.42\textwidth}
\includegraphics[scale=.27]{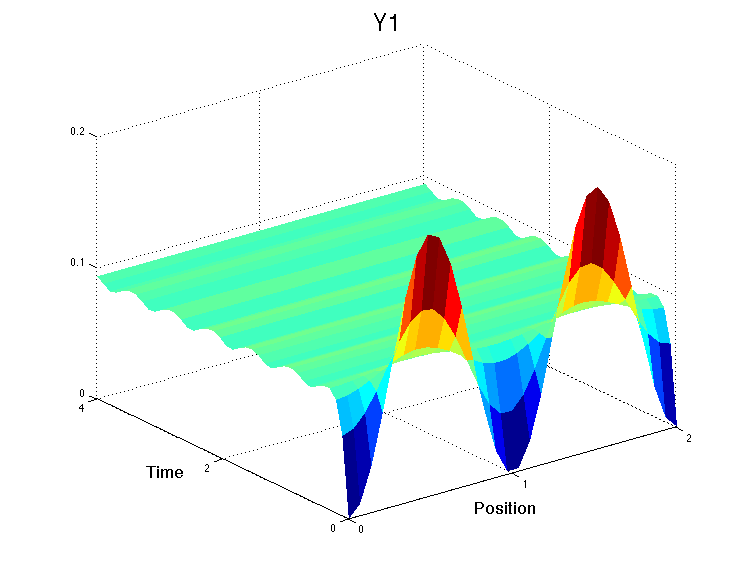}
\end{subfigure}

\begin{subfigure}[b]{0.42\textwidth}
\includegraphics[scale=.27]{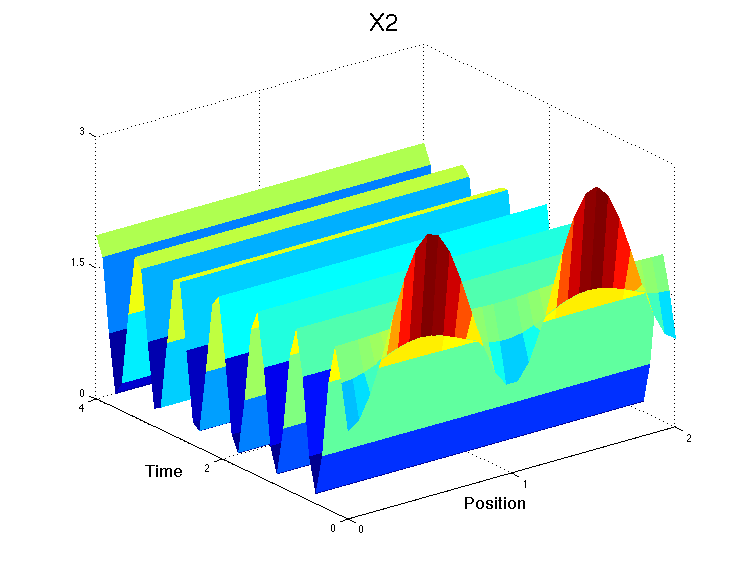}
\end{subfigure}
\qquad
\begin{subfigure}[b]{0.42\textwidth}
\includegraphics[scale=.27]{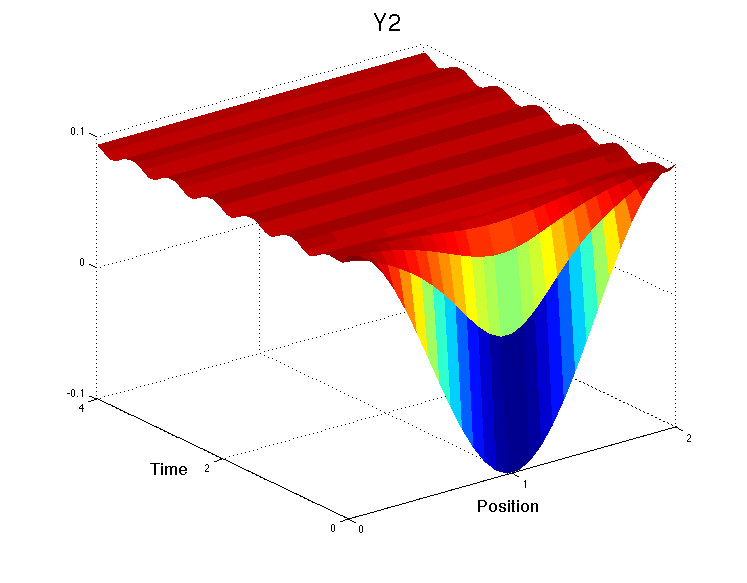}
\end{subfigure}

\begin{subfigure}[b]{0.42\textwidth}
\includegraphics[scale=.27]{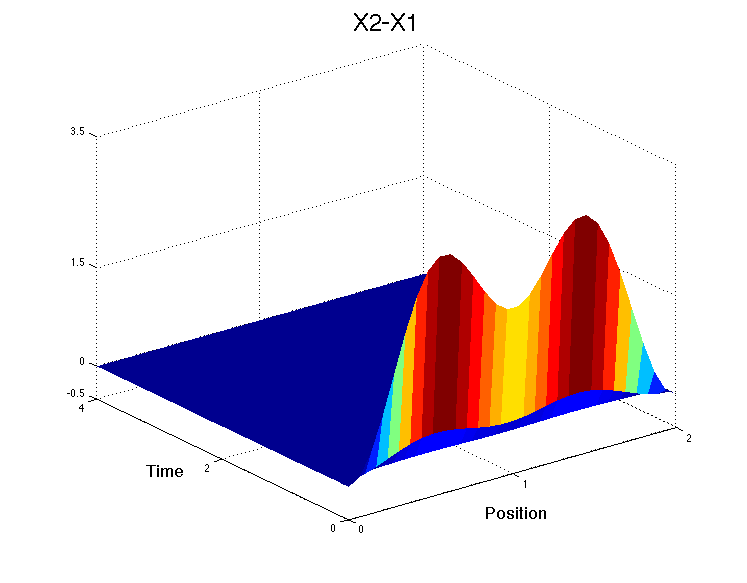}
\end{subfigure}
\qquad
\begin{subfigure}[b]{0.42\textwidth}
\includegraphics[scale=.27]{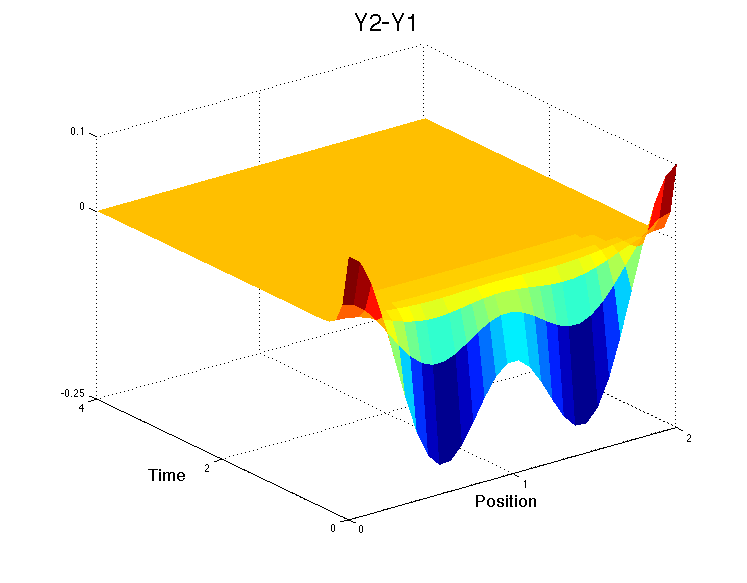}
\end{subfigure}
\quad

\label{biochemical_example_6figures}
\end{center}
\end{figure}
\eremark

{
In the following remark w compare our result, Theorem \ref{thm-1D-synch}, with the results in \cite{Arcak} and \cite{othmer-large-diffusion}.

\bremark\label{goodwin_compare_with_arcak_othmer}
Considering (\ref{Goodwin_PDE}), in (\cite{Arcak}, Equation 55), the following sufficient condition is given for synchronization:
\be{Arcak:Goodwin}
\dis\frac{\a\gamma a}{k(b+\l d_1)(\b+\l d_2)\l d_3}\;<\;4,
\ee
where $\l=\pi^2$. 

A simple calculations show that  for weighted matrix
$Q=\diag(1, 12, 11)$, and for $2.2/\pi^2<d_1$, and $d_2=d_3=0$,
\[
\dis\sup _{w=(x,y,z)^T}M_{1,Q}[J_F(w)-\pi^2 D]<0.
\]
Applying Theorem \ref{thm-1D-synch}, we conclude that for
${2.2}/{\pi^2}<d_1$, and $d_2=d_3=0$, (\ref{Goodwin_PDE}) synchronizes,
meaning that solutions tend to uniform solutions.

Note that when $d_3=0$, one cannot apply (\ref{Arcak:Goodwin}) directly to get
synchronization.

Figure~\ref{Goodwin_X_diffuces} shows the spatially uniformity of the
solutions of (\ref{Goodwin_PDE}), for the same parameter values and initial
conditions as in Figure \ref{Goodwin_no_diffusion}, when ${2.2}/{\pi^2}<d_1$,
and $d_2=d_3=0$.

In (\cite{othmer-large-diffusion}, Equation 3), Othmer provides a sufficient
condition for uniform behavior of the solutions of the reaction-diffusion
(\ref{re-di}) on $(0,L)$, subject to Neumann boundary conditions: 
 \be{othmer:condition}
 \sup_w\|J_F(w)\| < \pi^2/L^2 \min\{d_i\}.
 \ee
In Goodwin's example (\ref{Goodwin_PDE}), $\sup_w\|J_F(w)\|$ is positive and
 finite (the $\sup$ is taken at $z=0$), and $\min\{d_i\}=0$, hence
 (\ref{othmer:condition}) doesn't hold and this condition is not applicable
 for this example. 
\eremark
}

%------------------------------------------------------------------------------------------------
\smallskip
\section{Synchronization in a system of ODEs}\label{Synchronization in a system of ODEs}
%------------------------------------------------------------------------------------------------

In this section, we study a network of identical ODE models which are diffusively
interconnected.

The state of the system will be described by a vector $x$ which one may interpret as a vector collecting the states $x_i$ (each
of them itself possibly a vector) of identical ``agents'' which tend to
follow each other according to a diffusion rule, with interconnections
specified by an undirected graph.  Another interpretation, useful in the
context of biological modeling, is a set of chemical reactions among species
that evolve in separate compartments (e.g., nucleus, cytoplasm, membrane, in
a cell); then the $x_i$'s represent the vectors of concentrations of the species
in each separate compartment.

In order to formally describe the interconnections, we use the following concepts in this section:
\begin{itemize}

\item For a fixed convex subset of $\r^n$, say $V$, $\tilde{F}\colon V^{N} \times [0,\infty) \to\r^{nN}$ is a function of the form:
\be{Ftilde}\nonumber
\tilde{F}(x,t)=\left(F(x_1,t)^T, \ldots, F(x_N,t)^T\right)^T,
\ee
 where $x=\left(x_1^T,\ldots, x_N^T\right)^T$, with $x_i\in V$ for each $i$, and $F(\cdot, t):=F_t\colon V\to\r^n$ is a $C^1$ function. 
  
\item For any $x\in V^N$ we define $\normpQ{x}$ as follows:
\[\normpQ{x}=\left\|\left(\|Qx_1\|_p, \cdots, \|Qx_N\|_p\right)^T\right\|_p,\]
where $Q=\diag{(q_1,\ldots, q_n)}$ is a positive diagonal matrix and $1\leq p\leq\infty$.

With a slight abuse of notation, we use the same symbol for a norm in $\r^n$: \[\|x\|_{p,Q}:=\|Qx\|_p.\]

\item $D=\diag(d_1,\ldots, d_n)$ with $d_i\geq0$, and $d_j>0$ for some $j$, which we call the diffusion matrix.
\item $\mathcal{L}\in\r^{N\times N}$ is a symmetric matrix and $\mathcal{L}\mathbf{1}=0$, where $\mathbf{1}=(1,\ldots, 1)^T$. 
We think of $\mathcal{L}$ as the Laplacian of a graph that describes the interconnections among component subsystems. 
\item $\otimes$ denotes the Kronecker product of two matrices.  
\end{itemize}
 
\begin{definition}\label{}
 For any arbitrary graph $\mathcal{G}$ with the associated 
 (graph) Laplacian matrix $\mathcal{L}$, any diagonal
  matrix $D$, and any $F\colon V\times [0,\infty)\to\r^n$, the associated $\mathcal{G}-$compartment system, denoted by $(F, \mathcal{G}, D)$, is defined by
 \be{g-compartment}
 \dot{x}(t)=\tilde{F}(x(t), t)-(\mathcal{L}\otimes D)x(t),
 \ee
 where $x, \tilde{F},$ and $D$ are as defined above.
 \end{definition}
 
 The ``symmetry breaking'' phenomenon of diffusion-induced, or Turing,
instability refers to the case where a dynamic equilibrium $\bar u$ of the
non-diffusing ODE system $\dot x=F(x, t)$ is stable, but, at least for some diagonal
positive matrices $D$, the corresponding interconnected system (\ref{g-compartment}) is
unstable.

The following theorem (from \cite{aminzare-sontag}), shows that, for contractive reaction part $F$, 
 no diffusion instability will occur, no matter what is the size
of the diffusion matrix $D$.

\bthm\label{diffusive_ODE_contraction}
Consider the system $(\ref{g-compartment})$. Let
 \[c=\displaystyle\sup_{t\in[0, \infty)}\MpQ [F_t],\]
  where $\MpQ$ is the (lub) logarithmic Lipschitz constant induced by the norm $\|\cdot\|_{p,Q}$ on $\r^n$  defined by $\|x\|_{p,Q}:=\|Qx\|_p$. Then for any two solutions $x, y$ of $(\ref{g-compartment})$, we have 
\[
\normpQ{x(t)-y(t)}\leq e^{ct}\normpQ{x(0)-y(0)}.
\]
\ethm

\bremark
Under the assumptions of Theorem \ref{diffusive_ODE_contraction}, by Remark~\ref{M[f]:M[J]}, if for any $t\geq0$ and any $x$, 
$\MpQ[J_{F}(x,t)]\leq c$, then
 \[
 \normpQ{x(t)-y(t)}\leq e^{ct}\normpQ{x(0)-y(0)}.
 \]
\eremark 

\begin{definition}
We say that the $\mathcal{G}-$compartment system (\ref{g-compartment}) synchronizes, if for any solution $x=\lt(x_1^T,\ldots, x_N^T\rt)^T$ of (\ref{g-compartment}), and for all $i,j\in\{1,\ldots, N\}$, $(x_i-x_j)(t)\to 0$ as $t\to\infty$.
 \end{definition}

An easy first result is as follows.

\bprop
\label{contraction_to_synchronization}
Under the assumptions of Theorem \ref{diffusive_ODE_contraction}, if $c<0$, then the $\mathcal{G}-$compartment system (\ref{g-compartment}) synchronizes. 
\eprop
\bp
Note that $z(t):=\lt(z_1(t),\ldots, z_1(t)\rt)^T$ is a solution of
(\ref{g-compartment}), where $z_1(t)$ is a solution of $\dot{x}=F(x,t)$. By
Theorem \ref{diffusive_ODE_contraction}, if for any $t\geq0$ and any $x$,
$\MpQ[J_{F}(x,t)]\leq c$, then for any solution $x(t)$ of (\ref{g-compartment}),
\[
\normpQ{x(t)-z(t)}\leq e^{ct}\normpQ{x(0)-z(0)}.
\]
When $c<0$, $(x_i-z_1)(t)\to 0$, hence $(x_i-x_j)(t)\to 0$ as $t\to\infty$. 
\ep

In Proposition~\ref{contraction_to_synchronization}, we imposed a strong
condition on $F$, which in turn leads to the very strong conclusion that
all solutions should converge exponentially to a particular solution,
no matter the strength of the interconnection (choice of diffusion matrix). 
A more interesting and challenging problem is to provide a condition that
links the vector field, the graph structure, and the matrix $D$, so that
interesting dynamical behaviors  (such as oscillations in
autonomous systems, which are impossible in contractive systems) can be
exhibited by the individual systems, and yet the components synchronize.
The following example illustrates this question.

%------------------------------------------------------------------------------------------------
\smallskip
\subsubsection*{An example: synchronous autonomous oscillators}\label{Goodwin}
%------------------------------------------------------------------------------------------------

We consider the following three-dimensional system (all variables are
non-negative and all coefficients are positive):
 \be{ODE:Goodwin}
\bal
\dot{x}&\;=\;\dis\frac{a}{k+z}-bx\\
\dot{y}&\;=\;\a x-\beta y\\
\dot{z}&\;=\;\gamma y-\dis\frac{\delta z}{k_M+z}.
\eal
\ee
where $x,y,$ and $z$ are functions of $t$.
This system is a variation (\cite{Thron}) of a model, often called in
mathematical biology the ``Goodwin model,'' that was proposed in order to
describe a generic model of an oscillating autoregulatory gene, and
its oscillatory behavior has been well-studied \cite{Goodwin1965}.
It is sketched in Fig.~\ref{fig:Goodwin}.
In Goodwin's original formulation, $X$ is the mRNA transcribed from a given
gene, $Y$ an enzyme translated from this mRNA, and $Z$ a metabolite whose
production is catalyzed by $Y$.  It is assumed that $Z$, in turn, can
inhibits the expression of the original gene.
However, many other interpretations are possible. 
Fig.~\ref{fig:Goodwin_isolated}, shows non-synchronized
oscillatory solutions of
(\ref{ODE:Goodwin}) for $6$ different initial conditions, using the following
parameter values from the textbook \cite{Ingalls}: 
 \be{}\nonumber
 a=150,\; k=1,\; b=\a=\beta=\gamma=0.2,\; \delta=15,\; K_M=1.
 \ee
Fig.~\ref{fig:Goodwin_linear} shows the solutions of the same system ($6$
compartments, with the same initial conditions as in
Fig.~\ref{fig:Goodwin_isolated}) that are now interconnected diffusively by a
linear graph in which only $X$ diffuses, that is,
$D=\diag(d,0,0)$.
The following system of ODEs describes the evolution of the full system:
(in all equations, $i=1,\ldots, N$):
\be{}\nonumber
\bal
\dot{x}_i&\;=\;\dis\frac{a}{k+z_i}-b\;x_i  \;+d \lt(x_{i-1}-
2x_i +x_{i+1}\rt)
\\
\dot{y}_i&\;=\;\a\; x_i-\beta \;y_i\\
\dot{z}_i&\;=\;\gamma\; y_i- \dis\frac{\delta z_i}{k_M+z_i}
\eal
\ee
where for convenience we are writing 
$x_0=x_1$ and $x_N=x_{N+1}$.
In Fig.~\ref{fig:Goodwin_complete} we show solutions of the same system ($6$
compartments with the same initial conditions as in
Fig.~\ref{fig:Goodwin_isolated}) that are now interconnected, with the same $D$,
by a complete graph.
Observe that the second and ``more connected'' graph structure
(reflected, as discussed in the magnitude of its second Laplacian eigenvalue,
which is used in the conditions discussed below) leads to much faster
synchronization.

%fig
\begin{figure}[htdp]
\caption{Goodwin Oscillator}
\begin{center}
 \begin{subfigure}[b]{0.29\textwidth}
\includegraphics[scale=.24]{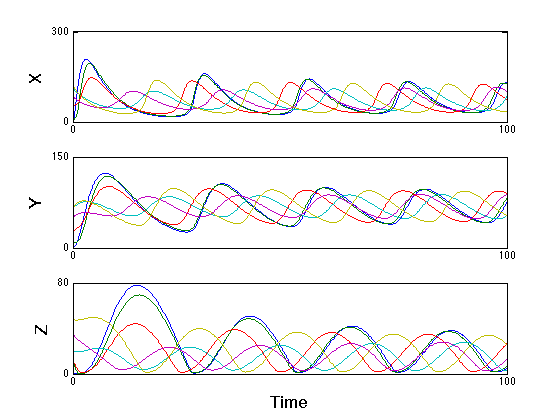}
\caption{{\scriptsize{$6$ isolated compartments}}}
\label{fig:Goodwin_isolated}
\end{subfigure}
\quad
\begin{subfigure}[b]{0.29\textwidth}
\includegraphics[scale=.24]{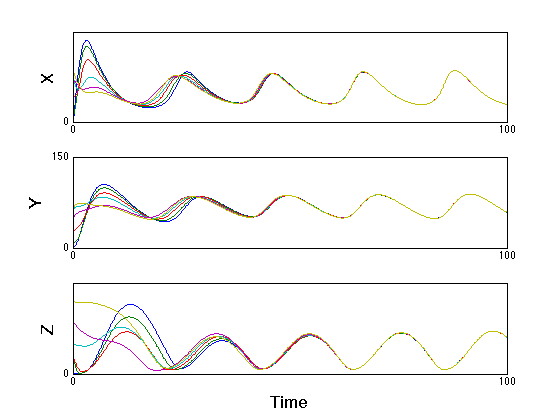}
\caption{{\scriptsize{Linear interconnection}}}
\label{fig:Goodwin_linear}
\end{subfigure}
\quad
\begin{subfigure}[b]{0.29\textwidth}
\includegraphics[scale=.24]{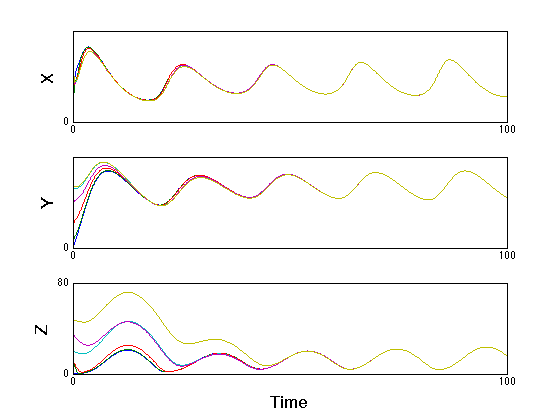}
\caption{{\scriptsize{Complete interconnection}}}
\label{fig:Goodwin_complete}
\end{subfigure}
\end{center}
\end{figure}

%------------------------------------------------------------------------------------------------
\smallskip
\subsubsection*{Synchronization conditions based on contractions}
%------------------------------------------------------------------------------------------------

In this section, we discuss several matrix measure based conditions that
guarantee synchronization of ODE systems; for additional results, see
\cite{Loh_Slo_98, slotine, Russo_Bernardo, slotine_wang, Chen2007}. 

We will use ideas from spectral graph theory, see for example \cite{Mesbahi_Egerstedt}.
Recall that a Laplacian matrix $\cal{L}$, with eigenvalues
$\l_1\leq\ldots\leq\l_N$, is always positive semi-definite
($0=\l_1\leq\ldots\leq\l_N$). In a connected graph, $\l_1$ is the only zero
eigenvalue and $v_1=(1, \ldots, 1)^T$ is the unique corresponding eigenvector
(up to a constant).  The second smallest eigenvalue, $\l_2$, is called the
\emph{algebraic connectivity} of the graph.  
This number helps quantify the ``how connected'' the graph is; for example, a
complete graph is ``more connected'' than a linear graph with the same number
of nodes, and this is reflected in the fact that
 the second eigenvalue of
the Laplacian matrix of a complete graph ($\l_2=N$) is larger that the second
eigenvalue of the Laplacian matrix of a line graph
($\l_2=4\sin^2\lt(\pi/2N\rt)$).

Consider a $\mathcal{G}-$compartment system, $(F, \mathcal{G}, D)$, where
$\mathcal{G}$ is any arbitrary graph. The following re-phasing of a theorem from
\cite{Arcak}, provides sufficient conditions on $F$ and $D$, based upon
contractions with respect to $L^2$ norms,
that guarantee synchrony of the associated $\mathcal{G}-$compartment system. 
We have translated the result to the language of contractions.
(Actually, the result in \cite{Arcak} is stronger, in that it allows for certain
non-diagonal diffusion and also certain non-diagonal weighting matrices $Q$, by
substituting these assumptions by a commutativity type of condition.)

\bthm
\label{theo:arcakODE}
Consider a $\cal{G}-$compartment as defined in Equation (\ref{g-compartment}) and
suppose that $V\subseteq \r^n$ is convex. For a given diagonal
positive matrix $Q$, let 
\be{}
c\;:=\;\dis\sup_{(x, t)}M_{2,Q} [J_F(x,t)-\l_2 D].
\ee
Then for every forward-complete solution $x$ that remains in $V$, the following inequality holds:
\be{}\nonumber
\lt\| \left(\begin{array}{c}x_1-\bar{x} \\\vdots \\x_N-\bar{x}\end{array}\right)(t)\rt\|_{2,I\otimes Q} \;\leq\; 
e^{ct} \lt\| \left(\begin{array}{c}x_1-\bar{x} \\\vdots \\x_N-\bar{x}\end{array}\right)(0)\rt\|_{2,I\otimes Q},
\ee
where $\bar{x}=(x_1+\ldots+x_N)/N$. 

In particular, if $c<0$, then
for any pair $i,j\in\{1,\ldots,N\}$, $(x_i-x_j)(t)\to 0$ exponentially as
$t\to\infty$. 
\ethm

Recall that a directed incidence matrix of a graph with $N$ nodes and $m$
edges, is an $N\times m$ matrix $E$ which is defined as follows, for any fixed
ordering of nodes and edges:
The $(i,j)-$entry of $E$ is $0$ if vertex $i$ and edge $e_j$ are not incident,
and otherwise, it is $1$ if $e_j$ originates at vertex $i$, and $-1$ if $e_j$
terminates at vertex $i$. In addition, the (graph) Laplacian matrix $\mathcal{L}$
of $\mathcal{G}$, is equal to $EE^T$.  
Observe that $E^T \mathcal{L} = E^T (E E^T)= (E^TE) E^T$, so this means that
$K:=E^TE$ satisfies
\be{K-eqn}
E^T\mathcal{L} = KE^T \,.
\ee
{Note that $E^TE$ is called edge Laplacian of $\mathcal{G}$. }
If $E^TE$ is nonsingular ({e.g. in linear graphs}), then $K=E^TE$ is the unique matrix satisfying
\eqref{K-eqn}.
However, in general, $K$ is not necessarily unique. 
For example, suppose that $\mathcal{G}$ is a complete
graph.
 Then 
  $E^TEE^T= N E^T$ 
  (see the proof of Proposition \ref{complete-prop}).
   So one can pick $K=NI$, where $I$
is the identity matrix.
Since $E^TE\neq NI$, in a complete graph,
this gives an alternative choice of
$K$.

The following theorem, provides a
sufficient condition on $F, D,$ and $\mathcal{G}$ that guarantees synchrony of
the associated $\mathcal{G}-$compartment system in any norm.  

%thm
\bthm\label{general:graph:thm}
Consider a $\mathcal{G}-$compartment system, $(F, \mathcal{G}, D)$, where
$\mathcal{G}$ is an arbitrary graph of $N$ nodes and $m$ edges, and a norm
$\|\cdot\|$ on $\r^{nm}$. 
Let $E$ be a directed incidence matrix of $\mathcal{G}$, and pick any
$m\times m$ matrix $K$ satisfying~\eqref{K-eqn}.
Denote:
\be{c:general_thm}
c\;:=\; \dis\sup_{(w,t)} M \lt[J(w,t) - K\otimes D\rt],
\ee
where $M$ is the logarithmic norm induced by $\|\cdot\|$, and $J(w,t)$ is
defined as follows: 
\[
J(w,t) = \diag \lt( J_F(w_1,t), \ldots, J_F(w_m,t)\rt),
\]
where
\[ 
w=\lt(w_1^T, \ldots,
w_m^T\rt)^T
\]
and $J_F$ denotes the Jacobian of $F$. 
Then
\be{}\nonumber
\lt\| \lt(E^T\otimes I\rt) x(t)\rt\| \; 
\leq \; e^{ct} \lt\| \lt(E^T\otimes I\rt) x(0)\rt\|.
\ee 
\ethm
Note that $\lt(E^T\otimes I \rt) x$ is a column vector whose entries are the
differences $x_i-x_j$, for each edge $e =\{i,j\}$ in $\mathcal{G}$. Therefore,
if $c<0$, the system synchronizes. 

\bp
Assume that $x$ is a solution of 
\[\dot{x}=\tilde{F}(x,t) -\lt(\mathcal{L}\otimes D\rt) x.\]
Let's define $y$ as follows: for any $t$,
\[y(t):= \lt(E^T\otimes I\rt)x(t) = \left(\begin{array}{c}x_{i_1}-x_{j_1} \\\vdots \\x_{i_m}-x_{j_m}\end{array}\right),\]
where for $k=1,\ldots, m$, $x_{i_k}-x_{j_k}$ indicates the $k$th edge of
$\cal{G}$, i.e., the difference between states associated to the two
  nodes that constitute the edge, and $I$ is the $n\times n$ identity matrix.
Then, using the Kronecker product identity 
\[(A\otimes B)(C\otimes D)=AC\otimes BD\] 
for matrices $A, B, C,$ and $D$ of appropriate dimensions, we have:
\be{y:general:graph:thm}\nonumber
\bal
\dot{y} &= \lt(E^T\otimes I\rt)\dot{x}\\
&=\lt(E^T\otimes I\rt)\lt(\tilde{F}(x,t) -\lt(\mathcal{L}\otimes D\rt) x\rt)\\
&= \lt(E^T\otimes I\rt)\tilde{F}(x,t) -\lt(E^T \mathcal{L}\otimes D\rt) x\\
&= \lt(E^T\otimes I\rt)\tilde{F}(x,t) -\lt(KE^T\otimes D\rt) x\\
&= \lt(E^T\otimes I\rt)\tilde{F}(x,t) -\lt(K\otimes D\rt) \lt( E^T\otimes I\rt)x\\
&= \lt(E^T\otimes I\rt)\tilde{F}(x,t) -\lt(K\otimes D\rt) y,
\eal
\ee 
where 
\be{}\nonumber
 \lt(E^T\otimes I\rt)\tilde{F}(x,t)= \left(\begin{array}{c}F(x_{i_1},t)-F(x_{j_1},t) \\\vdots \\F(x_{i_m},t)-F(x_{j_m},t)\end{array}\right).
\ee
Now let $u$, $v$, and $G$ be as follows:
\[u:=\left(\begin{array}{c}x_{i_1} \\\vdots \\x_{i_m}\end{array}\right), v:= \left(\begin{array}{c}x_{j_1} \\\vdots \\x_{j_m}\end{array}\right),\]
\[ G_t(u):= \left(\begin{array}{c}F(x_{i_1},t) \\\vdots \\F(x_{i_m},t)\end{array}\right)-\lt(K\otimes D\rt)\left(\begin{array}{c}x_{i_1} \\\vdots \\x_{i_m}\end{array}\right) ,\]
then 
\[\dot{u}-\dot{v} = G_t(u)-G_t(v).\]
By Remark \ref{M[f]:M[J]}, 
\be{}\nonumber
  \|u(t)-v(t)\|\;\leq\; e^{ct}\|u(0)-v(0)\|,
  \ee
  where 
  \[c \;=\; \dis\sup_{(w,t)}M\lt[J_{G_t}(w)\rt] \;=\;\dis\sup_{(w,t)}M\lt[J(w,t)- K\otimes D\rt].
   \]
\ep

{The following corollary is already known, see \cite{Arcak}, but we show here how it
follows from Theorem \ref{general:graph:thm} as a special case.

\begin{corollary}\label{tree:L2}
Consider a $\mathcal{G}-$compartment system, $(F, \mathcal{G}, D)$, where
$\mathcal{G}$ is a tree (graphs with no cycles) and denote
\[c\;:=\; \dis\sup_{(w,t)} M_{2,Q} \lt[J_F(w,t) - \l D\rt],\]
where $\l$ is the smallest nonzero eigenvalue of the Laplacian of $\mathcal{G}$ and $Q$ is a positive diagonal matrix. 
Then 
\beqn
\lt\| \lt(E^T\otimes I\rt) x(t)\rt\|_{2, I\otimes Q} &\leq& e^{ct} \lt\| \lt(E^T\otimes I\rt) x(0)\rt\|_{2, I\otimes Q}.
\eeqn
where $I$ is the identity matrix of appropriate size and $E$ is a directed incidence matrix of $\mathcal{G}$. 
\end{corollary}

We will need the following lemmas to prove Corollary \ref{tree:L2}. 

\blem\label{edge:graph:eigenvalues} \cite{Mesbahi2007} 
Let $\mathcal{G}$ be a connected directed graph with incidence matrix $E$ and edge Laplacian $\mathcal{K}=E^TE$ and (graph) Laplacian $\mathcal{L}=EE^T$. Then
\begin{enumerate}
\item The nonzero eigenvalues of $\mathcal{K}$ are equal to the nonzero eigenvalues of $\mathcal{L}$.
\item The null space of the edge Laplacian depends on the number of cycles in the graph. In particular, the null space of a tree is equal to $0$, i.e. all the eigenvalues are nonzero. 
\end{enumerate}
\elem

\blem\label{block_diagonal_formula_M_p}
Let $A$ be a block diagonal matrix with matrices $A_1,\ldots, A_n$ on its diagonal. Then for any $1\leq p\leq \infty$,
\[M_{p} [A] \leq \max\lt\{M_p[A_1], \ldots, M_p[A_n]\rt\}.\]
\elem
See the Appendix for the proof. 

\newtheorem*{proofofcorollarytree}{Proof of Corollary \ref{tree:L2}}
\begin{proofofcorollarytree}
Let $K=E^TE$ and $J=\diag\lt(J_F(w_1,t),\ldots, J_F(w_m,t)\rt)$, where $m$ is the number of edges of $\mathcal{G}$. By subadditivity of $M$, 
\be{inequality:tree:L2}
\bal
M_{2,I\otimes Q} \lt[J(w,t) - K\otimes D \rt]&\leq M_{2,I\otimes Q} \lt[J(w,t) - \l I\otimes D \rt]\\
&\;+\;M_{2,I\otimes Q} \lt[\l I\otimes D - K\otimes D \rt]
\eal
\ee 
We first show that the second term of the right hand side of the above inequality is zero. 
By Lemma \ref{edge:graph:eigenvalues}, $\l$ is the smallest eigenvalue of the edge Laplacian, $E^TE$, so the largest eigenvalue of $\l I-K$ and hence $(\l I-K)\otimes D$ is $0$. Therefore,
\beqn
M_{2, I\otimes Q} [(\l I-K)\otimes D]&=&
M_2\lt[\lt(I\otimes Q\rt)\lt((\l I-K)\otimes D\rt)\lt(I\otimes Q^{-1}\rt)\rt]\\
&=&M_2\lt[(\l I-K)\otimes D\rt]\\
&=& \mbox{largest eigenvalue of $(\l I-K)\otimes D$}=0.
\eeqn 
Next, we will show that the first term of the right hand side of Equation (\ref{inequality:tree:L2}) is $\leq c$. 

 By Lemma \ref{block_diagonal_formula_M_p},
\beqn
  M_{2, I\otimes Q} \lt[J(w,t) - \l I\otimes D\rt] 
  &\leq&\max_i\lt\{M_{2,Q}\lt[J_F(w_i,t)-\l D\rt]\rt\}.
\eeqn
By taking $\dis\sup$ over all $w=(w_1^T,\ldots, w^T_{m})^T$ and all $t\geq0$, we get 
  \be{}\nonumber
   \dis\sup_{(w,t)}M_{2, I\otimes Q}\lt[J(w,t)-K\otimes D\rt]= \dis\sup_t\sup_{x\in\r^n}M_{2,Q}\lt[J_F(x,t)-\l D\rt]=c.
   \ee
Now by applying Theorem~\ref{general:graph:thm}, we obtain the desired inequality. 

\qed
\end{proofofcorollarytree}
}

{In the following section, we will see the application of
Theorem~\ref{general:graph:thm} to complete graphs (Proposition \ref{complete-prop}) and linear graphs (Proposition \ref{thm:ode:linear}) in any weighted $L^p$ norm, $1\leq p\leq \infty$.}

We next specialize to the linear case, $F(x,t) = A(t) x$. 

% linear case
\bthm\label{linear_synchronization}
Consider a $\mathcal{G}-$compartment system, $(F, \mathcal{G}, D)$, and
suppose that $F(x,t) = A(t) x$, i.e.,
\be{linearF:diffusive_interconnected}
\dot{x}(t) \;=\; \lt(I \otimes A(t) - \mathcal{L}\otimes D\rt)x(t).
\ee
 For a given arbitrary norm in $\r^{n}$, $\|\cdot\|$, suppose that 
\[
\dis\sup_tM [A(t)-\l_2 D] < 0,
\]
where $\l_2$ is the smallest nonzero eigenvalue of the Laplacian matrix
$\mathcal{L}$ and $M$ is the logarithmic norm induced by $\|\cdot\|$. Then, for any $i, j\in \{1, \ldots, N\}$, $(x_i-x_j)(t) \to 0$, exponentially as $t\to\infty.$ 
\ethm

\bp 
Note that any solution $x$ of Equation (\ref{linearF:diffusive_interconnected})
can be written as follows:
\[
x(t) \;= \;
\dis\sum_{i=1,\ldots, N}\;\dis\sum_{j=1,\ldots, n} 
c_{ij}(t) \; (v_i\otimes e_j)
\]
where $v_i$'s, $v_i\in\r^N$ are a set of orthonormal eigenvectors of $\mathcal{L}$
(that make up a basis for $\r^N$), corresponding to the eigenvalues $\l_i$'s of
$\cal{L}$, 
where we assume that the eigenvalues are ordered, and $\l_1=0$,
and the $e_j$'s are the standard basis of $\r^n$. In addition, $c_{ij}$'s are
the coefficients that satisfy 
\[
\dot{C}(t) = \left(\begin{array}{ccc}A(t)-\l_1 D &  &  \\ & \ddots &  \\ &  & A(t)-\l_ND\end{array}\right) C(t),
\]
where $C= \lt(c_{11}, \ldots, c_{1n}, \ldots, c_{N1}, \ldots, c_{Nn}\rt)^T$,
with appropriate initial conditions.
By the definition of $y$, $y=(E^T\otimes I) x$, we have 
\be{}\nonumber
\bal
y(t) &= 
\dis\sum_{i=1,\ldots, N}\;\dis\sum_{j=1,\ldots, n} 
c_{ij}(t) 
\; \lt(E^Tv_i\otimes e_j\rt)\\
&= \dis\sum_{i=2,\ldots, N}\;\dis\sum_{j=1,\ldots, n} c_{ij}(t) \; \lt(E^Tv_i\otimes e_j\rt)
\eal
\ee
because
$E^T v_1= 0$ 
(where $v_1=(1/\sqrt{n})(1,\ldots,1)^T$).
Therefore, if $\dis\sup_tM[A(t)-\l_2D]<0$, then $\dis\sup_tM[A(t)-\l_iD]<0,
i=2,\ldots,N$, and by Lemma \ref{Dini_M}, the $c_{ij} (t)$'s, 
for $j\geq 2$, and hence
also $y(t)$, converge to $0$ exponentially as $t\to \infty$.
\ep

%------------------------------------------------------------------------------------------------
\smallskip
\subsection{Some special graphs}
%------------------------------------------------------------------------------------------------

While the results for measures based on Euclidean norm are quite general, for
$L^p$ norms, $p\not=2$, we only have special cases to discuss, depending on
the graph structure.
We present sufficient conditions for synchronization for some special graphs
(linear, complete, star),
and compositions of them (Cartesian product
graphs).  
See Table \ref{table:atomic} and Table \ref{table:cartesian} for a summary of the results that will be proved in this section. 
\begin{table}[htdp]

\caption{sufficient condition for synchronization in complete, line and star graphs of $N$ nodes.
If no subscript is used in $M$, the result has been proved for arbitrary norms.}
\begin{center}
\begin{tabular}{|c|c|c|}
\hline
graph & second eigenvalue, $\l_2$ & synchronization condition\\
\hline
complete &&\\
 &$N$ & $M[J_F-ND] <0$\\
 \begin{tikzpicture}

 \path node at ( 72:.8cm)(1) [circle,fill=blue!20,draw,scale=0.5] {};
  \path node at (144: .8cm)(2) [circle,fill=blue!20,draw,scale=0.5] {};
 \path node at ( 216: .8cm)(3) [circle,fill=blue!20,draw,scale=0.5] {};
 \path node at ( 288: .8cm)(4) [circle,fill=blue!20,draw,scale=0.5] {};
 \path node at ( 360: .8cm)(5) [circle,fill=blue!20,draw,scale=0.5] {};

   \path[every node/.style={font=\sffamily\small}]
    (1) edge node[] {} (2)
    edge node[] {} (3)
    edge node[] {} (4)
    edge node[] {} (5)
    (2) edge node[] {} (3)
     edge node[] {} (4)
    edge node[] {} (5)
    (3) edge node[] {} (4)
    edge node[] {} (5)
    (4) edge node[] {} (5)
      ;
\end{tikzpicture} 
&  & \\
&&\\
\hline
line &&\\
 
\begin{tikzpicture}[-,>=stealth',shorten >=1pt,auto,node distance=1.5cm,
  thin,main node/.style={circle,fill=blue!20,draw,scale=0.5}]

  \node[main node] (1) {};
  \node[main node] (2) [ right of=1] {};
  \node[main node] (3) [ right of=2] {};
  \node[main node] (4) [ right of=3] {};
  \node[main node] (5) [ right of=4] {};

  \path[every node/.style={font=\sffamily\small}]
    (1) edge node[] {} (2)
    (2) edge node[] {} (3)
    (3) edge node[] {} (4)
    (4) edge node[] {} (5)
    (5) 
      ;
\end{tikzpicture}
&$4\sin^2(\pi/2N)$ & $M_{p,Q}\lt[J_F-\l_2D\rt] <0$\\
& & \\
 &&\\
\hline

star &&\\
 &$1$ & $M\lt[J_F-\l_2D\rt] <0$\\
 \begin{tikzpicture}[-,>=stealth',shorten >=1pt,auto,node distance=1.5cm,
  thin,main node/.style={circle,fill=blue!20,draw,scale=0.5}]

  \node[main node] (1) {};
  \node[main node] (2) [ above right of=1] {};
  \node[main node] (3) [ below right of=1] {};
  \node[main node] (4) [ below left of=1] {};
    \node[main node] (5) [ above left of=1] {};

  \path[every node/.style={font=\sffamily\small}]
    (1) edge node[] {} (2)
    (1) edge node[] {} (3)
    (1) edge node[] {} (4)
    (1) edge node[] {} (5)   
       ;
\end{tikzpicture}

&  & \\
\hline
\end{tabular}
\end{center}
\label{table:atomic}

\end{table}%

%%%%%%%

\begin{table}[htdp]

\caption{Sufficient conditions for synchronization in cartesian products of $K$
  line and complete graphs.
If no subscript is used in $M$, the result has been proved for arbitrary norms.}
\begin{center}
\begin{tabular}{|c|c|c|}
\hline
graph & second eigenvalue, $\l_2$ & synchronization condition\\
\hline
&&\\
 hypercube && \\
 &   $4\dis\min_{1\leq i\leq K}\lt\{\sin^2(\pi/2N_i)\rt\}$&$M_{p,Q}[J_F-\l_2D]<0$\\
 \begin{tikzpicture}[-,>=stealth',shorten >=1pt,auto,node distance=1.5cm,
  thin,main node/.style={circle,fill=blue!20,draw,scale=0.5}]

  \node[main node] (1) {};
  \node[main node] (2) [ right of=1] {};
  \node[main node] (3) [ below left of=1] {};
  \node[main node] (4) [ below left of=2] {};
  \node[main node] (5) [ below  of=3] {};
    \node[main node] (6) [ below  of=4] {}; 
     \node[main node] (7) [ below  of=2] {};
   \path[every node/.style={font=\sffamily\small}]
    (1) edge node[] {} (2)
    edge node[] {} (3)
    (2) edge node[] {} (7)
     edge node[] {} (4)
    (3) edge node[] {} (1)
     edge node[] {} (5)
    edge node[] {} (4)
    (4) edge node[] {} (2)
     edge node[] {} (3)
    edge node[] {} (6)
    (5) edge node[] {} (3)
     edge node[] {} (6)
    (6) edge node[] {} (4)
     edge node[] {} (5)
    edge node[] {} (7)
    (7) edge node[] {} (2)
    edge node[] {} (6)
        ;
\end{tikzpicture} 
&&  \\
&&\\
\hline
 &&\\

Rook & $\min\{N_1,\ldots,N_K\}$ & $M[J_F-\l_2D]<0$ \\
 \begin{tikzpicture}
  \path node at ( -1.8,.5)(1) [circle,fill=blue!20,draw,scale=0.5] {}
        node at ( -1.5,1.5)(2) [circle,fill=blue!20,draw,scale=0.5] {}
        node at ( -1.5, -.5) (3)[circle,fill=blue!20,draw,scale=0.5] {}
 
  node at ( -.3, .35)(4) [circle,fill=blue!20,draw,scale=0.5] {}
        node at ( 0,1.8)(5) [circle,fill=blue!20,draw,scale=0.5] {}
        node at ( 0, -.8) (6)[circle,fill=blue!20,draw,scale=0.5] {}
  
   node at ( 1,.5)(7) [circle,fill=blue!20,draw,scale=0.5] {}
        node at ( 1.3,1.5)(8) [circle,fill=blue!20,draw,scale=0.5] {}
        node at ( 1.3, -.5)(9) [circle,fill=blue!20,draw,scale=0.5] {};
   \path[]
 (1) edge node[] {} (2)
      edge node[] {} (3)
      edge node[] {} (4)
      edge node[] {} (7)
 (2) edge node[] {} (3)
      edge node[] {} (5)
      edge node[] {} (8)
 (3) edge node[] {} (6)
      edge node[] {} (9)
(4)  edge node[] {} (5)
      edge node[] {} (6)
      edge node[] {} (7)
(5) edge node[] {} (6)
      edge node[] {} (8)
(6) edge node[] {} (9)
(7) edge node[] {} (8)
      edge node[] {} (9)
(8) edge node[] {} (9)
(9) ;
        \end{tikzpicture}
 &&\\
\hline

\end{tabular}
\end{center}
\label{table:cartesian}

\end{table}%

%------------------------------------------------------------------------------------------------   Two compartments
\smallskip
\subsubsection*{Two compartments}
%------------------------------------------------------------------------------------------------

We first study the relatively trivial case of a system with two compartments,
$N=2$, shown in this graph:
\begin{center} 
 \begin{tikzpicture}[-,>=stealth',shorten >=1pt,auto,node distance=2cm,
  thick,main node/.style={circle,fill=blue!20,draw,font=\sffamily\Large\bfseries}]

  \node[main node] (1) {};
  \node[main node] (2) [ right of=1] {};

  \path[every node/.style={font=\sffamily\small}]
    (1) edge node[] {} (2)
         ;
\end{tikzpicture}
\end{center} 

Since it makes no difference in the proof, we allow in this case a
  ``nonlinear diffusion'' term represented by a function $h$ which need not be
  linear:
\be{2-comp:ode}
\bal
\dot{x}_1\;&=\;F(x_1, t)+h_1(x_2)-h_1(x_1)\\
\dot{x}_2\;&=\;F(x_2, t)+h_2(x_1)-h_2(x_2)\\
\eal
\ee

%prop
\bprop\label{2-comp}
Let $c=\dis\sup_{(x,t)} M\lt[J_F(x,t)-(J_{h_1}+J_{h_2})(x)\rt]$, and $(x_1,x_2)^T$ be a solution of (\ref{2-comp:ode}). Then 
\[\|x_1(t)-x_2(t)\|\;\leq\; e^{ct}\|x_1(0)-x_2(0)\|,\]
where $\|\cdot\|$ is an arbitrary norm in $\r^n$ and $M$ is the logarithmic norm induced by $\|\cdot\|$.
\eprop

%Proof of Proposition \ref{2-comp}
\bp
Note that $\dot{x}_1-\dot{x}_2 = G_t(x_1)- G_t(x_2)$ where $G_t(x)=F(x,t)-(h_1+h_2)(x)$. By Remark \ref{M[f]:M[J]}, 
\[
\|x_1(t)-x_2(t)\|\;\leq\; e^{ct}\|x_1(0)-x_2(0)\|,
\]
where $c= \dis\sup_{(x,t)} M\lt[J_{G_t}(x)\rt] = \dis\sup_{(x,t)} M\lt[J_F(x,t)- (J_{h_1}+J_{h_2})(x)\rt]$.
\ep

%------------------------------------------------------------------------------------------------     Linear Graphs
\smallskip
\subsubsection*{Linear Graphs}
%------------------------------------------------------------------------------------------------

Now consider a system of $N\geq3$ compartments, $x_1,\ldots, x_N$, that are
connected to each other by a linear graph $\mathcal{G}$.

Assuming $x_0=x_1,\; x_{N+1}=x_N$, 
the following system of ODEs describes the evolution of the individual agent
$x_i$, for $i=1, \ldots, N$: 
\be{ODE:linear}
\bal
\dot{x}_i\;=\;F(x_i,t)+ D(x_{i-1}-x_i+x_{i+1}-x_i).\\
\eal
\ee
The following $N\times N$ matrix indicates the Laplacian matrix of a linear graph of $N$ nodes:
\be{Laplacian_linear_graph}
\cal{L} = \left(\begin{array}{ccccc}1 & -1 &  &  &  \\-1 & 2 & -1 &  &  \\ &  & \ddots &  &  \\ &  & -1 & 2 & -1 \\ &  &  & -1 & 1\end{array}\right).
\ee

Before stating and proving the main result of this section, we state the following lemma about the eigenvalues of tridiagonal matrices. For more details see \cite{tridiagonal_eigenvalue}.

\blem\label{tridiagonal_eigenvalue}
Denote by $M=M(v,a,b,s,t)$ the $n\times n$ tridiagonal matrix 
\[M = \left(\begin{array}{ccccc}a+v & t &  &  &  \\s & v & t &  &  \\ &  \ddots& \ddots &\ddots  &  \\ & & s & v & t \\ &  &  & s & b+v\end{array}\right),\]
where $v,a,b,s,t\in\r$. Let $\sigma=\sqrt{st}$, and assume that $\l_1\leq\ldots\leq\l_n$ are the eigenvalues of $M$.
Then
\begin{enumerate}
\item 
For $a=b=0$, and $k=1,\ldots,n$, $\l_k= v-2\sigma\cos\lt(\dis\frac{(n+1-k)\pi}{n+1}\rt)$.
\item 
For $a=b=\sigma$, and $k=1,\ldots,n$, $\l_k= v-2\sigma\cos\lt(\dis\frac{(n+1-k)\pi}{n}\rt)$.
\end{enumerate}
\elem

Note that in $-\mathcal{L}$, as defined in (\ref{Laplacian_linear_graph}), $v=-2$, and $a=b=s=t=\sigma=1$. Therefore, by Lemma \ref{tridiagonal_eigenvalue}, 
\[\l_2(\mathcal{L}) = -\l_2(-\mathcal{L}) = 2+2\cos\lt({(N-1)\pi}/{N}\rt) = 4\sin^2(\pi/2N).\]

We next state the Perron-Frobenius Theorem which we will use to prove the main result of this section.
\bthm
Let $A$ be an $n\times n$ Metzler (meaning that its off-diagonal entries are non-negative) matrix. Then the following statements hold.
\begin{enumerate}
\item There is a real number $\l^*$, called the Perron-Frobenius eigenvalue, such that $\l^*$ is an eigenvalue of $A$ and $\abs{\l^*}>\l$ for any other eigenvalue $\l$ of $A$.
\item The Perron-Frobenius eigenvalue is simple. Consequently, the left and right eigenspace associated to $\l^*$ is one-dimensional. 
\item There exist a left and a right eigenvector $v = (v_1, \ldots, v_n)$ of $A$ corresponding to eigenvalue $\l^*$ such that all components of $v$ are positive.
\item There are no other positive left and right eigenvectors except positive multiples of $v$.\end{enumerate}
\ethm

The following result is an application of Theorem~\ref{general:graph:thm} to linear graphs.

%prop
\bprop\label{thm:ode:linear}
Let $\{x_i\}$ be solutions of (\ref{ODE:linear}), and let 
\be{c:linear}
c=\dis\sup_{(x,t)} M_{p,Q}\lt[J_F(x,t)- 4\sin^2\lt({\pi}/{2N}\rt)D\rt],
\ee
for $1\leq p\leq\infty$.
 Then
\be{equ:linear:thm}
\left\|\left(\begin{array}{c}e_1 \\
e_2\\
\vdots \\
e_{N-1}\end{array}\right)(t)\rt\|_{p, Q_p\otimes Q}
\leq
e^{ct}
\left\|\left(\begin{array}{c}e_1 \\
e_2
\\
\vdots 
\\
e_{N-1}\end{array}\right)(0)\rt\|_{p, Q_p\otimes Q},
\ee
where $e_i=x_i-x_{i+1}$ denotes the $i$th edge of the linear graph, and $\|\cdot\|_{p, Q_p\otimes Q}$ denotes the weighted $L^p$ norm with the weight $Q_p\otimes Q$, where for any $1\leq p\leq\infty$,
\beqn
Q_p =\diag\lt(p_1^{\frac{2-p}{p}},\ldots, p_{N-1}^{\frac{2-p}{p}}\rt)
\eeqn

and for $1\leq k\leq N-1$, $p_k= \sin(k \pi/N)$, and $Q$ is a positive diagonal matrix. In addition, $4\sin^2\lt({\pi}/{2N}\rt)$ is the smallest nonzero eigenvalue of the Laplacian matrix of $\mathcal{G}$. 
Note that $Q_{\infty}=\diag\lt(1/p_1,\ldots,1/p_{N-1}\rt).$
\eprop

 Before we prove Proposition \ref{thm:ode:linear}, we will explain where $(p_1,\ldots, p_{N-1})$ and 
 $4\sin^2\lt({\pi}/{2N}\rt)$ come from. 
 
For a linear graph with $N$ nodes, consider the following directed incidence matrix: 
\[E= \left(\begin{array}{cccc}-1&  &  &  \\1 & -1 &  &  \\ & 1 & \ddots &  \\ &  & \ddots & -1 \\ &  &  & 1\end{array}\right)_{N\times N-1},\]
and the following $N-1\times N-1$ edge Laplacian $K:=E^TE$, 
 \be{Discrete:Linear:edge:Laplacian}
 K=\left(\begin{array}{cccc}2& -1 &  &  \\-1 & 2 & -1 &  \\ & \ddots & \ddots &  \\ &  -1& 2 & -1 \\ &  &  -1& 2\end{array}\right).
 \ee
  Note that since $-K$ is a Metzler matrix, 
it follows by the
Perron-Frobenius Theorem that
it has a positive eigenvector $(v_1,\ldots, v_{N-1})$ corresponding to $-\gamma_1$, the largest eigenvalue of $-K$, ($\gamma_1$ is the smallest eigenvalue of $K$), i.e., 
 \be{PF_eigenvalue_K}
 \lt(v_1,\ldots, v_{N-1}\rt) (-K) = -\gamma_1\lt(v_1,\ldots, v_{N-1}\rt).
 \ee
A simple calculations shows that $v_k= p_k=\sin(k \pi/N)$ and $\gamma_1=4\sin^2\lt({\pi}/{2N}\rt)$ (Apply Lemma \ref{tridiagonal_eigenvalue}, part $1$, to matrix $K$).

To prove Proposition \ref{thm:ode:linear}, we first prove the following Lemma:

\begin{lemma}\label{linear:graph:lemma:Mp}
Let $K$ be the edge Laplacian of a linear graph with $N\geq3$ nodes as shown in (\ref{Discrete:Linear:edge:Laplacian}). Then for any $1\leq p\leq\infty$, 
\be{linear:graph:eq:Mp}
M_{p, Q_p\otimes Q} \lt[ 4\sin^2\lt({\pi}/{2N}\rt)I\otimes D - K\otimes D\rt] \leq0,
\ee
where $Q$ and $Q_p$ are as in Proposition \ref{thm:ode:linear}. 
\end{lemma}

\bp
To show (\ref{linear:graph:eq:Mp}), we will show that $M_p[\mathcal{A}]\leq0$, where 
\[\mathcal{A}:= \lt(Q_p\otimes Q\rt)\lt(4\sin^2\lt({\pi}/{2N}\rt)I\otimes D-K\otimes D\rt)\lt(Q_p^{-1}\otimes Q^{-1}\rt).\]
(Recall that $M_{p,Q} \lt[A\rt]= M_{p} \lt[QAQ^{-1}\rt],$
and $A^{-1}\otimes B^{-1}= (A\otimes B)^{-1}$.)

We first show for $p=1$, $M_p[\mathcal{A}]=0$. 
A simple calculation shows that, for $p=1$
\beqn
\mathcal{A}&=&\left(\begin{array}{cccc}(\l-2)D&   \frac{p_1}{p_2}D&  &  \\  \frac{p_2}{p_1}D& (\l-2)D& \frac{p_2}{p_3}D  &  \\ &\ddots & \ddots &  \\ &  & &  \\ &  &  \frac{p_{N-1}}{p_{N-2}}D & (\l-2)D\end{array}\right),
\eeqn
where $\l=4\sin^2\lt({\pi}/{2N}\rt)$.
  For $\mathbf{1} = (1,\ldots,1)^T$, and $p=1$, since $\mathbf{1}^TQ_p= (p_1,\ldots,p_{N-1})$, it follows by Equation (\ref{PF_eigenvalue_K}) that
$\mathbf{1}^T Q_p(-K)Q_p^{-1}= -\lambda\mathbf{1}^T$, 
therefore,
\be{M1:linear:graph}
-2+\dis\frac{p_2}{p_1}=-2+\dis\frac{p_1}{p_2}+\frac{p_3}{p_2}=\cdots=-2+\dis\frac{p_{N-2}}{p_{N-1}}=-\lambda.
\ee
Hence, by the definition of $M_1$, $M_1[A]=\max_{j}\left(a_{jj}+\sum_{i\neq j}\abs{a_{ij}}\right)$ \cite{Desoer}, and because $D$ is diagonal, $M_1[\mathcal{A}]=0$.

Now, we show that $M_{\infty}[\mathcal{A}]=0$. 
A simple calculation shows that, for $p=\infty$, since $Q_{\infty}=\diag\lt(1/p_1,\ldots,1/p_{N-1}\rt)$,
\beqn
\mathcal{A}&=&\left(\begin{array}{cccc}(\l-2)D&   \frac{p_2}{p_1}D&  &  \\  \frac{p_1}{p_2}D& (\l-2)D& \frac{p_3}{p_2}D  &  \\ &\ddots & \ddots &  \\ &  & &  \\ &  &  \frac{p_{N-2}}{p_{N-1}}D & (\l-2)D\end{array}\right).
\eeqn
Therefore, by the definition of $M_{\infty}$, $M_{\infty}[A]=\max_{i}\left(a_{ii}+\sum_{i\neq j}\abs{a_{ij}}\right)$, and because $D$ is diagonal,
 $M_{\infty}[\mathcal{A}]=\max\lt\{\l-2+\dis\frac{p_2}{p_1}, \ldots, \l-2+\dis\frac{p_{N-2}}{p_{N-1}} \rt\}=0$.

 Next we show for $1<p<\infty$, $M_p[\mathcal{A}]\leq0$. 
 {
A simple calculation shows that $\mathcal{A}$ can be written as follows:
\beqn
\mathcal{A}&=&\left(\begin{array}{cccc}(\l-2)D&   \a_1^{-1}D&  &  \\  \a_1D& (\l-2)D& \a_2^{-1}D  &  \\ &\ddots & \ddots &  \\ &  & &  \\ &  &  \a_{N-2}D & (\l-2)D\end{array}\right),
\eeqn
where $\a_i=\lt(\frac{p_{i+1}}{p_i}\rt)^{\frac{2-p}{p}}$.
}
 To show $M_p[\mathcal{A}]\leq0$, using Lemma \ref{M+Dini-linear} and the definition of $M$, it suffices to show that $D^+\|u\|_p\leq0$, where 
\[u=(u_{11},\ldots,u_{1n},\ldots, u_{N-1 1},\ldots, u_{N-1n})^T\]
  is the solution of 
 $\dot{u}=\mathcal{A}u$, or equivalently, $\frac{d\Phi}{dt}(u(t))\leq0$, where $\Phi(t)= \|u(t)\|^p_p$.
In the calculations below, we use the following simple fact:
 For any real $\alpha$ and $\beta$ and $1\leq p$:
\[\lt(|\alpha|^{p-2}+|\beta|^{p-2}\rt)\alpha\beta\leq |\alpha|^p+|\beta|^p.\]

In the calculations below, we let $\b_i=\a_i^{\frac{2}{2-p}}$.
  We also use the fact that $|x|^p$ is differentiable for $p>1$ and 
   \[\displaystyle\frac{d \Phi}{d u_i}=\displaystyle\frac{d}{d u_i}|u_i|^{p}=p|u_i|^{p-1}\displaystyle\frac{u_i}{|u_i|}=p|u_i|^{p-2}u_i.\]

Observe that
  \beqn
&&\!\!\!\!\!\!\!\!\!\!\!\!
\frac{d\Phi}{dt}(u(t))
=\dis\sum_{i,k} \frac{d\Phi}{d u_{ik}}\frac{d u_{ik}}{d t}
=\triangledown \Phi \cdot \dot{u}
=\triangledown \Phi \cdot \mathcal{A}u\\
&=&
p\lt(|u_{11}|^{p-2}u_{11},\ldots, |u_{nN-1}|^{p-2}u_{nN-1}\rt)\mathcal{A} (u_{11}, \ldots, u_{nN-1})^T\\
&=&p \sum_{k=1}^nd_k\mathcal{Q}_k
\eeqn 
where $\mathcal{Q}_k$ is the following expression:
{\small{\beqn
&&\!\!\!\!\!\!\!\!\!\!\!\!
\sum_{i=1}^{N-1} (\l-2)\abs{u_{ik}}^p
+\sum_{i=1}^{N-2}\lt(\a_i\abs{u_{i+1k}}^{p-2}u_{i+1k}u_{ik}+\a_i^{-1}\abs{u_{ik}}^{p-2}u_{i+1k}u_{ik}\rt)\\
&&\!\!\!\!\!\!\!\!\!\!\!\!
=\sum_{i=1}^{N-1} (\l-2)\abs{u_{ik}}^p 
+\sum_{i=1}^{N-2}\frac{\a_i}{\b_i}\lt(\abs{u_{i+1k}}^{p-2}u_{i+1k}(\b_iu_{ik})+\abs{\b_iu_{ik}}^{p-2}u_{i+1k}(\b_iu_{ik})\rt)\\
&&\!\!\!\!\!\!\!\!\!\!\!\!
\leq\sum_{i=1}^{N-1} (\l-2)\abs{u_{ik}}^p %\\
%&&\!\!\!\!\!\!\!\!\!\!\!\!
+\sum_{i=1}^{N-2}\frac{\a_i}{\b_i}\lt(\abs{u_{i+1k}}^{p}+\abs{\b_iu_{ik}}^{p}\rt)\\
&&\!\!\!\!\!\!\!\!\!\!\!\!
=\sum_{i=1}^{N-1} (\l-2)\abs{u_{ik}}^p
+\sum_{i=1}^{N-2}\frac{\a_{i}}{\b_i}\abs{u_{i+1k}}^{p}+\a_i\b_i^{p-1}\abs{u_{ik}}^{p}\\
&&\!\!\!\!\!\!\!\!\!\!\!\!
=\sum_{i=1}^{N-1} (\l-2)\abs{u_{ik}}^p
+\sum_{i=1}^{N-2}\frac{p_{i}}{p_{i+1}}\abs{u_{i+1k}}^{p}+\frac{p_{i+1}}{p_{i}}\abs{u_{ik}}^{p}\\
&&\!\!\!\!\!\!\!\!\!\!\!\!
=  \abs{u_{1k}}^p\lt(\l-2+\frac{p_2}{p_1}\rt)+\ldots+\abs{u_{N-1k}}^p\lt(\l-2+\frac{p_{N-2}}{p_{N-1}}\rt)
\eeqn}}
%\end{@twocolumnfalse}
%]
and this last term vanishes
by Equation (\ref{M1:linear:graph}).
\ep

\newtheorem*{proofofpropositionlineargraph}{Proof of Proposition \ref{thm:ode:linear}}

\begin{proofofpropositionlineargraph}
Let $K$ be as defined in (\ref{Discrete:Linear:edge:Laplacian}) and for $w=(w_1,\ldots,w_{N-1})^T$, let $J(w,t)=\diag\lt(J_F(w_1,t), \ldots, J_F(w_{N-1},t)\rt)$.
By subadditivity of $M$, and Lemma \ref{linear:graph:lemma:Mp}, for any $1\leq p\leq\infty$, 
\beqn
&& M_{p,Q_p\otimes Q} \lt[J(w,t) - K\otimes D\rt]\\
&\leq&  M_{p,Q_p\otimes Q} \lt[J(w,t) - \l I\otimes D\rt] 
 \;+\; M_{p,Q_p\otimes Q} \lt[\l I\otimes D - K\otimes D\rt]\\
 &\leq&  M_{p,Q_p\otimes Q} \lt[J(w,t) - \l I\otimes D\rt] \\
  &\leq&\max\lt\{M_{p,Q}\lt[J_F(w_1,t)-\l D\rt], \ldots, M_{p,Q}\lt[J_F(w_{N-1},t)-\l D\rt]\rt\}.
\eeqn
The last equality holds by Lemma \ref{block_diagonal_formula_M_p}. Note that $Q_p$ does not appear in the last equation. 
Now by taking $\dis\sup$ over all $w=(w_1^T,\ldots, w^T_{N-1})^T$ and all $t\geq0$, we get 
  \be{eq2:pf:thm:ode:linear}
   \dis\sup_{(w,t)}M_{p,Q_p\otimes Q}\lt[J(w,t)-K\otimes D\rt]\leq \dis\sup_t\sup_{x\in\r^n}M_{p,Q}\lt[J_F(x,t)-\l D\rt],
   \ee
Now by applying Theorem~\ref{general:graph:thm}, we obtain the desired inequality, (\ref{equ:linear:thm}). 
 \qed
  \end{proofofpropositionlineargraph}

 The significance of Proposition~\ref{thm:ode:linear} is as follows: since the numbers
$p_k=\sin(k \pi/N)$ are nonzero,
we have, when $c<0$, exponential convergence to uniform solutions in a
weighted $L^p$ norm, the weights being specified in each compartment
by the matrix $Q$ and the relative weights among compartments being weighted
by the numbers $p_k=\sin(k\pi/N)$.

\bremark\label{linear:inequality_for_edges}
Under the conditions of Proposition \ref{thm:ode:linear}, since the norms are
equivalent (here $L^p$ weighted and unweighted norms)
on $\r^{N-1}$, there exists 
$\a>0$ such that the following inequality holds: 
\[
\dis\sum_{i=1}^{N-1} \|e_i(t)\|_{p,Q} 
\;\leq\; \alpha e^{ct} \dis\sum_{i=1}^{N-1} \|e_i(0)\|_{p,Q}.
\]
\end{remark}
\bp
Using Equation (\ref{equ:linear:thm}) and the following inequality for $L^p$ norms, $p\geq1$, on $\r^{N-1}$:
\be{equivalence} 
\|\cdot\|_p \leq \|\cdot\|_1 \leq  (N-1)^{1-1/p}\|\cdot\|_p,
\ee
we will get the desired result. 
\ep

%------------------------------------------------------------------------------------------------  Complete Graphs
\smallskip
\subsubsection*{Complete Graphs}
%------------------------------------------------------------------------------------------------

Consider a $\mathcal{G}-$compartment system with 
an undirected complete graph $\cal{G}$.
 The following system of ODEs describes the evolution of the interconnected
agents $x_i$'s:
\be{complete}
\bal
\dot{x}_i\;&=\;F(x_i,t)+ D\dis\sum_{j=1}^N(x_j-x_i)
\eal
\ee
The following $N\times N$ matrix indicates the Laplacian matrix of a complete graph of $N$ nodes, 
\[\cal{L} = \left(\begin{array}{cccc}N-1 & -1 & \ldots & -1 \\-1 & N-1 & \ldots & -1 \\ &   \ddots & &  \\-1 & \ldots & -1 & N-1\end{array}\right),
\]
with $\l_1=0$ and $\l_2=N$.

{The following result is an application of Theorem~\ref{general:graph:thm} to complete graphs and one possible
  generalization to arbitrary norms of Theorem~\ref{theo:arcakODE} (but
  restricted to complete graphs).
}

%prop1
\bprop\label{complete-prop}
Let $\|\cdot\|$ be an arbitrary norm on $\r^{n}$. Suppose $x$ is a solution of Equation (\ref{complete}) and let
\[
c:=\dis\sup_{(x,t)} M [J_F(x,t)-ND]
\]
 where $M$ is the logarithmic norm induced by $\|\cdot\|$.
  Then 
\be{complete_estimate}
\dis\sum_{i=1}^m \|e_i(t)\| \;\leq \; e^{ct} \;\dis\sum_{i=1}^m \|e_i(0)\|,
\ee
where $e_i$, for $i=1,\ldots, m$ are the edges of $\mathcal{G}$,
meaning the differences $x_i(t)-x_j(t)$ for $i<j$.
\eprop
  
\bp 
The following $N\times N$ matrix indicates the (graph) Laplacian matrix of a complete graph of $N$ nodes, 
\[\mathcal{L} = \left(\begin{array}{cccc}N-1 & -1 & \ldots & -1 \\-1 & N-1 & \ldots & -1 \\ &   \ddots & &  \\-1 & \ldots & -1 & N-1\end{array}\right),
\]
with $\l_1=0$ and $\l_2=N$.
Let $E$ be an incidence matrix of $\mathcal{G}$. 
We first show that 
$E^T E E^T = N E^T$.
For any orientation of $\mathcal{G}$, $E^T$ is an ${N\choose 2}\times N$ matrix such that its $i-$th row looks like $(\e_{i1}, \ldots, \e_{iN})$, where for exactly one $j$, $\e_{ij}=1$, for exactly one $j$, $\e_{ij}=-1$, and for the rest of $j$'s, $\e_{ij}=0$. Observe that for any row $i$, $\sum_j \e_{ij} =0$, and 
\[
\lt(E^T \mathcal{L}\rt)_{(ij)} =  \lt(E^T\rt)_{r_i} \lt(\mathcal{L}\rt)_{c_j},
\]
where $(A)_{(ij)}$ denotes the $(i,j)-$th entry of matrix $A$, and $(A)_{r_i}$ and $(A)_{c_i}$ denote the $i$th row and $i$th column of $A$, respectively. Hence, 
\beqn
\lt(E^T \mathcal{L}\rt)_{(ij)} &=& \lt(\e_{i1}, \ldots, \e_{iN}\rt) \left(\begin{array}{c}-1 \\\vdots \\N-1 \\\vdots \\-1\end{array}\right)\leftarrow \mbox{$j^{th}$}\\
&=& -\e_{i1} -\ldots +(N-1) \e_{ij} -\ldots -\e_{iN}\\
&=& N \e_{ij} -\sum_k \e_{ik}= N \e_{ij}.
\eeqn
This proves $E^T\mathcal{L} = E^T E E^T = N E^T.$
Thus we may apply Theorem~\ref{general:graph:thm} with
$K=NI$. Then $\mathcal{J} := J(w,t)
 -K\otimes D$ can be written as follows: 
\[\mathcal{J} = \left(\begin{array}{ccc}J_F(w_1,t)-ND &  &  \\ & \ddots &  \\ &  & J_F(w_m,t)-ND\end{array}\right).\]
For $u=(u_1,\ldots, u_m)^T$, with $u_i\in\r^n$, let $\|u\|_* := \lt\|\lt(\|u_1\|, \ldots, \|u_m|\|\rt)^T\rt\|_1$, where $\|\cdot\|_1$ is $L^1$ norm on $\r^m$, and let $M_*$ be the logarithmic norm induced by $\|\cdot\|_*$. Then by the definition of $M_*$ and Lemma \ref{block_diagonal_formula_M_p}, 
\[
M_*[J(w,t)- K\otimes D] \leq \dis\max_i \lt\{M[J_F(w_i,t)-ND]\rt\}.
\]
Therefore, by taking $\sup$ over all possible $w$'s 
in both sides of the above inequality, we get:
\[
\sup_wM_*[J(w,t)- K\otimes D] \leq  \sup_{(x,t)}M[J_F(x,t)-ND] = c.
\]
Applying Theorem~\ref{general:graph:thm}, we 
conclude the desired result.
\ep

The estimate~\eqref{complete_estimate} can also be established as an
  application of Lemma \ref{Dini_M}, as follows. 

\bprop\label{complete:inequality_for_edges}
For complete graphs, and
under the conditions of Proposition \ref{complete-prop}, 
for any $i=1,\ldots, m$,
\be{complete_estimate_2}\nonumber
\|e_i(t)\| \leq e^{ct} \|e_i(0)\|\,.
\ee
\eprop
\bp
For any fixed $i,j\in\{1,\ldots, N\}$, using Equation (\ref{complete}), we have:
 \be{complete_proof}\nonumber
 \bal
\dot{x}_i-\dot{x}_j\;&= \;G_t(x_i)-G_t(x_j),
\eal
\ee
where $G_t(x):=F(x,t) - N D x$. Remark \ref{M[f]:M[J]} gives the desired result. 
\ep

%------------------------------------------------------------------------------------------------ Star Graphs
\smallskip
\subsubsection*{Star Graphs}
%------------------------------------------------------------------------------------------------

Now consider a $\mathcal{G}-$compartment system, where $\mathcal{G}$ is a
star graph of $N+1$ nodes.  

The following system of ODEs describes the evolution of the complete system:
\be{star}
\bal
\dot{x}_i\;&=\;F(x_i,t)+ D\lt(x_0-x_i\rt),\; i=1,\ldots,N\\
\dot{x}_0\;&=\;F(x_0,t)+ D\dis\sum_{i\neq 0}(x_i-x_0)
\eal
\ee

The following $(N+1)\times (N+1)$ matrix indicates the Laplacian matrix of a star graph of $N+1$ nodes:
\[\cal{L} = \left(\begin{array}{ccccc}1 &  &  &  & -1 \\ & 1 &  &  &-1  \\ &  & \ddots &  &  \\ &  &  & 1 & -1 \\ -1&  &  & -1 & N\end{array}\right).\]
Note that $\l_1=0$, and $\l_2=1.$

%prop
\bprop\label{star-prop}
Let $\{x_i, \; i=0.\ldots, N\}$ be solutions of (\ref{star}) and 
\[c:=\dis\sup_{(x,t)} M[J_F(x,t)-D],\]
where $M$ is the logarithmic norm induced by an arbitrary norm $\|\cdot\|$ on $\r^n$.
 Then for any $i\in\{1,\ldots, N\}$,
\be{star-prop:ineq}\nonumber
\lt\| (x_i-x_0)(t)\rt\| \;\leq\; (1+\a_i t)e^{ct}\, \lt\| (x_i-x_0)(0)\rt\|
\ee
where $\a_i=\dis\sum_{j\neq i,0}\lt\| (x_j-x_i)(0)\rt\|$. 

In particular, when $c<0$, for any $i=1,\ldots, N$,
 $(x_i-x_0)(t)\to0$ exponentially as $t\to\infty.$
\eprop

 \bp
Using (\ref{star}),
$\dot{x}_i-\dot{x}_j = \lt(F(x_i,t)-Dx_i\rt) - \lt(F(x_j,t)-Dx_j\rt),$
for any $i,j=1,\ldots, N$.
Applying Lemma~\ref{Dini_M}, we get
\be{star:xi-xj}
\|(x_i-x_j)(t)\| \;\leq\; e^{ct}\;\|(x_i-x_j)(0)\|.
\ee
For any $i=1,\ldots, N$, we have:
\beqn
\dot{x}_i-\dot{x}_0
&=& F(x_i,t)-F(x_0,t)- D(x_i-x_0) - D\dis\sum_{j=1}^N\lt( x_j-x_0\rt)\\
&= &F(x_i,t)-F(x_0,t)- D(x_i-x_0) - D\dis\sum_{j=1}^N\lt( x_j-x_0\;+\;x_i-x_i\rt)\\
&=& F(x_i,t)-F(x_0,t)- D(N+1)(x_i-x_0) -D\dis\sum_{j=1}^N\lt( x_j-x_i\rt)
\eeqn
Now using the Dini derivative for $\|x_i-x_0\|$ and using the upper bound for $\|x_i-x_j\|$ derived in (\ref{star:xi-xj}), we get:
\beqn
D^+\|(x_i-x_0)(t)\| &\leq& \tilde{c} \|(x_i-x_0)(t)\| + \a_i e^{ct},
\eeqn
where, $\a_i=\sum_{j\neq i,0} \|(x_j-x_i)(0)\|$ and by subadditivity of $M$,
  \beqn
 \tilde{c}&:=& \dis\sup_x M[J_F(x,t)-(N+1)D] \\
 & \leq&\dis\sup_x M[J_F(x,t)-D] +\dis\sup_x M[-ND]\\
 &\leq& \dis\sup_{(x,t)} M[J_F(x,t)-D]=c \quad\mbox{since $M[-ND]<0$}
 \eeqn
Applying Gronwall's inequality to the above inequality, we get Equation (\ref{star-prop:ineq}).
\ep

\bcor
Under the conditions of Proposition \ref{star-prop}, the following inequality holds:
\be{star-cor:ineq}\nonumber
\dis\sum_{i\neq0}\lt\| (x_i-x_0)(t)\rt\| \;\leq\; P(t)e^{ct}\, \dis\sum_{i\neq0}\lt\| (x_i-x_0)(0)\rt\|
\ee
where $P(t)= 1+2(N-1)\;t\;\dis\sum_{i\neq0}\lt\| (x_i-x_0)(0)\rt\|$. 
\ecor

\bp
For any $i\neq0$, using the triangle inequality, we have
\be{}\nonumber
\bal
\a_i=\dis\sum_{j\neq i,0} \|(x_j-x_i)(0)\| &\leq \dis\sum_{j\neq i,0} \|(x_j-x_0)(0)\| +\dis\sum_{j\neq i,0} \|(x_i-x_0)(0)\|\\
&= \dis\sum_{j\neq i,0} \|(x_j-x_0)(0)\| +(N-1) \|(x_i-x_0)(0)\|,
\eal
\ee
taking sum over all $i\neq0$, we get 
\be{}\nonumber
\bal
\dis\sum_{i\neq0}\a_i &\;\leq\; (N-1)  \dis\sum_{j\neq0} \|(x_j-x_0)(0)\| + (N-1)\dis\sum_{i\neq 0} \|(x_i-x_0)(0)\|.
\eal
\ee
Therefore, since $\a_i$'s are nonnegative, for any $i$,
\be{}\nonumber
\bal
1+\a_i t \;\leq\; 1+ t \dis\sum_{i\neq0}\a_i &\leq 1+ 2(N-1)\;t\; \dis\sum_{i\neq 0} \|(x_i-x_0)(0)\| :=P.
\eal
\ee
and hence,
\[ \|(x_i-x_0)(t)\| \leq Pe^{ct} \|(x_i-x_0)(t)\|.\]
Now taking sum over all $i\neq0$, we get Equation (\ref{star-cor:ineq}) as we wanted.
\ep

%------------------------------------------------------------------------------------------------       Cartesian product Graphs
\smallskip
\subsubsection*{Cartesian products}
%------------------------------------------------------------------------------------------------

For $k=1,\ldots, K$, let $G_k = \lt(V_k, E_k\rt)$ be an arbitrary graph,
with $|V_k| = N_k$ and Laplacian matrix $L_{G_k}$. 

Consider a system of $N=\dis\Pi_{k=1}^{K} N_k$ compartments $x_{i_1,\ldots,
  i_K}\in\r^n$, $i_j=1,\ldots, N_j$, which are interconnected by
$\mathcal{G}=G_1\times\ldots\times G_{K}$, where $\times$ denotes the
Cartesian product. The following system of ODEs describe the evolution of the $x_{i_1,\ldots, i_K}$'s:
\be{Cartesian:ODE}
\bal
\dot{x}\;=\; \tilde{F}(x,t) -\lt(\mathcal{L}\otimes D\rt)x
\eal
\ee
where $x=\lt(x_{i_1,\ldots, i_K}\rt)$ is the vector of all $N$ compartments,
$\tilde{F}(x,t)=\lt(F(x_{i_1,\ldots, i_K},t)\rt)$, and $\mathcal{L}$ is defined as
follows: 
\[
\dis\sum_{i} I_{N_K}\otimes\ldots\otimes L_{G_i}\otimes\ldots\otimes I_{N_1},
\]
and $D=\diag(d_1,\ldots,d_n)$ is the diffusion matrix. 

Note that Laplacian spectrum of the Cartesian product $\mathcal{G}$ is the set:
\[
\lt\{ \l_{i_1} (G_1) +\ldots + \l_{i_K} (G_K) \;|\; i_j=1,\ldots, N_j\rt\}. 
\]
Therefore, $\l_2(\mathcal{G}) = \min \lt\{ \l_2(G_1), \ldots, \l_2(G_K)\rt\}.$

%prop
\bprop\label{Cartesian:thm}
Given graphs $G_k$, $k=1,\ldots, K$ as above, suppose that for each $k$, there are a norm $\|\cdot\|_{(k)}$ on $\r^n$, a real nonnegative number $\l_{(k)}$, and a polynomial $P_{(k)}(z,t)$ on $\r^2_{\geq0}$, with the property that for each $z$, $P_{(k)}(z,0)\geq1$, such that 
for every solution $x$ of (\ref{Cartesian:ODE}), 
{\small{
\be{G_k:inequlaty}
\dis\sum_{e\in E_k}\lt\|e(t)\rt\|_{(k)}\; \leq\; P_{(k)} \lt(\dis\sum_{e\in E_k}\lt\|e(0)\rt\|_{(k)}, t\rt)\;e^{c_kt}\dis\sum_{e\in E_k}\lt\|e(0)\rt\|_{(k)},
\ee
}}
holds, where $c_k:=\dis\sup_{(x,t)} M_{(k)}\lt[J_F(x,t)-\l_{(k)}D\rt]$, 
and $M_{(k)}$ is the logarithmic norm induced by $\|\cdot\|_{(k)}$.
Then for any norm $\|\cdot\|$ on $\r^n$, there exists a polynomial $P(z, t)$ on $\r^2_{\geq0}$, with the property that for each $z$, $P(z,0)\geq1$, such that 
\be{}\nonumber
\dis\sum_{e\in \mathcal{E}}\lt\|e(t)\rt\| \;\leq\; P\lt(\dis\sum_{e\in \mathcal{E}}\lt\|e(0)\rt\|, t\rt)\; e^{ct}\dis\sum_{e\in \mathcal{E}}\lt\|e(0)\rt\|,
\ee
where $c:=\max\{c_1,\ldots, c_K\}$, and $\mathcal{E}$ is the set of the edges of $\mathcal{G}$.
\eprop

{
Note that for $K=1$, Remark~\ref{linear:inequality_for_edges}, Proposition \ref{complete-prop}, and Proposition \ref{star-prop} show that (\ref{G_k:inequlaty}) holds when $G_k$ is a line, complete or star graph, for $P_{(k)}(z,t)=\a, 1, 1+2(N-1)tz$, respectively. 
Therefore, for a hypercube (cartesian product of $K$ line graphs) with $N_1\times\ldots\times N_K$ nodes, if for $1\leq p\leq\infty$, and $Q$ a positive diagonal matrix, and $\l_2=4\min\lt\{\sin^2(\pi/2N_i)\rt\}$,
\[\dis\sup_{(x,t)}M_{p,Q}\lt[J_F(x,t) - \l_2D\rt] <0,\]
then the system synchronizes. See Table \ref{table:cartesian}.

Also, for a Rook graph (cartesian product of $K$ complete graphs) of $N_1\times\ldots\times N_K$ nodes, if for any given norm, and $\l_2=\min\lt\{N_i\rt\}$,
\[\dis\sup_{(x,t)}M\lt[J_F(x,t) - \l_2D\rt] <0,\]
then the system synchronizes. See Table \ref{table:cartesian}.
}

The idea of the proof of Proposition \ref{Cartesian:thm} is exactly the same as the proof of Proposition \ref{grid2D:thm} below. For ease of notations and explanation, we will give a proof for Proposition \ref{grid2D:thm} and skip the proof of Proposition \ref{Cartesian:thm}.

%------------------------------------------------------------------------------------------------ Grid Graphs
\smallskip
\subsubsection*{Grid Graphs}
%------------------------------------------------------------------------------------------------

Consider a network of $N_1\times N_2$ compartments that are connected to each other by a $2$-D, $N_1\times N_2$ lattice (grid) graph $\mathcal{G}=\lt(\mathcal{V}, \mathcal{E}\rt)$, where 
\[\mathcal{V}=\lt\{x_{ij}, \; i=1,\ldots, N_1, \; j=1,\ldots, N_2\rt\}\]
is the set of all vertices and $\mathcal{E}$ is the set of all edges of $\mathcal{G}$.

\begin{figure}[ht]
\begin{center} 
\begin{tikzpicture}
[>=stealth',shorten >=1pt,auto,node distance=1.6cm,
  thick,main node/.style={circle,fill=blue!14,draw,font=\sffamily\small\bfseries}]

  \node[main node] (1) {$x_{11}$};
  \node[main node] (2) [ right of=1] {$x_{12}$};
  \node[main node] (3) [ right of=2] {$x_{13}$};
  \node[main node] (4) [ below of=1] {$x_{21}$};
  \node[main node] (5) [ right of=4] {$x_{22}$};
  \node[main node] (6) [ right of=5] {$x_{23}$};
  \node[main node] (7) [ below of=4] {$x_{31}$};
  \node[main node] (8) [ right of=7] {$x_{32}$};
  \node[main node] (9) [ right of=8] {$x_{33}$};
  \node[main node] (10) [ right of=3] {$x_{14}$};
  \node[main node] (11) [ right of=6] {$x_{24}$};
  \node[main node] (12) [ right of=9] {$x_{34}$};

  \path[every node/.style={font=\sffamily\small}]
    (1) edge node[] {} (2)
         edge node[] {} (4)
    (2) edge node[] {} (3)
         edge node[] {} (5)
    (3) edge node[] {} (6)
        edge node[] {} (10)
    (4) edge node[] {} (5)
        edge node[] {} (7)

     (5) edge node[] {} (6)
        edge node[] {} (8)

    (6)edge node[] {} (9)
    edge node[] {} (11)
    (7)edge node[] {} (8)
    (8)edge node[] {} (9)
    (9)edge node[] {} (12)
    (10)edge node[] {} (11)
    (11)edge node[] {} (12)
    (12)
      ;
\end{tikzpicture}
\end{center} 
\caption{An example of a grid graph: $3\times4$ nodes}
\label{fig:grid}
\end{figure}
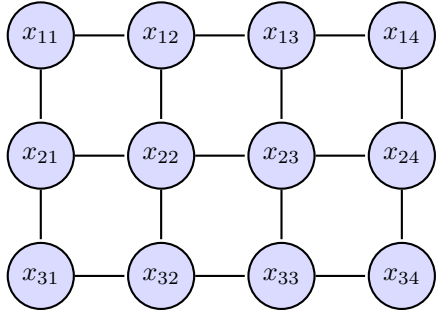

The following system of ODEs describes the evolution of the $x_{ij}$'s: 
for any $i=1,\ldots, N_1$, and $j=1,\ldots, N_2$
\be{grid2D:ode}
\bal
\dot{x}_{i,j}\;=\; F(x_{ij},t) &+ D\lt(x_{i-1,j}-2x_{i,j}+x_{i+1,j}\rt) \\
&+D\lt(x_{i,j-1}-2x_{i,j}+x_{i,j+1}\rt), 
\eal
\ee
assuming Neumann boundary conditions, i.e. $x_{i,0}=x_{i,1}$,
$x_{i,N_2}=x_{i,N_2+1}$, etc. 
 
% prop
\bprop\label{grid2D:thm}
Let $x=\{x_{ij}\}$ be a solution of Equation (\ref{grid2D:ode}) and $c=\max\{c_1, c_2\}$, where for $i=1,2$, 
\[c_i:=\dis\sup_{(x,t)}M_{p, Q}\lt[J_F(x,t)-4\sin^2\lt(\pi/2N_i\rt) D\rt],\]
 and $1\leq p\leq\infty$. 
 Then, there exist positive constants $\a\geq1$, and $\b$ such that 
\be{grid2D_inequality_thm}
 \dis\sum_{e\in\mathcal{E}}\|e(t)\|_{p, Q} \;\leq\; \lt(\a + \b t\rt)e^{ct}\dis\sum_{e\in\mathcal{E}}\|e(0)\|_{p, Q}.
\ee

In particular, when $c<0$, the system (\ref{grid2D:ode}) synchronizes, i.e., for all $i,j,k,l$
\[(x_{ij}-x_{kl})(t)\to0, \quad \mbox{exponentially as $t\to\infty$}.\]
\eprop

\bp
For $i=1,\ldots, N_1$, let $x_i=\lt(x_{i1}, \ldots, x_{iN_2}\rt)^T$, and assume that $x_i$'s are diffusively interconnected by a linear graph of $N_1$ nodes. 

For ease of notation, we assume that for $i=1,\ldots, N_1$, $\mathcal{E}_{(i)}$ is the set of all edges in the compartment $i$, i.e., all the edges in each row in Figure \ref{fig:grid}. In addition, we let $\mathcal{E}_h=\dis\bigcup_{i=1}^{N_1} \mathcal{E}_{(i)}$ denote all the horizontal edges in $\mathcal{G}$.
Also we assume that for $i=1,\ldots, N_1$, $\mathcal{E}^{(i)}$ is the set of all edges that connect the compartment $i$ to the other compartments. In addition, we let $\mathcal{E}_v=\dis\bigcup_{i=1}^{N_1} \mathcal{E}^{(i)}$ denote all the vertical edges in $\mathcal{G}$. 

For each $i=1,\ldots, N_1$, and fixed $t$, let
\[G(x_i, t):=\tilde{F}(x_i, t)- L_2\otimes D x_i,\]
 where $L_2$ is the Laplacian matrix of the linear graph of $N_2$ nodes; and $\tilde{F}(x_i, t)= (F(x_{i1}, t), \ldots, F(x_{iN_2}, t))^T$. 
 We can think of $G$ as the reaction operator acts in each compartment $x_i$. 
 
 Then the system (\ref{grid2D:ode}) can be written as:
\be{grid2D_reduced_to_linear}\nonumber
\bal
\dot{x}_1\;&=\;G(x_1, t)+\lt(I_{N_2}\otimes D\rt)(x_2-x_1)\\
\dot{x}_2\;&=\;G(x_2, t)+\lt(I_{N_2}\otimes D\rt)(x_1-2x_2+x_3)\\
\vdots\\
\dot{x}_{N_1}\;&=\;G(x_{N_1}, t)+\lt(I_{N_2}\otimes D\rt)(x_{{N_1}-1}-x_{N_1})
\eal
\ee
By Remark \ref{linear:inequality_for_edges}, if for $1\leq p\leq\infty$, $c_1$ is defined as follows
\[c_1=\dis\sup_{(x,t)}M_{p, I_{N_2}\otimes Q}\lt[J_{G}(x,t)-4\sin^2\lt(\pi/2N_1\rt)\lt(I_{N_2}\otimes D\rt)\rt],\]
 then:
\be{grid2D_reduced_to_linear_inequality_c1}
\bal
\dis\sum_{e\in \mathcal{E}_v} \lt\|e(t) \rt\|_{p,Q} \;\leq\; \a_{1}\; e^{c_1t}\; \dis\sum_{e\in \mathcal{E}_v} \lt\|e(0) \rt\|_{p,Q},
\eal
\ee
where
\[\a_{1}= \max_k\lt\{\sin{\lt(\dis\frac{k\pi}{{N_1}}\rt)}\rt\} \big/ \min_k\lt\{\sin{\lt(\dis\frac{k\pi}{{N_1}}\rt)}\rt\}(N_1-1)^{1-1/p}.\]
By Lemma \ref{block_diagonal_formula_M_p}, for any $p$,
\be{c1<c}
\bal
c_1&=\dis\sup_{(x,t)}M_{p, I_{N_2}\otimes Q}\lt[J_{G}(x,t)-4\sin^2\lt(\pi/2N_1\rt)\lt(I_{N_2}\otimes D\rt)\rt]\\
&\leq \dis\sup_{(x,t)}M_{p, Q}\lt[J_{F}(x,t)-4\sin^2\lt(\pi/2N_1\rt)D\rt]\leq c.
\eal
\ee
Therefore, using Equations (\ref{grid2D_reduced_to_linear_inequality_c1}) and (\ref{c1<c}), we have 
\be{grid2D_reduced_to_linear_inequality}
\bal
\dis\sum_{e\in \mathcal{E}_v} \lt\|e(t) \rt\|_{p,Q} \;\leq\; \a_{1}\; e^{ct}\; \dis\sum_{e\in \mathcal{E}_v} \lt\|e(0) \rt\|_{p,Q}.
\eal
\ee
Now let's look at each compartment $x_i$ which contains $N_2$ sub-compartment that are connected by a linear graph. For example, for $i=1$:
\be{grid2D_x1}
\bal
\dot{x}_{11}&=F(x_{11},t)+D(x_{12}-x_{11}+x_{21}-x_{11})\\
\dot{x}_{12}&=F(x_{12},t)+D(x_{11}-2x_{12}+x_{13}+x_{22}-x_{12})\\
\vdots\\
\dot{x}_{1N_2}&=F(x_{1N_2},t)+D(x_{1N_2-1}-x_{1N}+x_{2N_2}-x_{1N_2}).
\eal
\ee
 Let
$u:=(x_{11}, \ldots, x_{1{N_2}-1})^T$, $v:=(x_{12}, \ldots, x_{1{N_2}})^T,$
and for any fixed $t$, define $\tilde{G}$ as follows:
\be{tildeG:grid}
  \tilde{G}(u,t):=\left(\begin{array}{c}F(x_{11}, t) \\F(x_{12}, t)\\\vdots \\F(x_{1{N_2}-1},t)\end{array}\right)- K\otimes D \left(\begin{array}{c}x_{11} \\x_{12}\\\vdots \\x_{1{N_2}-1}\end{array}\right),
  \ee
 where $K$ is as defined in (\ref{Discrete:Linear:edge:Laplacian}). Then
\beqn
 \dot{u}-\dot{v}&=&\tilde{G}(u,t)-\tilde{G}(v,t)+\left(\begin{array}{c}(x_{21}-x_{11})-(x_{22}-x_{12}) \\\vdots \\(x_{2{N_2}-1}-x_{1{N_2}-1})-(x_{2{N_2}}-x_{1{N_2}})\end{array}\right)\otimes D.
 \eeqn
Using the Dini derivative, for any $p$, and $Q_p$ as defined in Proposition \ref{thm:ode:linear}, we have: (for ease of the notation let $\|\cdot\|:= \|\cdot\|_{p, Q_p\otimes Q}$.)
\beqn
D^{+}\|(u-v)(t)\|
&=&\displaystyle\limsup_{h\to 0^+}\frac{1}{h}\;\left(\|(u-v)(t+h)\|-\|(u-v)(t)\|\right)\\
&=&\displaystyle\limsup_{h\to 0^+}\frac{1}{h}\left(\|(u-v+h(\dot{u}-\dot{v}))(t)\|-\|(u-v)(t)\|\right)\\
&\leq&\displaystyle\lim_{h\to 0^+}\frac{1}{h}\left(\|(u-v)(t)+h(\tilde{G}(u,t)-\tilde{G}(v,t))\|-\|(u-v)(t)\|\right)\\
&&+\lt\|\left(\begin{array}{c}(x_{21}-x_{11})-(x_{22}-x_{12}) \\\vdots \\(x_{2{N_2}-1}-x_{1{N_2}-1})-(x_{2{N_2}}-x_{1{N_2}})\end{array}\right)\otimes D\rt\|\\
&\leq& \dis\sup_{(w,t)}M_{p, P\otimes Q}\lt[J_{\tilde{G}}(w,t)\rt]\|(u-v)(t)\|\\
&&+\lt\|\left(\begin{array}{c}(x_{21}-x_{11})-(x_{22}-x_{12}) \\\vdots \\(x_{2{N_2}-1}-x_{1{N_2}-1})-(x_{2{N_2}}-x_{1{N_2}})\end{array}\right)\otimes D\rt\|. 
\eeqn

 Note that the last term is the difference between some of the vertical edges of $\mathcal{G}$. Therefore by Equation (\ref{grid2D_reduced_to_linear_inequality}), and the triangle inequality, we can approximate the last term as follows:
\beqn
\lt\|\left(\begin{array}{c}(x_{21}-x_{11})-(x_{22}-x_{12}) \\\vdots \\(x_{2N_2-1}-x_{1N_2-1})-(x_{2N_2}-x_{1N_2})\end{array}\right)\otimes D\rt\|_{p, Q_p\otimes Q}
\leq 2d a \a_1 \dis\sum_{e\in \mathcal{E}^{(1)}} \lt\|e(t) \rt\|_{p,Q}
\eeqn
 where $a= \max_i\lt\{(Q_p)_i\rt\}$, 
$d=\max\{d_1,\ldots,d_n\}$, and $\mathcal{E}^{(1)}$ is the set of edges of $\mathcal{G}$ which connect the compartment $x_1$ to the compartment $x_2$.

By Equation (\ref{eq2:pf:thm:ode:linear}), for any $1\leq p\leq \infty$
\beqn
\dis\sup_{(u,t)}M_{p, Q_p\otimes Q}\lt[J_{\tilde{G}}(u,t)\rt]
&\leq&\dis\sup_{(x,t)}M_{p, Q}\lt[J_F(x,t)-4\sin^2\lt(\pi/2N_2\rt)D\rt]\leq c.
\eeqn
Therefore for $x_1$, we have:
\beqn
 D^+ \dis\sum_{e\in \mathcal{E}_{(1)}} \lt\|\phi_ee(t) \rt\|_{p,Q} 
 &\leq&  c\dis\sum_{e\in \mathcal{E}_{(1)}} \lt\|\phi_ee(t) \rt\|_{p,Q} 
 + 2d \;a \;\a_1 \dis\sum_{e\in \mathcal{E}^{(1)}} \lt\|e(t) \rt\|_{p,Q},
\eeqn
where $\phi_e = (Q_p)_k$, when $e=e_k$ is the $k$-th edge of the $N_2$-linear graph. 
 
Repeating the same process for other compartments, $x_2, \ldots, x_{N_1}$, and adding them up, we get the following inequality
\beqn
 D^+ \dis\sum_{e\in \mathcal{E}_h} \lt\|\phi_e e(t) \rt\|_{p,Q} 
 &\leq&  c\dis\sum_{e\in \mathcal{E}_h} \lt\|\phi_e e(t) \rt\|_{p,Q} + 2\times2d \;a \;\a_1 \dis\sum_{e\in \mathcal{E}_v} \lt\|e(t) \rt\|_{p,Q}\\
 &\leq&  c\dis\sum_{e\in \mathcal{E}_h} \lt\|\phi_e e(t) \rt\|_{p,Q} + 4d \;a \;\a_1 e^{ct} \dis\sum_{e\in \mathcal{E}^{(1)}} \lt\|e(0) \rt\|_{p,Q}
\eeqn
Note that in the first inequality, the coefficient $2$ appears because each edge $e$ that connects the $i$th compartment to the $j$th compartment is counted twice: once when we do the process for $x_i$ and once when we do it for $x_j$.

Applying Gronwall's inequality implies:
\beqn
  \dis\sum_{e\in \mathcal{E}_h} \lt\| \phi_e e(t) \rt\|_{p,Q} 
&\leq&  e^{ct}\dis\sum_{e\in \mathcal{E}_h} \lt\| \phi_e e(0) \rt\|_{p,Q} + 4d \;a \;\a_1 t e^{ct}\dis\sum_{e\in \mathcal{E}_v} \lt\|e(0) \rt\|_{p,Q}\\
   &\leq&  e^{ct}\dis\sum_{e\in \mathcal{E}_h} \lt\|\phi_ee(0) \rt\|_{p,Q} + 4d \;a \;\a_1 t e^{ct}\dis\sum_{e\in \mathcal{E}} \lt\|e(0) \rt\|_{p,Q}.
\eeqn
Now using Equation (\ref{equivalence}) and the following inequalities:
\beqn
&&\min_k\lt\{(Q_p)_k\rt\} \lt\| e(t) \rt\|_{p,Q} \leq \lt\| \phi_e e(t) \rt\|_{p,Q} \\
\\
&&\lt\| \phi_e e(0) \rt\|_{p,Q} \leq  \max_k\lt\{(Q_p)_k\rt\} \lt\| e(0) \rt\|_{p,Q},
\eeqn
we get
\be{grid2D_inequality}
\bal
  \dis\sum_{e\in \mathcal{E}_h} \lt\|  e(t) \rt\|_{p,Q} 
 \leq \a_2  e^{ct}\dis\sum_{e\in \mathcal{E}_h} \lt\| e(0) \rt\|_{p,Q} + \b t e^{ct}\dis\sum_{e\in \mathcal{E}} \lt\|e(0) \rt\|_{p,Q}.\\
\eal
\ee
where $\a_2= \dis\frac{\max_k\lt\{(Q_p)_k\rt\} }{\min_k\lt\{(Q_p)_k\rt\}} (N_2-1)^{1-1/p}$, and $\b=\dis\frac{4d \;a \;\a_1}{\a_2}$.
Let $\a= \max\lt\{\a_1, \a_2\rt\}$, then Equations (\ref{grid2D_reduced_to_linear_inequality}) and (\ref{grid2D_inequality}), imply (\ref{grid2D_inequality_thm}). 
\ep

%------------------------------------------------------------------------------------------------
\smallskip
\section{Appendix}
%------------------------------------------------------------------------------------------------

%------------------------------------------------------------------------------------------------
\smallskip
\subsection*{Proof of Lemma \ref{claim2:thm:npde:1d}}
%------------------------------------------------------------------------------------------------

To prove Lemma \ref{claim2:thm:npde:1d}, we need the following lemma:

\blem\label{Delta_abs}
For any $u\colon\Omega\subset\r^m\to\r$, assume that $\Delta u$ is defined on $\Omega$. Then, there exists a set $I\subset\Omega$ such that:
\bi
\item $\mu(I)=0$, where $\mu$ denote the measure; and 
\item $\Delta\abs{u}$ is defined on $\Omega\setminus I$.
\ei
In fact, $I=\lt\{\o\in\Omega\;:\; u(\o)=0, \;\nabla u(\o)\neq 0\rt\}$.
\elem

\bp
We only prove for the especial case $\Omega=(a,b)$. The proof for the general $\Omega$ is analogue. We show that $I$ is countable, and hence of measure zero: 

Fix $\o^*\in I$ such that $\frac{\partial u}{\partial \o}(\o^*)\neq0$. Since $u$ is continuous and $\frac{\partial u}{\partial \o}(\o^*)\neq0$, there exists an open subinterval $I^*$ around $\o^*$ such that $u(\o)\neq0$ for all $\o\neq\o^*\in I^*$. Pick a rational number in $I^*$. Since the intersection of two such subintervals are empty (if not, there exists e sequence $\{\o_n\}$, $u(\o_n)=0$ and $\o_n\to\o^*$. By Mean Value Theorem, there exists a sequence $\{\nu_n\}$, $\o_n<\nu_n<\o_{n+1}$, such that $\frac{\partial u}{\partial \o}(\nu_n)=0$.  Since $\nu_n\to\o^*$, and $\frac{\partial u}{\partial \o}(\nu_n)=0$, by continuity, $\frac{\partial u}{\partial \o}(\o^*)=0$, that contradicts the choice of $\o^*$), every member of $I$ is in one of these subinterval. Hence $I$ is countable. 

If $u>0$ or $<0$, then it is trivial that $\Delta\abs{u}=\abs{\Delta u}$. Suppose that $u(\o^*)=0$ and $\dis\frac{\partial u}{\partial \o}(\o^*)=0$. 
Then $u(\o)=(\o-\o^*)^2 v(\o)$ for some function $v$.
Then 
\be{Del}
\Delta u(\o)=2v(\o)+(\o-\o^*)^2\Delta v(\o)+4(\o-\o^*)\dis\frac{\partial v}{\partial \o}(\o).
\ee

On the other hand, 
\[\dis\frac{d}{d\o}\abs{u}(\o)=\dis\left\{\begin{array}{ccc}\abs{2(\o-\o^*)v(\o)+(\o-\o^*)^2\dis\frac{\partial v}{\partial \o}(\o)} &  & v(\o)\neq0 \\0 &  & v(\o)=0\end{array}\right.\]
Therefore,
\be{DelAbs}
\Delta \abs{u}({\o})=\dis\left\{\begin{array}{ccc}\abs{2v(\o)+(\o-\o^*)^2\Delta v(\o)+4(\o-\o^*)\dis\frac{\partial v}{\partial \o}(\o)} &  & v(\o)\neq0 \\\dis\lim_{\nu\to\o}\dis\frac{1}{\nu-\o}\abs{2(\o-\o^*)v(\o)+(\o-\o^*)^2\dis\frac{\partial v}{\partial \o}(\o)} &  & v(\o)=0\end{array}\right.
\ee
Hence, by computing (\ref{Del}) and (\ref{DelAbs}) at $\o=\o^*$, we get:
 \[\Delta \abs{u}(\o^*)=\abs{2v(\o^*)}=\abs{\Delta u(\o^*)}.\] 
\ep

Now we are ready to prove Lemma \ref{claim2:thm:npde:1d}:
\bp 
By definition of $M^+_{\mathcal{V}}$ induced by $\|\cdot\|_{1,\phi, Q}$, we have:
\[
M^+_{\mathcal{V}}[\mathcal{A}+\Lambda^{(d)}]=\dis\sup_{u\in Y^{(d)}}\dis\lim_{h\to0^+}\frac{1}{h}\lt\{\dis\frac{\dis\sum_i q_i\dis\int_{\Omega}\phi(\o)\abs{u_i+hd_i(\Delta+\l^{(d)}_1)u_i(\o)}\;d\o}{\dis\sum_iq_i\dis\int_{\Omega}\phi(\o)\abs{u_i(\o)}\;d\o}-1\rt\},
\]
it is enough to show that for a fixed $u\neq0\in Y^{(d)}$ and a fixed $i=1,\ldots, n$:
\be{ineq1:claim2}
\dis\lim_{h\to0^+}\frac{1}{h}\lt\{{\dis\int_{\Omega}\phi(\o)\abs{u_i(\o)+hd_i(\Delta+\l^{(d)}_1) u_i(\o)}\;d\o}-{\dis\int_{\Omega}\phi(\o)\abs{u_i}\;d\o}\rt\}=0.
\ee
Or equivalently, after dividing by $d_i{\dis\int_{\Omega}\phi(\o)\abs{u_i}\;d\o}$, (note that if $d_i=0$, then the left hand side of (\ref{ineq1:claim2}) is zero, so we assume that $d_i\neq0$) and renaming $d_ih$ as $h$, and dropping $i$, we need to show that:
\be{ineq2:claim2}
\dis\lim_{h\to0^+}\frac{1}{h}\lt\{\dis\frac{\dis\int_{\Omega}\phi(\o)\abs{u(\o)+h(\Delta+\l^{(d)}_1) u(\o)}\;d\o}{\dis\int_{\Omega}\phi(\o)\abs{u}\;d\o}-1\rt\}=0.
\ee
 Let $I$ be as in Lemma \ref{Delta_abs}: the set of points of $\Omega$ such that for any $\o\in I$, $u(\o)=0$ and $\nabla u(\o)\neq0$. 

To show (\ref{ineq2:claim2}), we add and subtract $\phi(\o)\lt(\abs{u}+h\Delta\abs{u}+\l^{(d)}_1\abs{u}\rt)$ in the integral of the numerator of the left hand side of (\ref{ineq2:claim2}), and get:
\be{ineq3:claim2}
\bal
&\dis\lim_{h\to0^+}\frac{1}{h}\lt\{\dis\frac{\dis\int_{\Omega}\phi(\o)\abs{u+h(\Delta+\l^{(d)}_1) u}\;d\o}{\dis\int_{\Omega}\phi(\o)\abs{u}\;d\o}-1\rt\}\\
&=\dis\lim_{h\to0^+}\frac{1}{h}\lt\{\dis\frac{\dis\int_{\Omega}\phi(\o)\lt(\abs{u}+h(\Delta+\l^{(d)}_1) \abs{u}\rt)\;d\o}{\dis\int_{\Omega}\phi(\o)\abs{u}\;d\o}-1\rt\}\\
&\qquad+\dis\lim_{h\to0^+}\frac{1}{h}\lt\{\dis\frac{\dis\int_{\Omega}\phi(\o)\lt(\abs{u+h(\Delta+\l^{(d)}_1) u}-\abs{u}-h(\Delta+\l^{(d)}_1)\abs{u}\rt)\;d\o}{\dis\int_{\Omega}\phi(\o)\abs{u}\;d\o}\rt\}
\eal
\ee
First, we show that the first term of the right hand side of (\ref{ineq3:claim2}) is $0$.
By Divergence Theorem and Dirichlet boundary conditions, we have (recall that $\phi=\phi_1^{(d)}$):
 \beqn
 \dis\int_{\Omega}  \phi_1^{(d)} \Delta\abs{u} 
&=& \dis\int_{\partial\Omega} \phi_1^{(d)}\nabla \abs{u}\cdot \mathbf{n} - \int_{\Omega} \nabla\abs{u}\cdot\nabla\phi_1^{(d)} \quad\mbox{($\phi_1=0$ on $\partial\Omega$)}\\
&=& -\dis\int_{\partial\Omega} \nabla\phi_1^{(d)}\abs{u}\cdot \mathbf{n} +  \int_{\Omega}  \abs{u}\Delta\phi_1^{(d)} \quad\mbox{($u=0$ on $\partial\Omega$)}\\
&=& \int_{\Omega}  \abs{u}\Delta\phi_1^{(d)} 
= -\int_{\Omega}  \abs{u}\l_1^{(d)}\phi_1^{(d)}\\
\eeqn
 Therefore,
 \[\dis\int_{\Omega}\phi(\o)\lt(\l_1^{(d)}+\Delta\rt)\abs{u}(\o)\;d\o=0,\] 
and so:
\beqn
&&\dis\lim_{h\to0^+}\frac{1}{h}\lt\{\dis\frac{\dis\int_{\Omega}\phi(\o)\lt(\abs{u}+h(\Delta+\l^{(d)}_1)\abs{u}\rt)\;d\o}{\dis\int_{\Omega}\phi(\o)\abs{u}\;d\o}-1\rt\}=0.
\eeqn

Next, we show that the second term of the right hand side of (\ref{ineq3:claim2}) is $0$:
\be{ineq4:claim2}
\dis\lim_{h\to0^+}\frac{1}{h}\lt\{\dis\frac{\dis\int_{\Omega}\phi(\o)\lt(\abs{u+h\lt(\Delta+\l^{(d)}_1\rt) u}-\abs{u}-h\lt(\Delta+\l^{(d)}_1\rt) \abs{u}\rt)\;d\o}{\dis\int_{\Omega}\phi(\o)\abs{u}\;d\o}\rt\}=0.
\ee
In this part, we drop the superscript $(d)$ for the ease of notation: $\l_1=\l_1^{(d)}.$
For the fixed $u\in Y^{(d)}$, we define $F_h$, for any $0<h$, as follows:
\be{F_h}
F_h(\o)\;:=\;\dis\frac{1}{h}\lt\{\phi(\o)\lt(\abs{u+h(\Delta+\l_1) u}-\abs{u}-h(\Delta+\l_1) \abs{u}\rt)(\o)\rt\}.\nonumber
\ee
\ben
\item First, we will show that there exist $M>0$ such that for all $h$ positive, $\abs{F_h}<M$ almost everywhere:

 We study $F_h$, for any $0<h$, on the following possible subsets of $W:=\Omega\setminus I$:
 \begin{itemize}[leftmargin=*]
 \item $W_1:=\{\o:u(\o)>0,\; (\Delta+\l_1) u(\o)\geq0\}$. By definition, 
 \[F_h(\o)=\dis\frac{\phi(\o)}{h}\lt(u+h(\Delta+\l_1) u-u-h(\Delta+\l_1) u\rt)(\o)=0.\]
 \item $W_2:=\{\o:u(\o)<0,\; (\Delta+\l_1) u(\o)\leq0\}$. By definition, 
 \[F_h(\o)=\dis\frac{\phi(\o)}{h}\lt(-u-h(\Delta+\l_1) u+u+h(\Delta+\l_1) u\rt)(\o)=0.\]
  \item $W_3:=\{\o:u(\o)>0,\; (\Delta+\l_1) u(\o)<0, \; u>\abs{(\Delta+\l_1) u}h\}$. By definition, 
 \[F_h(\o)=\dis\frac{\phi(\o)}{h}\lt(u+h(\Delta+\l_1) u-u-h(\Delta+\l_1) u\rt)(\o)=0.\]
 \item $W_4:=\{\o:u(\o)<0, \;\Delta u(\o)>0,\; \abs{u}>{(\Delta+\l_1) u}h\}$. By definition, 
 \[F_h(\o)=\dis\frac{\phi(\o)}{h}\lt(-u-h(\Delta+\l_1) u+u+h(\Delta+\l_1) u\rt)(\o)=0.\]
 \item $W_5:=\{\o:u(\o)=0, u_{\o}(\o)=0\}$. In this case, by definition of $\Delta\abs{u}$, $\Delta\abs{u}(\o)=\abs{\Delta u(\o)}$. Therefore, $F_h(\o)=0$.
  \item $W_6:=\{\o:u(\o)>0, \; (\Delta+\l_1) u(\o)<0,\; u<\abs{(\Delta+\l_1) u}h\}$. By definition, 
 \[F_h(\o)=\dis\frac{\phi(\o)}{h}\lt(-u-h(\Delta+\l_1) u-u-h(\Delta+\l_1) u\rt)(\o).\]
 Using the triangle inequality and the assumption $u<\abs{(\Delta+\l_1) u}h$, we get:
 \be{M:bound}
 \bal
 \abs{F_h}&<\dis\frac{2}{h}\dis\max_{\Omega}\abs{\phi}\lt(\abs{u}+h\abs{(\Delta+\l_1) u}\rt)\\
 &< 4\dis\max_{\Omega}\abs{\phi} \abs{(\Delta+\l_1) u}\\
 &\leq4\dis\max_{\Omega}\abs{\phi}\lt(\max_{\Omega}\abs{\Delta u}+\l_1\max_{\Omega}\abs{u}\rt)=:M.
 \eal
 \ee
 (Note that, without loss of generality, we assume that $M\neq0$; otherwise, $u=0$. Therefore $F_h=0$ on $\Omega$.)
 \item $W_7:=\{\o:u(\o)<0, \;(\Delta+\l_1) u(\o)>0,\; \abs{u}<h{(\Delta+\l_1) u}\}$. Similar to the previous case, $\abs{F_h}<M.$
\ei
  \item Next, we will show that as $h\to 0$, $F_h\to 0$ almost everywhere.
 Fix $\o\in \Omega\setminus I$ and consider the following cases:
 \bi[leftmargin=*]
 \item $u(\o)>0$. We can choose $h$ small enough, such that 
 \[\abs{u(\o)+h(\Delta+\l_1)u(\o)}=u(\o)+h(\Delta+\l_1) u(\o).\] 
 Therefore, 
 \[F_h(\o)=\frac{1}{h}\phi(\o)(u(\o)+h(\Delta+\l_1) u(\o)-u(\o)-h(\Delta+\l_1) u(\o))=0.\]
 \item $u(\o)<0$. We can choose $h$ small enough, such that 
 \[\abs{u(\o)+h(\Delta+\l_1) u(\o)}=-u(\o)-h(\Delta+\l_1) u(\o).\] 
 Therefore, 
 \[F_h(\o)=\frac{1}{h}\phi(\o)(-u(\o)-h(\Delta+\l_1) u(\o)+u(\o)+h(\Delta+\l_1) u(\o))=0.\]
 \item $u(\o)=0$. Then as we discussed before, on $W_5$, $F_h(\o)=0$.
 \ei
 \een
 Using $1$ and $2$, and the Dominated Convergence Theorem, we can conclude (\ref{ineq4:claim2}). 
 \ep

%------------------------------------------------------------------------------------------------
\smallskip
\subsection*{Proof of Lemma \ref{M+:reaction} } 
%------------------------------------------------------------------------------------------------

By the definition of $c:=M_V[G]$, we have 
\[
\lim_{h\to0^+}\frac{1}{h}\sup_{x\neq y\in V}\left(\frac{\normpQ{x-y+h(G(x)-G(y))}}{\normpQ{x-y}}-1\right)= c.
\]
Fix an arbitrary $\epsilon>0$. Then there exists $h_0>0$ such that for all $0<h<h_0$, 
\[\frac{1}{h}\sup_{x\neq y\in V}\left(\frac{\normpQ{x-y+h(G(x)-G(y))}}{\normpQ{x-y}}-1\right)<c+\epsilon.
\]
Therefore, for any $x\neq y$, and $0<h<h_0$
\begin{equation}\label{h0}
\frac{\normpQ{x-y+h(G(x)-G(y))}}{\normpQ{x-y}}<(c+\epsilon)h+1.
\end{equation}
For fixed $u\neq v\in Y^{(d)}$, let $\Omega_1=\{\omega\in\bar\Omega:u(\omega)\neq v(\omega)\}$. 
Fix $\omega\in\Omega_1$, and let $x=u(\omega)$ and $y=v(\omega)$. We give a proof for the case $p<\infty$; the case $p=\infty$ is analogous. Using equation $(\ref{h0})$, we have:
\begin{equation}\label{normdef}
\displaystyle\frac{\left(\displaystyle\sum_iq_i^p\abs{u_i(\omega)-v_i(\omega)+h(G_i(u(\omega))-G_i(v(\omega)))}^p\right)^{\frac{1}{p}}}{\left(\displaystyle\sum_iq_i^p\abs{u_i(\omega)-v_i(\omega)}^p\right)^{\frac{1}{p}}}<(c+\epsilon)h+1.
\end{equation}
Multiplying both sides by the denominator and raising to the power $p$, we have:
\begin{equation}\label{forw}
\displaystyle\sum_iq_i^p\left|{(u_i-v_i)(\omega)+h\left(G_i(u)-G_i(v)\right)(\omega)}\right|^p<((c+\epsilon)h+1)^p\displaystyle\sum_iq_i^p\abs{(u_i-v_i)(\omega)}^p.
\end{equation}
Since $\tilde{G}(u)(\omega)=G(u(\omega))$, Equation (\ref{forw}) can be written as:
\begin{equation}\label{ftilde}
\displaystyle\sum_iq_i^p\left|{(u_i-v_i)(\omega)+h\left(\tilde{G}_i(u)-\tilde{G}_i(v)\right)(\omega)}\right|^p<((c+\epsilon)h+1)^p\displaystyle\sum_iq_i^p\abs{(u_i-v_i)(\omega)}^p.
\end{equation}
Now by multiplying both sides of the above inequality by $\phi_1(\o)$ which is nonnegative, and taking the integral over $\bar\Omega$, we get:
\[\|{u-v+h\left(\tilde{G}(u)-\tilde{G}(v)\right)}\|_{p,\phi,Q}<((c+\epsilon)h+1)\|{u-v}\|_{p,\phi,Q}.
\]  
(Note that for $\omega\notin\Omega_1$, 
\[((c+\epsilon)h+1)^p\displaystyle\sum_iq_i^p\abs{u_i(\omega, t)-v_i(\omega, t)}^p=0\] which we can add to the right hand side of $(\ref{ftilde})$, and also 
\[
\displaystyle\sum_iq_i^p\abs{u_i(\omega)-v_i(\omega)+h(G_i(u(\omega))-G_i(v(\omega)))}^p=0
\]
which we can add to the left hand side of $(\ref{ftilde})$, and hence we can indeed take the integral over all $\bar\Omega$.)

Hence,
\[\displaystyle\lim_{h\to0^+}\frac{1}{h}\left(\frac{\lt\|{u-v+h\left(\tilde{G}(u)-\tilde{G}(v)\right)}\rt\|_{p,\phi,Q}}{\|{u-v}\|_{p,\phi,Q}}-1\right)\leq c+\epsilon.
\]
Now by letting $\epsilon\to0$ and taking $\sup$ over $u\neq v\in Y^{(d)}$, we get $M^+_{V}[\tilde{G}]\leq c$.

%------------------------------------------------------------------------------------------------
\smallskip
\subsection*{Proof of Lemma \ref{block_diagonal_formula_M_p}} 
%------------------------------------------------------------------------------------------------

By the definition, for $p\neq\infty$, $M_*[A]$ can be written as follows:   
%{\scriptsize{
\beqn
M_*[A]=\dis\sup_{e\neq0}\lim_{h\to0^+}\dis\frac{1}{h}\lt\{\lt(\dis\frac{\dis\sum_{i=1}^m\|(I+hA_i)e_i\|^p}{\dis\sum_{i=1}^m\|e_i\|^p}\rt)^{\frac{1}{p}}-1\rt\}.
\eeqn
%}}
For a fixed $e=(e_1^T,\ldots, e_m^T)^T\neq0$, there exists some $k\in\{1,\ldots, m\}$, depends on $e$, such that for all $i\in\{1,\ldots, m\}$ 
\beqn
\|(I+hA_i)e_i\| \leq \dis\frac{\|(I+hA_k)e_k\|}{\|e_k\|}\|e_i\|,
\eeqn
by taking $\sum$ over all $i$'s, we get
\beqn
\dis\sum_{i=1}^m\|(I+hA_i)e_i\|^p \leq \frac{\|(I+hA_k)e_k\|^p}{\|e_k\|^p}\dis\sum_{i=1}^m\|e_i\|^p.
\eeqn
Therefore
\be{}\nonumber
\bal
\dis\lim_{h\to0^+}\dis\frac{1}{h}\lt\{\lt(\dis\frac{\dis\sum_{i=1}^n\|(I+hA_i)e_i\|^p}{\dis\sum_{i=1}^n\|e_i\|^p}\rt)^{\frac{1}{p}}-1\rt\}
&\leq \lim_{h\to0^+}\dis\frac{1}{h}\lt\{\dis\frac{\|(I+hA_k)e_k\|}{\dis\|e_k\|}-1\rt\}\\
&\leq M[A_k]
\leq\max\lt\{M[A_1], \ldots, M[A_m]\rt\}.
\eal
\ee
Now by taking $\sup$ over all $e\neq0$, we get the desired result.

For $p=\infty$, 
\beqn
M_*[A]=\dis\sup_{e\neq0}\lim_{h\to0^+}\dis\frac{1}{h}\lt\{\dis\frac{\dis\max_i\|(I+hA_i)e_i\|}{\dis\max_i\|e_i\|}-1\rt\}.
\eeqn
Note that 
\beqn
\dis\frac{\dis\max_i\|(I+hA_i)e_i\|}{\dis\max_i\|e_i\|} =\dis\max_i \dis\frac{\|(I+hA_i)e_i\|}{\dis\max_i\|e_i\|}
\leq\dis\max_i \dis\frac{\|(I+hA_i)e_i\|}{\|e_i\|}\\
\eeqn
Therefore, 
\beqn
\dis\frac{\dis\max_i\|(I+hA_i)e_i\|}{\dis\max_i\|e_i\|}-1 
\leq\dis\max_i \dis\frac{\|(I+hA_i)e_i\|}{\|e_i\|}-1
=\dis\max_i\left\{\dis\frac{\|(I+hA_i)e_i\|}{\|e_i\|}-1\right\}\\
\eeqn
dividing both sides by $h>0$, taking $\lim$ as $h\to0^+$, and taking $\sup$ over all $e\neq0$, we get 
\beqn
M_*[A] &\leq& \dis\sup_{e\neq0} \max_i\lim_{h\to0^+}\dis\frac{1}{h}\left\{\dis\frac{\|(I+hA_i)e_i\|}{\|e_i\|}-1\right\}\\
&=& \dis \max_i\sup_{e\neq0}\lim_{h\to0^+}\dis\frac{1}{h}\left\{\dis\frac{\|(I+hA_i)e_i\|}{\|e_i\|}-1\right\}\\
&=&\max\lt\{M[A_1], \ldots, M[A_m]\rt\}.
\eeqn

%------------------------------------------------------------------------------------------------
\smallskip
\subsection*{Another proof of Theorem \ref{thm-1D-synch} }
%------------------------------------------------------------------------------------------------

 Proof by discretization:

Let $0=\omega_0<\omega_1<\ldots<\omega_{N+1}=1$ be the mesh points of the closed interval $[0,1]$ with equal mesh size $\Delta \o=\dis\frac{1}{N+1}$. For $i=0,\ldots, N+1$, define
 \[x_i(t):=u(\omega_i,t),\]
 By the Neumann boundary condition, we have:
 \[0=u_{\omega}(0,t)\simeq\dis\frac{u(\omega_1,t)-u(\o_0,t)}{\Delta\o}\quad\Rightarrow\quad u(\omega_1,t)=u(\o_0,t),\]
where $u_{\o}=\dis\frac{\partial u}{\partial \o}$. Therefore for any $t$:
 \[\quad x_0(t)=x_1(t),\]
 and similarly 
 \[x_N(t)=x_{N+1}(t).\]
 Now using the definition of $u_{\o\o}$, at mesh points:
 \be{u_ww}
 \bal
 u_{\o\o}(\o_i,t)&=&\dis\lim_{\Delta\o\to0}\dis\frac{u(\o_{i-1},t)-2u(\o_i,t)+u(\o_{i+1},t)}{{\Delta\o}^2}\\
 &=&\dis\lim_{N\to\infty}(N+1)^2\lt(x_{i-1}-2x_i+x_{i+1}\rt)(t),
 \eal
 \ee
 we can write Equation (\ref{re-di}) for the mesh points as follows:
 \be{1D-discretization_sys}
\bal
\dot{x}_1\;&=\;F(x_1)+(N+1)^2D(x_2-x_1)\\
\dot{x}_2\;&=\;F(x_2)+(N+1)^2D(x_1-2x_2+x_3)\\
\vdots\\
\dot{x}_N\;&=\;F(x_N)+(N+1)^2D(x_{N-1}-x_N)
\eal
\ee
By Proposition \ref{thm:ode:linear}, if $c_N:=\dis\sup_{(x,t)} M_{1}\lt[J_{F_t}(x)-4(N+1)^2\sin^2\lt(\dis\frac{\pi}{2N}\rt) D\rt]$, where $-4\sin^2\lt(\dis\frac{\pi}{2N}\rt)$ is the smallest nonzero eigenvalue of (graph) Laplacian of linear graph, then 
\be{1D-discretization_ineq}
\dis\sum_{k=1}^{N-1}\sin{\lt(\dis\frac{k\pi}{N}\rt)}\|(x_k-x_{k+1})(t)\|_{1}\leq e^{c_Nt}\dis\sum_{k=1}^{N-1}\sin{\lt(\dis\frac{k\pi}{N}\rt)}\|(x_k-x_{k+1})(t)\|_{1}. 
\ee

Now divide both sides of (\ref{1D-discretization_ineq}) by $\Delta\omega=\dis\frac{1}{N+1}$ and let $N\to\infty$, we get:
\be{1D_ineq}
\dis\int_{0}^{1}\sin{(\pi \o)}\abs{\dis\frac{\partial u}{\partial \o}(t)}\;d \o\leq e^{ct}\dis\int_{0}^{1}\sin{(\pi \o)}\abs{\dis\frac{\partial u}{\partial \o}(t)}\;d\o,
\ee
where 
\[
\begin{array}{lcl}
\dis\lim_{N\to\infty}c_N&=&\dis\lim_{N\to\infty}\sup_{(x,t)}M_{1}\lt[J_{F_t}(x)-4(N+1)^2\sin^2\lt(\dis\frac{\pi}{2N}\rt) D\rt]\\
&=&\dis\sup_{(x,t)}M_{1}\lt[J_{F_t}(x)-4\lim_{N\to\infty}(N+1)^2\sin^2\lt(\dis\frac{\pi}{2N}\rt) D\rt]\\
&=&\dis\sup_{(x,t)}M_{1}\lt[J_{F_t}(x)-\pi^2D\rt]\\
&=&c.
\end{array}
\]

%------------------------------------------------------------------------------------------------

\end{document}